\documentclass[journal]{IEEEtran}

\usepackage{cite}
\usepackage[pdftex]{graphicx}
\usepackage{mathrsfs}
\usepackage{amsmath}
\usepackage[square, comma, sort&compress, numbers]{natbib}
\usepackage[ruled]{algorithm2e}
\usepackage{algpseudocode}
\usepackage{algpseudocode}

\usepackage{enumerate}
\usepackage{multirow}

\usepackage{amsthm}
\usepackage{amsmath} 
\usepackage{amssymb}
\renewcommand{\vec}[1]{\mathbf{#1}} 
\usepackage{color}
\usepackage[square,numbers,sort&compress]{natbib}

\usepackage{array}
\newcommand{\PreserveBackslash}[1]{\let\temp=\\#1\let\\=\temp}
\newcolumntype{C}[1]{>{\PreserveBackslash\centering}p{#1}}
\newcolumntype{R}[1]{>{\PreserveBackslash\raggedleft}p{#1}}
\newcolumntype{L}[1]{>{\PreserveBackslash\raggedright}p{#1}}

\usepackage{marvosym}
\usepackage{CJK}
\usepackage{amsfonts}
\usepackage{subfigure}

\usepackage{multirow}

\newtheorem{myDef}{Definition}

\newtheorem{myproof}{Proof}[section]

\newtheorem{mypro}{Proposition}[section]
\newtheorem{proofintuition}{Proof Intuition}[section]

\usepackage[breaklinks,colorlinks,linkcolor=green,anchorcolor=black,citecolor=blue,urlcolor=black]{hyperref}

\begin{document}

\title{Privacy-Preserving Machine Learning Training\\in IoT Aggregation Scenarios}

\author{Liehuang Zhu,~\IEEEmembership{Member,~IEEE,}
        Xiangyun Tang,
        Meng Shen,~\IEEEmembership{Member,~IEEE,}\\
        Jie Zhang,
        Xiaojiang Du,~\IEEEmembership{Fellow Member,~IEEE}
\thanks{L. Zhu, X. Tang, M. Shen and J. Zhang are with School of Computer Science, Beijing Institute of Technology, Beijing, China. Email: \{liehuangz, xiangyunt, shenmeng, 3120181068\}@bit.edu.cn. Prof. Meng Shen is the corresponding author.}
\thanks{X. Du is with the Department of Computer and Information Sciences, Temple University, Philadelphia, USA. Email: dxj@ieee.org.}}

\maketitle

\begin{abstract}
To develop Smart City, the growing popularity of Machine Learning (ML) that appreciates high-quality training datasets generated from diverse IoT devices raises natural questions about the privacy guarantees that can be provided in such settings.
Privacy-preserving ML training in an aggregation scenario enables a model demander to securely train ML models with the sensitive IoT data gathered from personal IoT devices.
Existing solutions are generally server-aided, cannot deal with the collusion threat between the servers or between the servers and data owners, and do not match the delicate environments of IoT.
We propose a privacy-preserving ML training framework named \texttt{Heda} that consists of a library of building blocks based on partial homomorphic encryption (PHE) enabling constructing multiple privacy-preserving ML training protocols for the aggregation scenario without the assistance of untrusted servers and defending the security under collusion situations.
Rigorous security analysis demonstrates the proposed protocols can protect the privacy of each participant in the honest-but-curious model and defend the security under most collusion situations.
Extensive experiments validate the efficiency of \texttt{Heda} which achieves the privacy-preserving ML training without losing the model accuracy.
\end{abstract}


\section{Introduction}
\IEEEPARstart{I}{nternet}
of Things (IoT) plays a remarkable role in all aspects of our daily lives,
covering various fields including healthcare, industrial appliances, sports, homes, etc \cite{IoTsurvey1,IoTsurvey2}.
The large data collected from IoT devices with machine learning (ML) technologies have been accelerating Smart City step and improving our daily lives.
For a personal healthcare example,
fitness records monitored by wearable IoT sensors can be feeded to a ML model provided by medical research institutions, for self-rated health measurement.

ML features data-driven \cite{61}.
The comprehensive training dataset is one of the critical factors for accomplishing accurate ML models.
While the data collected from a single type of IoT devices is not comprehensive.
Or some companies (e.g., the medical research institutions), desiring to develop a ML model to provide the smart services to users, own no training data generated from IoT devices.
Mostly, the companies (model demanders) have to develop ML models on the training dataset gathered from multiple IoT data owners.

However, the personal IoT data contains users' sensitive information.
Privacy concerns and data protection laws surrounding data sovereignty and jurisdiction prevent IoT data owners from openly sharing users' data \cite{GDPR,115}.
Besides, the model demanders are unwilling to let others know about their personal models.
This paper focuses on ML training phase and targets at the setting where a model demander learns nothing but a final model from the gathered IoT dataset, and each IoT data owner cannot learn any useful thing after sharing a protected version data with the model demander.

\begin{table}[t]
  \centering
  \caption{Differences between Collaborative Scenario and Aggregation Scenario}\label{table:Differences between Scenarios}
  \footnotesize{
  \begin{tabular}{||c|p{6.3cm}||}
  \hline
  \textbf{Aspects} &\textbf{Differences Details}\\
  \hline
  \hline
  \multirow{3}{*}{\shortstack{Model\\Demander}}&\textbf{In aggregation scenarios}: the company that desires ML models but has no training data is the model demander.\\
  \cline{2-2}
   &\textbf{In collaborative scenarios}: every data owner is a model demander.\\
  \hline
  \hline
  \multirow{3}{*}{\shortstack{Computation\\Mode}}&\textbf{In aggregation scenarios}: the model demander interacts with the data owners for initiative training models.\\
  \cline{2-2}
   &\textbf{In collaborative scenarios}: the data owners cooperating with each other to training models, servers is auxiliary.\\
  \hline
  \hline
  \multirow{3}{*}{\shortstack{Privacy\\Guarantee}}&\textbf{In aggregation scenarios}: the model demander learns nothing but the model; the data owners learn neither the model nor the data of other data owners.\\
  \cline{2-2}
   &\textbf{In collaborative scenarios}: the data owners' datasets are confidential; the model is known by every participant.\\
  \hline
  \hline
  \end{tabular}}
\end{table}

To train ML models upon gathered datasets securely,
most existing works have explored the solutions under collaborative scenarios (e.g., Federated Learning \cite{FederatedLearning,40} and Multi-party Collaborative Learning \cite{112,8}),
where a set of data owners share protected versions of their data with each other and jointly train a global model over these data for themselves.
Our setting is different from them in some aspects.
To distinguish from them, we name our target setting as Aggregation Scenario,
the differences between the two scenarios are summarized in Table \ref{table:Differences between Scenarios}.
Specifically,
in most of solutions for collaborative scenarios, the model is known by every data owner, which ignores the security requirement of model demanders \cite{SALTDSC2020}.
PrivFL \cite{PrivFL} fits Federated Learning setting and considers model privacy.
But since it delegates most of heavy cryptographic computations to data owners,
and IoT data owners are generally resource-constrained and cannot support the heavy cryptographic computations generally.
PrivFL dose not match IoT setting.
Besides, even though a method provides a separate privacy-preserving solution for training one type of ML model in collaborative scenarios such as for
Support Vector Machine (SVM) \cite{8,28},
Linear Regression \cite{26},
Naive Bayesian (NB) \cite{30},
and K-means \cite{31},
other types of ML models remain unsolved, as well as the solutions for the aggregation scenario.

In our aggregation scenario,
IoT data owners share protected versions of their data with the model demander and only allow to leak the information of the final model.
To reduce the computing load at the IoT data owner side, the model demander shoulders most of the cryptographic computations.
During the whole training process, the model demander performs computations on ciphertext to learn the model.
Few works explored privacy-preserving ML training in the aggregation scenario.
These works generally introduce untrusted servers to facilitate secure training and cannot defend the security of collusion situations \cite{SALTDSC2020,115}.
Once model demanders collude with data owners \cite{SALTDSC2020} or the servers \cite{115}, the sensitive data in the training dataset will be leaked in these methods.
Privacy-preserving data aggregation is a kind of fundamental algorithm widely applied in IoT \cite{aggregationIoT,aggregationsmartgrid,aggregationsensingsystems}, which handles the scenario that resembles the aggregation scenario.
Since privacy-preserving data aggregation solutions just attain simple operations such as secure summation and secure maximum, it cannot cope with the complex computations in ML training algorithms directly.

Given the limitations of existing schemes,
designing a privacy-preserving ML training scheme in the aggregation scenario of IoT faces two main challenges.
\textit{First},
how to obtain the correct model without loss of accuracy on the premise of ensuring the security requirement in the aggregation scenario.
Without server-aid, the model demander is forced to only get in touch with the ciphertext version of training data throughout the training process for developing ML models.
While most ML training algorithms require iteratively updating model parameters using training datasets,
and updating the model parameters using the protected version of datasets directly tend to cause a accuracy loss in models.
\textit{Second},
how to guarantee the security when the model demander colludes with data owners or data owners collude with each other.
Most existing works limit the collusion situation such as prohibiting the collusion between the model demander and data owners, to meet the security requirements.
While each participant has the motives for collusion and inferring the sensitive information of other participants.
The complexity of the form of IoT privacy data in collusion situations increases the difficulty of privacy-preserving.

In this paper, we design a general framework that supports training three ML models,
Logistic Regression (LR), SVM and NB,
and allows the collusion between the model demander and IoT data owners and the collusion between data owners.
Although practical deep learning models have sprung up and achieved excellent results,
it cannot replace these traditional ML algorithms that are able to obtain accurate models from small training datasets \cite{DPvsML,DPvsML2,DPvsML3}.
And some deep learning applications employ traditional ML models (e.g., SVM and LR) for final prediction outputs, where neural networks are used to extract features \cite{DPSVM,DPSVM2}.

\begin{table*}[!t]
\centering
\caption{Summary of Existing Privacy-Preserving Machine Learning Schemes}\label{table:Summary of Existing Privacy-Preserving Machine Learning Schemes}
\footnotesize{
\begin{tabular}{||c|p{1.7cm}|c|c|c||}
\hline
ML Phases&\multicolumn{2}{c|}{Target Algorithms}&Solutions&Ref.\\
\hline
\hline
\multirow{6}{*}{Prediction}&\multicolumn{2}{c|}{Deep Learning}&Differential Privacy &\cite{40}\\
\cline{2-4}
&\multicolumn{2}{c|}{Neural Network}&Secure Multi-party Computation &\cite{58}\\
\cline{2-4}
&\multicolumn{2}{c|}{Linear Regression}&Homomorphic Encryption &\cite{26}\\
\cline{2-4}
&\multicolumn{2}{c|}{Decision Trees, SVM and LR}&Homomorphic Encryption and Secret Sharing &\cite{20}\\
\cline{2-4}
&\multicolumn{2}{c|}{Hyperplane Decision, NB, and Decision Trees}&Homomorphic Encryption and Secret Sharing  &\cite{2}\\
\cline{2-4}
\hline
\hline
\multirow{16}{*}{Training}&\multirow{8}{*}{\shortstack{Collaborative \\Scenario}}&Deep Learning&Differential Privacy &\cite{41}\\
\cline{3-4}
& &Decision Trees &Differential Privacy &\cite{relatedwork-MCS-DPtrainig}\\
\cline{3-4}
& &Linear Regression &Differential Privacy &\cite{105}\\
\cline{3-4}
& &SVM& Homomorphic Encryption &\cite{8}\\
\cline{3-4}
& &LR& Homomorphic Encryption &\cite{36}\\
\cline{3-4}
& &NB& Homomorphic Encryption &\cite{30}\\
\cline{3-4}
& &Linear Regression& Homomorphic Encryption &\cite{9}\\
\cline{3-4}
& &Linear Regression, LR and Neural Network&Garbled Circuit and Secret Sharing  &\cite{115}\\
\cline{2-5}
&\multirow{4}{*}{\shortstack{Aggregation \\Scenario}}&Aggregate Statistics&Differential Privacy &\cite{Introduction-aggregator-scenario-1,Introduction-aggregator-scenario-3}\\
\cline{3-4}
& &Multilayer Perceptron &Homomorphic Encryption &\cite{SALTDSC2020}\\
\cline{3-4}
& &Sums of vectors &Secret Sharing &\cite{Relatedwork-aggregator-scenario-ccs17}\\
\cline{3-4}
& &Quadratic Optimization Problem &Homomorphic Encryption &\cite{Relatedwork-aggregator-scenario-quadraticoptimization}\\
\cline{3-4}
& &k-means& Homomorphic Encryption &\cite{Relatedwork-aggregator-scenario-SASkmeans}\\
\hline
\end{tabular}}
\end{table*}

\noindent
\textbf{\textit{Contributions.}}
To this end,
we propose \texttt{Heda}, a general privacy-preserving framework supporting multiple ML training protocols that satisfy the security requirements in the aggregation scenario and the delicate environments of IoT.
A thorough security analysis is provided, demonstrating the security of the proposed protocols in the honest-but-curious model.
Extensive experiments on real-world datasets validate the proposed protocols achieve the privacy-preserving ML training without losing the
model accuracy.
The contributions of our paper can be summarized as follows.

\textbf{\textit{1.}}
In order to train ML models upon ciphertext without accuracy loss,
we design a library of building blocks,
based on the additively homomorphic encryption Paillier \cite{1} and multiplicative homomorphic encryption {\small Cloud-RSA} \cite{cloud-RSA}.
Inspired by the idea of permutations and combinations that complex algorithms can be decomposed into several primitive operations,
after identifying a set of core operations that underlie many ML training algorithms,
we carefully design a library of building blocks supporting each of these core operations.

\textbf{\textit{2.}}
To guarantee the security under the collusion situations and meet the security requirements in the aggregation scenario,
we design other two building blocks that enable an algorithm's output to become the input of another algorithm.
All building blocks are designed in a composable way and satisfy both functionality and security,
where the security of the combinations between building blocks is ensured by Modular Sequential Composition \cite{15}.

\textbf{\textit{3.}}
We instantiate \texttt{Heda} to three privacy-preserving ML training protocols: LR, SVM, and NB,
without the assistance of untrusted servers.
To the best of our knowledge, we are the first to solve the non-linear function in privacy-preserving LR training without any approximate equation.

The rest of our paper is organized as follows.
Section \ref{sec:related_work} provides our related work.
Section \ref{sec:System Overview} provides the system overview.
In Section \ref{sec:preliminary}, we describe the background of PHE.
Section \ref{sec:Building Blocks} details the building blocks of \texttt{Heda}.
Section \ref{sec:Privacy-Preserving Machine Learning Models Training with Heda} presents the three training protocols instantiated from \texttt{Heda}.
The evaluation results are provided in Section \ref{sec:evaluation}.
Section \ref{sec:conclusion} concludes this paper.

\section{Related Work}\label{sec:related_work}
We devote to the privacy-preserving ML training in the aggregation scenario,
while many studies exploring the privacy-preserving ML training fall in collaborative scenarios.
More details about the differences between the aggregation scenario and collaborative scenarios can be found in Appendix \ref{app:related_work-Problem Description}.

Our work is related to privacy-preserving ML which can be broadly divided into privacy-preserving prediction and privacy-preserving training.
A series of excellent works have been developed for privacy-preserving prediction \cite{40,2,20,58}.
Some representative works for privacy-preserving prediction are summarized in Table \ref{table:Summary of Existing Privacy-Preserving Machine Learning Schemes}.
In this section, we give the literature review of privacy-preserving training in the aggregation scenario and collaborative scenarios.

\textit{In collaborative scenarios.}
Collaborative scenarios often occurs when multiple organizations have similar types of data, and they want to train a more accurate model on their joint their data \cite{115,Introduction-18sec-GarbledCircuit,8,36}.
Mohassel et al. proposed a privacy-preserving training scheme based on Garbled Circuit and Secret Sharing where two untrusted non-colluding servers are introduced \cite{115}.
Many secure PHE-based algorithms have been developed for different specialized ML training algorithms such as SVM \cite{8}, LR \cite{36}, Linear Regression \cite{9} and NB \cite{30}.
In order to handle complex non-linear function, PHE-based schemes usually depend on the untrusted servers \cite{8}, and use an approximate equation to simplify the complex iteration formula into a simple one\footnote{$\log (\frac{1}{1+\exp (u)})\approx \sum\limits_{j=0}^{k}{a_j\cdot u^j}$} \cite{36,115}.
Gonzlez et al. \cite{8} developed a secure addition algorithm and a secure subtraction algorithm for constructing a secure SVM training algorithm, while some operations that are not supported by Paillier have to be implemented with the assistance of the untrusted servers.
Mandal et al. proposed PrivFL \cite{PrivFL}, which guarantees data and model privacya and fits Federated Learning setting.
But it delegates most of heavy cryptographic computations to data owners,
which cannot match IoT setting where IoT data owners are usually resource-constrained.

\textit{In aggregation scenarios.}
Several works explored privacy-preserving aggregate statistics employing DP in the aggregation scenario \cite{Introduction-aggregator-scenario-1,Introduction-aggregator-scenario-3}.
Training algorithms generally contain multiple iterations, thus noise-based DP is not a good choice for accuracy concern \cite{40,104}.
Although, many efforts have been done for making Fully Homomorphic Encryption (FHE) practical,
FHE is still unsuitable for general purpose applications \cite{Introduction-FHE-survey}.
Keith et al. \cite{Relatedwork-aggregator-scenario-ccs17} designed a protocol for secure aggregation of high-dimensional data.
Their protocol allows a server to compute the sum of user-held data vectors from mobile devices.
Shoukry et al. \cite{Relatedwork-aggregator-scenario-quadraticoptimization} considered a problem where multiple agents participate in solving a quadratic optimization problem.
They proposed a PHE based protocol finding the optimal solution, where the privacy of the proposed protocol was analyzed by the zero-knowledge proofs.
Mittal et al. \cite{Relatedwork-aggregator-scenario-SASkmeans} proposed a secure k-means data mining approach based on PHE.

Li et al. \cite{SALTDSC2020} proposed a server-aid framework for training multilayer perceptron in the aggregation scenarios.
They introduce an untrusted server to assist the secure training and ignore the collusion situation.
Once the untrusted server colludes with model demanders or data owners, the confidentiality of models and datasets is not guaranteed.
Mohassel et al. \cite{115} proposed a privacy-preserving machine learning scheme with two untrusted but non-colluding servers.
Their setting is similar to ours if we treat one server as the model demander and authorize it as the final model.
But when an untrusted server is treated as the model demander, it will collude with the other server and figure out the plaintext training dataset. 

Privacy-preserving data aggregation resembles the aggregation scenario,
which is a kind of fundamental algorithm widely applied in Internet of Things \cite{aggregationIoT}, smart grid \cite{aggregationsmartgrid}, wireless sensor networks \cite{aggregationsensingsystems}, etc.
Since privacy-preserving data aggregation achieves simple operations compared to ML training such as secure summation, it cannot handle the privacy-preserving ML training problem directly.

Although there are respectable studies on privacy-preserving training as listed in Table \ref{table:Summary of Existing Privacy-Preserving Machine Learning Schemes},
solutions for the aggregation scenario mainly rely on server-aid mode and cannot resist the threat of the collusion between the model demander and data owners.
In this paper, we propose a general privacy-preserving framework, which can construct multiple privacy-preserving ML training protocols (including LR without simplification) that can guarantee security under the collusion situations.

\begin{figure*}[!t]
\centering
    \centering
    \includegraphics[height=6cm]{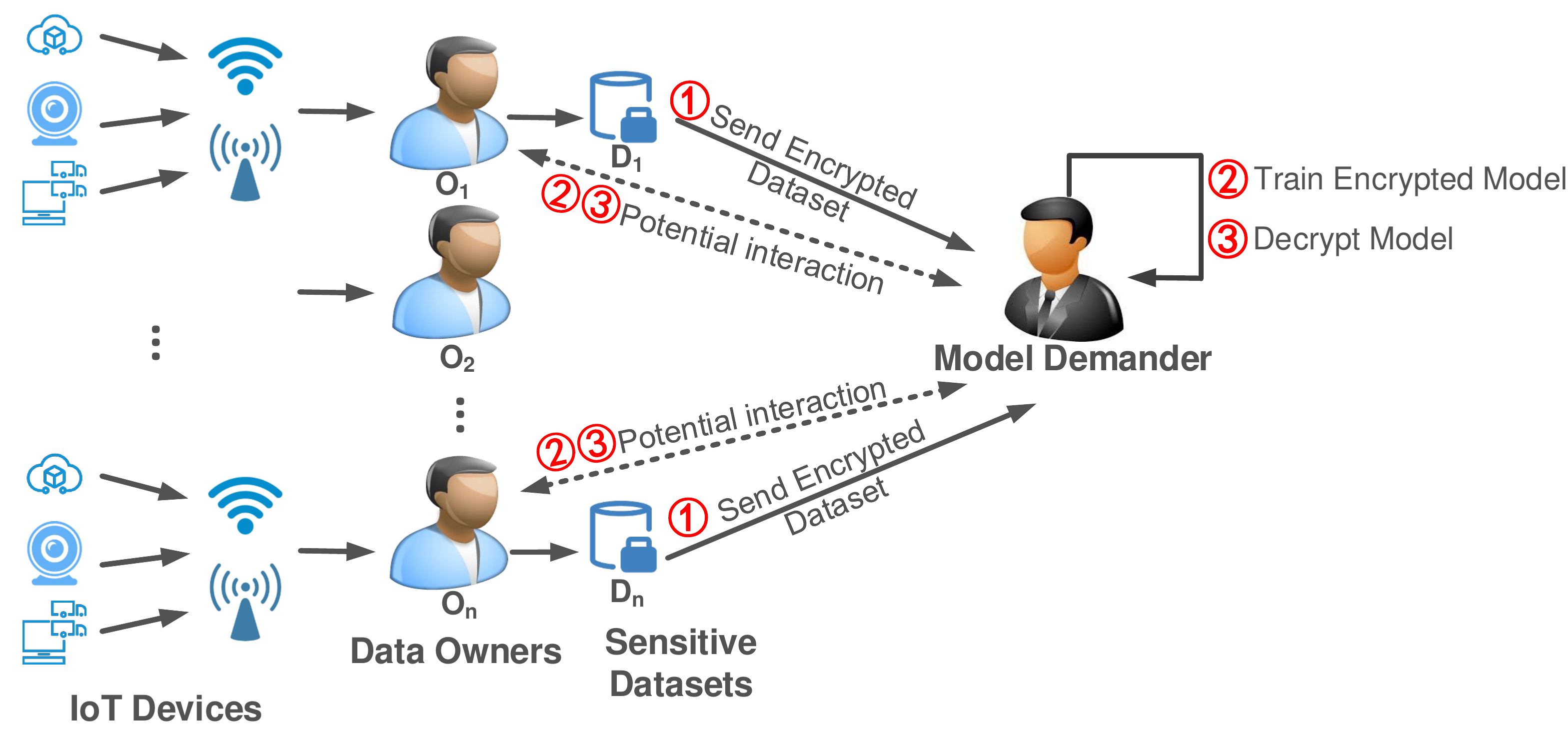}
\caption{Aggregation Scenario: a model demander learns the model only, and data owners neither know the model information nor other data owners' data.}
\label{fig:twoscenarios}
\end{figure*}

\section{System Overview}\label{sec:System Overview}
This section presents the system model, the threat model, and the security definitions used in this paper.

\subsection{System Model}\label{sec:System Model}
We envision a data-driven IoT ecosystem, shown in Fig. \ref{fig:twoscenarios}, including IoT devices, IoT data owners and a model demander:
\begin{itemize}
  \item \textbf{IoT devices} are responsible for sensing and transmitting valuable IoT data through wireless or wired networks.
  \item \textbf{Data owners} collect all pieces of the IoT data from the IoT devices within their own domains.
  \item \textbf{Model demander} wants to training a ML model upon the dataset gathered from multiple data owners.
\end{itemize}

We assume that all participants have agreed upon the system of \texttt{Heda} to jointly train a model, and each data owner consents to releasing the model to the server.
Every data owner generates its personal key pair to encrypted its data locally,
and the model demander also holds his own key pair for model encryption.

Formally, our system consists of $n$ data owners $\mathcal{O}_i$ $(i \in \{1,...,n\})$ and one untrusted model demander $\mathcal{A}$.
Each data owner $\mathcal{O}_i$ holds a dataset $\mathcal{D}_{i}$ that contains users' sensitive information.
This work considers \emph{horizontal data sharing} \cite{41,46,123}, that is, the $n$ datasets $\{\mathcal{D}_{i}\}_{i=1}^{n}$ share the same feature space but different in samples.
The model demander sequentially gathers the $n$ encrypted data,
and trains a ML model $\mathcal{M}$ upon the dataset $\mathcal{D}:= (\mathcal{D}_1 \cup ...\cup \mathcal{D}_n) $,
where $|\mathcal{D}| = \sum_{i=1}^n|D_i|$.
After executing a privacy-preserving training protocol $\mathcal{F}$ instantiated from \texttt{Heda},
the model demander obtains the desired model, i.e., the model parameters $\theta$.

\textbf{Security Goals:}
The privacy-preserving training protocol $\mathcal{F}$ satisfies the following security requirements:
\begin{itemize}
  \item The model demander cannot learn any sensitive information in the dataset $\mathcal{D}$.
  \item Every data owner cannot know the model parameters $\theta$.
  \item Each data owner $\mathcal{O}_i$ know nothing about other data owners' sensitive information.
\end{itemize}

\subsection{Threat Model}\label{sec:Threat Model}
All participants in our model do not trust one another.
Each data owner may try to learn as much other data owners' sensitive data and the model demander's model information as possible by honestly executing the pre-defined protocol.
And the model demander follows the protocol honestly, but it tries to infer data owners' sensitive data as much as possible from the values he learns.
Therefore, for any participant, we assume it is the passive (or honest-but-curious) adversary \cite{10}, that is,
it does follow the protocol, but it tries to infer other participants' privacy as much as possible from the values they learn.

The proposed building blocks (except secure summation) are in two-party computation setting which should be ensured secure in the honest-but-curious model.
While the privacy-preserving ML training protocols constructed from building blocks are multi-party protocols,
where participants may collude with each other to acquire more information for inferring other participants' privacy.
In our solution, we allow $(n-1)$ data owners at most collude with each other to steal the privacy of other participants,
and $(n-2)$ data owners at most collude with the model demander to steal the privacy of other participants.
Notice that when $n$ data owners collude with each other, they can compute the model results on their joint dataset straightforward,
and defending this extreme situation is meaningless.

Our model may encounter external adversaries that illegally obtain the data in the transmission process for their purposes by Internet eavesdropping or other means.
While external adversary can be controlled by setting up a confidential and authentic channel using existing common technology like TSL.

\noindent\textbf{Complementary directions.}
Since we assume every data owner agrees on the \texttt{Heda} to release the global model to the server,
\texttt{Heda} dose not consider information leakage from the model parameters at the server.
Employing DP could alleviate the information leakage, as the suggestion of existing works \cite{dpbased2,dpbased1}.
While there are multiple attacks against ML.
Protecting against these attacks is a complementary problem to that solved by \texttt{Heda}, and the corresponding solutions can be plugged into \texttt{Heda}.

\subsection{Security Definition}\label{sec:Security Definition}
To justify the security of two-party computation, we adapt Secure Two-party Computation framework \cite{2,9,10}.
To enable us to compose the building blocks into a privacy-preserving ML training protocol in a modular way securely, we invoke Modular Sequential Composition \cite{15}.

\textbf{Secure Two-party Computation.}
For two-party protocols, to ensure security,
we have to show that whatever A (B) can compute from its interactions with B (A) can be computed from its input and output,
which leads to a commonly used definition, i.e., secure two-party computation \cite{2,9,10}.
Let {\small$F=( {{f}_{A}},{{f}_{B}} )$} be a (probabilistic) polynomial function.
$\pi$ is a protocol computing $F$.
A and B want to compute {\small$F( a, b)$} where $a$ is A's input and $b$ is B's input.
The view of party A during the execution of $\pi$ is the tuple $view_{\text{A}}^{\pi }(a,b)=( a,r,{m}_{1},{{m}_{2}},...,m_{n})$ where ${m_{1}},{m_{2}},...,{{m}_{n}}$ are the messages received from B, $r$ is A's random tape.
The view of B is defined similarly.
Secure Two-party Computation is stated formally as follows:
\begin{myDef}[Secure Two-Party Computation \cite{10}]
\label{def:Secure Two-Party Computation}
A two-part protocol $\pi $ privately computes $f$ if for all possible inputs $(a,b)$ and simulators ${{S}_{A}}$ and ${{S}_{B}}$ hold the following properties:
\[S_{\text{A}}(a,{f_A}(a,b))\ {\equiv}_c\ view_{A}^{\pi }(a,b)\]
\[S_{\text{B}}(b,{f_B}(a,b))\ {\equiv}_c\ view_{B}^{\pi }(a,b)\]
where ${\equiv}_c $ denotes computational indistinguishability against Probabilistic Polynomial Time (PPT) adversaries with negligible advantage in the security parameter $\lambda$  \cite{1}.
\end{myDef}
More details of computational indistinguishability can be found in Appendix \ref{app:Security Definition}.

\textbf{Modular Sequential Composition.}
Since all our protocols are designed and constructed in a modular way, we employ Modular Sequential Composition \cite{15} for justifying the security proofs of our protocols, the detailed idea of which can be found in Appendix \ref{app:Security Definition}.

\begin{myDef}[Modular Sequential Composition \cite{15}]
\label{theorem:Modular sequential composition}
Let $f_1,\ldots,f_n$ be two-party probabilistic polynomial time functionalities and $\rho_1,\ldots,\rho_n$ protocols that securely compute respectively $f_1,\ldots,f_n$ in the presence of semi-honest adversaries.
Let $F$ be a probabilistic polynomial time functionality and $\pi$ a protocol that securely computes $F$ with $f_1,\ldots,f_n$ in the presence of semi-honest adversaries.
Then ${{\pi }^{{{\rho }_{1}},{{\rho }_{2}},\ldots,{{\rho }_{n}}}}$ securely computes $F$ in the presence of semi-honest adversaries.
\end{myDef}

\section{Preliminary}\label{sec:preliminary}
This section presents the PHE preliminaries used in this paper.
ML preliminaries used in this paper can be found in Appendix \ref{app:Machine Learning Preliminary}.
Before presenting the main content, we summarize the notations used in paper as follows.
A dataset $\mathcal{D}$ is an unordered set with the size of $|\mathcal{D}|$.
$x$ denotes a scalar.
{\small$\vec{x}\in {\mathbb{R}}^d$} denotes a vector.
{\small$\vec{x}_i=[ x_{i1},x_{i2},\ldots,x_{id} ]$} is the i-th record in dataset $\mathcal{D}$.
Each record has $d$ features.
${y}_i$ is the class label correspond to $\vec{x}_i$.
{\small$X = \{\vec{x}_1,\vec{x}_2,\ldots,\vec{x}_{|\mathcal{D}|} \}$}, {\small$Y = \{y_1,y_2,\ldots,y_{|\mathcal{D}|} \}$}.
$\theta$ are the learnable parameters of ML models.
{\small$[m]_P$} denotes a message $m$ encrypted by Paillier.
{\small$[m]_R$} denotes a message $m$ encrypted by {\small Cloud-RSA}.
{\small$[m]$} denotes an encrypted message $m$ that is encrypted with a non-specified cryptosystems.
Important notations are present in Table \ref{table:List of Notations}.

\begin{table}[hbt]
\centering
\caption{List of Notations}\label{table:List of Notations}
\footnotesize{
\begin{tabular}{r|l}
\hline
Notation&Meaning\\
\hline
\hline
$n$ & the number of data owners participating in scenario\\
$\mathcal{D}_i$   & dataset from the i-th data owner\\
$\mathcal{D}$    & training dataset gathered from multiple data owners\\
$d$& the number of features in $\mathcal{D}$ \\
$\vec{x}_i$ & the i-th record in training dataset $D$\\
$y_i$   & the class label correspond to $\vec{x}_i$\\
$\theta$ & learnable parameters of ML models\\
$[m]_P$ & a message encrypted by Paillier\\
$[m]_R$ & a message encrypted by {\footnotesize Cloud-RSA}\\
$[m]$   & a message encrypted by a non-specified cryptosystems\\
\hline
\end{tabular}}
\end{table}

Public-key cryptosystems employ a pair of keys ($\sf{PK}$, $\sf{SK}$), the public key ($\sf{PK}$, the encryption key) and the private key ($\sf{SK}$, the decryption key).
Some cryptosystems are gifted with a homomorphic property that can perform a set of operations on encrypted data without knowledge of the decryption key.
Formalized definition of homomorphic can be found in Appendix \ref{app:Homomorphic Definition}.

Two public-key cryptosystems are employed in this paper: Paillier \cite{1} and {\small Cloud-RSA} \cite{cloud-RSA}.
Paillier possesses additively homomorphic properties, and Cloud-RSA possesses multiplicative.
Ciphertext indistinguishability against chosen plaintext attacks \cite{1} ensures that no bit of information is leaked from ciphertexts.
We prove the security of the proposed algorithms and protocols based on the semantic security \cite{1} of Paillier and {\small Cloud-RSA}.
Let $\sf{GenModulus}$ be a polynomial-time algorithm that, on input $1^n$, outputs $(N, p, q)$ where $N = pq$ and $p$ and $q$ are n-bit primes \cite{1}.
We restate the definitions of Paillier and {\small Cloud-RSA} as following.

\textbf{Paillier.}
The security of Paillier is based on the Decisional Composite Residuosity assumption.
\begin{itemize}
  \item $\sf{Gen}$: run $\sf{GenModulus}(1^n)$ to obtain $(N, p, q)$.
The public key is {\small$N$}, and the private key is {\small$( N, \phi (N))$}.
  \item $\sf{Enc}$: on input a public key {\small$N$} and a message $m\in\mathbb{Z}_N$, choose an uniform $r \leftarrow \mathbb{Z}_N^*$ and output the ciphertext
{\small$ c:=[{( 1+N )}^{m}{r}^{N} mod {{N}^{2}}] $}.
  \item $\sf{Dec}$: on input a private key {\small$( N, \phi (N))$} and a ciphertext $c$, output the message
{\small$ m:=  [\frac{ [{c}^{\phi ( N )}mod{{N}^{2}}]-1 }{N}$}\\{\small$ \phi (N )^{-1} mod N] $}.
\end{itemize}

Assuming a pair of ciphertext {\small$( {{c}_{1}},{{c}_{2}} )$} is {\small$( {{m}_{1}},{{m}_{2}} )$} under the same Paillier encryption scheme, we have {\small${{c}_{1}}\times c_2= ( 1+N)^{m_1+m_2}{r^N}mod{N^2},\ m_1 + m_2 \in\mathbb{Z}_N$},
i.e.,
\\$\sf{Paillier.Enc}$ $(m_1+m_2)$ = $\sf{Paillier.Enc}$$(m_1)$ $*$ $\sf{Paillier.Enc}$$(m_2)$

\textbf{Cloud-RSA.}
RSA relies on the hardness of factoring assumption.
Although RSA is commonly used for ensuring the authenticity of digital data,
many attacks still threaten the security of RSA,
such as Factorization attacks and Low private exponent attacks \cite{RSA-attack-1,RSA-attack-1,RSA-attack-2}.
To strengthen RSA security, Optimal Asymmetric Encryption Padding scheme has been proposed in order to convert the RSA into a probabilistic encryption scheme and to achieve semantic security.
But Optimal Asymmetric Encryption Padding scheme removes the multiplicative homomorphic property from RSA.
To address the drawbacks of RSA,
Khalid et al. \cite{cloud-RSA} has proposed an enhanced encryption scheme named {\small Cloud-RSA} which keeps the multiplicative homomorphic property of the plain RSA and can resist the well-known attacks.
{\small Cloud-RSA} encryption scheme $(\sf{Gen, Enc, Dec})$ is given:
\begin{itemize}
  \item $\sf{Gen}$: run $\sf{GenModulus}(1^n)$ to obtain $(N, p, q)$.
choose $e > 1$ such that {\small$gcd(e,\phi(N) ) = 1$}.
compute {\small$d :=[{{e}^{-1}} mod\phi ( N )]$}.
The public key is {\small$(N)$}, and private key is {\small$(N, e, d)$}.
Both the public key and private key of Cloud-RSA are hold by data owners.
  \item $\sf{Enc}$: on input a message $m\in\mathbb{Z}_N^*$, output the ciphertext
{\small$c:=  [m^e modN ]$}.
  \item $\sf{Dec}$: on input a ciphertext $c\in\mathbb{Z}_N^*$, output the message
{\small$m:=  [c^d modN ]$}.
\end{itemize}

For security proof details about Cloud-RSA, we refer the reader to \cite{cloud-RSA}.
Assuming a pair of ciphertext {\small$( {{c}_{1}},{{c}_{2}} )$} is {\small$( {{m}_{1}},{{m}_{2}})$} under the same {\small Cloud-RSA} encryption scheme,
we have {\small${{c}_{1}}\times c_2= {(c_1c_2)^e}mod N$}, $c_1c_2 \in \mathbb{Z}_N^*$,
i.e.,
\\{\small$\sf{CloudRSA.Enc}$ $(m_1m_2)$=$\sf{CloudRSA.Enc}$$(m_1)$$*$$\sf{CloudRSA.Enc}$$(m_2)$}

\begin{table*}[hbt]
\centering
\caption{The Cases of Building Blocks}\label{table:The Cases of Building Blocks}
\footnotesize{
\begin{tabular}{|c|p{3.8cm}<{\centering}|p{3cm}<{\centering}|p{3cm}<{\centering}|}
\hline
\multirow{2}{*}{Algorithms}&\multicolumn{2}{c|}{Input}&\multicolumn{1}{c|}{Output}	
\\
\cline{2-4}	
 & \multicolumn{1}{c|}{Data Owner}&	 \multicolumn{1}{c|}{Model Demander} & \multicolumn{1}{c|}{Model Demander}
\\
\hline
\hline
Secure addition &$(\sf{SK},\sf{PK})_{Paillier}$,\ $[a]_P$ &	$[b]_P$	&  $[a+b]_P$
\\\hline
Secure subtraction&$(\sf{SK},\sf{PK})_{Paillier}$,\ $[a]_P$ &	$[b]_P$	&  $[a-b]_P$
\\\hline
Secure plaintext-ciphertext multiplication &$(\sf{SK},\sf{PK})_{Paillier}$,\ $[a]_P$& $b$	  &   $[a*b]_P$
\\\hline
Secure plaintext-ciphertext dot product &$(\sf{SK},\sf{PK})_{Paillier}$,\ $[\vec{a}]_P$& $\vec{b}$	  &   $[\vec{a}\cdot\vec{b}]_P$
\\\hline
Secure ciphertext-ciphertext multiplication&$(\sf{SK},\sf{PK})_{CloudRSA}$,\ $[a]_R$ &	$[b]_R$	&  $[a*b]_R$
\\\hline
Secure power function&$(\sf{SK},\sf{PK})_{CloudRSA}$,\ $[e^{\vec{a}}]_R$ &	$\vec{b}$	&  $[e^{\vec{a\cdot b}}]_R$
\\\hline
Secure summation	&$(\sf{SK},\sf{PK})_{Paillier}^i$, $a_i$&	$(\sf{SK},\sf{PK})_{Paillier}^\mathcal{A}$&	 $\sum_{i=1}^na_i$
\\\hline
Converting $[e^{\vec{a\cdot b}}]_R$ to $[e^{\vec{a\cdot b}}]_P$& $(\sf{SK},\sf{PK})_{Paillier}$,$(\sf{SK},\sf{PK})_{CloudRSA}$ &	$[e^{\vec{a\cdot b}}]_R$& $[e^{\vec{a\cdot b}}]_P$
\\\hline
Converting $[m]_P^1$ to $[m]_P^2$&$(\sf{SK},\sf{PK})_{Paillier}^1$, $(\sf{PK})_{Paillier}^2$ &	$[m]_P^1$	&	 $[m]_P^2$
\\
\hline
\end{tabular}}
\end{table*}

\section{Building Blocks based on Partial Homomorphic Encryption}\label{sec:Building Blocks}
This section details the design idea of \texttt{Heda} and the PHE-based library of building blocks.
Table \ref{table:The Cases of Building Blocks} lists the different cases for each building block.
For all building blocks, both parties cannot obtain other useful information except for the legal information.

\noindent\textbf{Security Proofs.}
Formal cryptographic proofs of the building blocks requires space beyond the page limit.
Hence, we provide the intuition behind the proofs, and delegate formal proofs to Appendix \ref{app:Security Analysis for Building Blocks}.

\subsection{Design Idea}

Just as a mathematical formula can be decomposed into the permutation and combination of addition, subtraction, multiplication, and division,
ML training algorithms can also be decomposed into a set of core primitive operations.
Intuitively, we can design secure building blocks for each of core primitive operations and appropriately combine them into the desire ML training protocols that satisfy the security requirements in the aggregation scenario.

After analyzing the typical supervised learning algorithms (i.e., LR, SVM, and NB),
we include addition, subtraction, multiplication, power function, and summation as the core primitive operations, see Table \ref{table:the core operations for machine learning training algorithm} for detail.
Hence, we design secure building blocks targeting each of these core operations.

\begin{table}[!h]
\centering
\caption{Core Operation of Machine Learning Training}\label{table:the core operations for machine learning training algorithm}
\footnotesize{
\begin{tabular}{c|cccccc}
\hline
Operations & Add. & Sub. & Multi. & Pow & Comp.&Sum.\\
\hline
LR & \checkmark  & \checkmark &  \checkmark & \checkmark && \\
\hline
SVM & \checkmark  & \checkmark &\checkmark & & \checkmark& \\
\hline
NB & \checkmark  &  &\checkmark&&&\checkmark\\
\hline
\end{tabular}}
\end{table}

Nevertheless, using the above secure building blocks is inadequate for the security requirements in the aggregation scenario.
Since there are $n$ data owners in \texttt{Heda},
and each data owner has their own encryption schemes (i.e., a certain plaintext-ciphertext space with a pair of keys {\small($\sf{PK}$,$\sf{SK}$))}.
Data owners send the ciphertext data encrypted by its public key to the model demander.
Homomorphic operations upon the ciphertext can only be operated in the same plaintext-ciphertext space.
The model demander has to update the model parameters upon the encrypted data under different plaintext-ciphertext spaces.

To supplement and enable a building block's output to become the input of another, while maintaining the underlying plaintext,
we design two building blocks for converting from the ciphertext under Cloud-RSA to under Paillier and converting the ciphertext under one data owner's Paillier encryption to the other data owner's Paillier encryption.

\subsection{Building Blocks for Primitive Operations}
Since it is plain to obtain the secure addition, subtraction and multiplication,
we include them in the library of building blocks directly:
\begin{itemize}
  \item Building Block 1: secure addition with Paillier.
  \item Building Block 2: secure subtraction with Paillier.
  \item Building Block 3: secure plaintext-ciphertext multiplication with {\small Cloud-RSA}.
  \item Building Block 4: secure ciphertext-ciphertext multiplication with {\small Cloud-RSA}.
\end{itemize}

\begin{mypro}\label{Propo:four building blocks}
Building Block 1-4 is secure in the
honest-but-curious model.
\end{mypro}

\begin{proofintuition}[for Proposition \ref{Propo:four building blocks}]
\emph{
The security property of the four building blocks is straightforward by ciphertext indistinguishability against chosen plaintext attacks of Paillier and {\small Cloud-RSA}.
See Appendix \ref{app:Security Analysis for Building Blocks} for a complete proof.
}
\qed
\end{proofintuition}

Designing secure power function and secure summation is not straightforward.
This paper novelly proposes secure power function and secure summation.
From here onwards, we introduce our particularly designed building blocks.

$\sf{Building\ Block\ 5:\ Secure\ Power\ Function.}$
The key to securely computing {\small$Sigmoid( \beta^{\sf{T}}\vec{x} )=\frac{e^{\beta^{\sf{T}}\vec{x}}}{1+e^{\beta^{\sf{T}}\vec{x}}}$} is to compute {\small$e^{\beta^{\sf{T}}\vec{x}}$}.

Supposing a data owner encrypts {\small$e^{\vec{a}}=\{ e^{a_1},e^{a_2},\ldots ,e^{a_d}$\\$ \}$} by his {\small CLoud-RSA} public key and sends the ciphertext $[e^{\vec{a}}]_R$ to a model demander.
The model demander has {\small$\vec{b}=\{ {{b}_{1}},{{b}_{2}}, \ldots,{{b}_{d}} \}$},
and he wants to obtain {\small$[e^{\vec{a \cdot b}}]_R$}.
\begin{equation}\label{eq:pow}
\small
\begin{split}
  [e^{\vec{a\cdot b}}]_R = &[e^{a_1b_1+a_2b_2+\ldots+a_db_d}]_R\\
   =&\prod_{i=1}^d[e^{a_ib_i}]_R  =\prod_{i=1}^d([e^{a_i}]_R)^{b_i}
\end{split}
\end{equation}
Equation \eqref{eq:pow} specifies our secure power function.
The key point is that {\small$e^{a_i b_i}$} equals multiplying {\small$e^{a_i}$} multiplied by itself $b_i$ times.
Along this way, the model demander is able to obtain {\small$[e^{\vec{a \cdot b}}]_R$} by {\small($\sum_{i=1}^d b_i + d-1$)} times multiplication.

\begin{mypro}\label{Propo:secure power function}
Secure power function is secure in the
honest-but-curious model.
\end{mypro}

\begin{proofintuition}[for Proposition \ref{Propo:secure power function}]
\emph{
The data owner does not any message, his view only consists in its input.
The model demander do computation on the ciphertext encrypted by the data owner's public key.
By ciphertext indistinguishability against chosen plaintext attacks of {\small Cloud-RSA}, the model demander cannot learn any useful information.
See Appendix \ref{app:Security Analysis for Building Blocks} for a complete proof.
}
\qed
\end{proofintuition}

$\sf{Building\ Block\ 6:\ Secure\ Summation.}$
There are $n$ data owners.
Each data owner $\mathcal{O}_i$ holds a value $a_i$.
Without decryption and revealing the values to the model demander, the model demander desire to obtain the summation of the $n$ values $\{a_i\}_{i=0}^n$.
Our secure summation solution is described Building Blocks \ref{algorithm:secure summation}.

\begin{algorithm}[!h]
\renewcommand{\algorithmcfname}{Building Block}
\small
\setcounter{algocf}{5}
\caption{Secure Summation}\label{algorithm:secure summation} 
\LinesNumbered
\SetKwInOut{KIN}{Participants:}
\KIN{$n$ data owners, one model demander}
\SetKwInOut{KIN}{Each Data Owner $(\mathcal{O})$ Input}
\KIN{{\footnotesize$(\sf{PK}, \sf{SK})_{Paillier}^i$ and $a_i$}}
\SetKwInOut{KIN}{Model Demander $(\mathcal{A})$ Input}
\KIN{{\footnotesize$(\sf{PK}, \sf{SK})_{Paillier}^{\mathcal{A}}$}}
\SetKwInOut{KOUT}{Model Demander $(\mathcal{A})$ Output}
\KOUT{$\sum_{i=1}^{n}a_i$}
\For{$i=1$ to $n$}{
$\mathcal{O}_i$ uniformly picks $r_i \in\mathbb{Z}_N $\;
$\mathcal{O}_i$: $\hat{a_i}=a_i+r_i$\;
$\mathcal{O}_i$: $[r_i]_P^i:=\sf{Paillier^i.Enc}$$(r_i)$\;
$\mathcal{O}_i$ sends $[r_i]_P^i$ and $a_i+r_i$ to $\mathcal{A}$\;
}
$\mathcal{A}$: $\hat{res}=\sum_{i=1}^n(a_i+r_i)$\;
$\mathcal{A}$: $[\hat{res}]_P^1:=\sf{Paillier^1.Enc}$$(\hat{res})$\;
\For{$i=1$ to $n-1$}{
   $\mathcal{A}$: $[\hat{res}]_P^i:=[\hat{res}]_P^i*([r_i]_P^i)^{(-1)}$ \;
   $\mathcal{A}$: converts $[\hat{res}]_P^i$ to $[\hat{res}]_P^{i+1}$ by Building Block \ref{algorithm:Converting Ciphertext 2};
}
$\mathcal{A}$: $[\hat{res}]_P^n:=[\hat{res}]_P^n*([r_n]_P^n)^{(-1)}$ \;
$\mathcal{A}$: converts $[\hat{res}]_P^n$ to $[\hat{res}]_P^{\mathcal{A}}$ by Building Block \ref{algorithm:Converting Ciphertext 2}\;
$\mathcal{A}$: $res:=\sf{Paillier^{\mathcal{A}}.Dec}$$([\hat{res}]_P^{\mathcal{A}})$ \Comment{$res=\sum_{i=1}^{n}a_i$}\;
\Return $\sum_{i=1}^{n}a_i$ to $\mathcal{A}$.
\end{algorithm}

Specifically, $\mathcal{A}$ removes the encrypted noise $\sum_{i=1}^{n}{r_i}$ from $res'$ that is obtained at the step 7 of Building Blocks 6 as following:\\
\noindent{\small\emph{(i)
From the first data owner to the last,
$\mathcal{A}$ sequentially computes ${[res']_P^{i}}:={[res'-r_i]}_P^{i}$ using secure subtraction,
and converts ${[res']_P^{i}}$ to ${[res']_P^{i+1}}$  by Building Block \ref{algorithm:Converting Ciphertext 2}.\\
(ii) Finally, using Building Block \ref{algorithm:Converting Ciphertext 2}, $\mathcal{A}$ converts $[res]_P^n:=res'- \sum_{c=1}^{n}{r_i}$ which is encrypted by the last data owner's Paillier public key to $[res]_P^\mathcal{A}$ that is encrypted by itself.
}
}

\begin{mypro}\label{Propo:secure summation}
Secure summation is secure in the honest-but-curious model.
\end{mypro}

\begin{proofintuition}[for Proposition \ref{Propo:secure summation}]
\emph{
The data owners add random noise $r_i$ to $a_i$, the random noise hides $a_i$ in an information-theoretic way (it is an one-time pad).
Even though the model demander receives $[r_i]_P^i$, the ciphertext indistinguishability against chosen plaintext attacks of Paillier prevents it figure out the $a_i$ from $r_i+a_i$.
Besides, the Building Block \ref{algorithm:Converting Ciphertext 2} is secure in the honest-but-curious model, we obtain the security of secure summation using the definition of Modular Sequential Composition.}

\emph{As for the collusion situations,
when $(n-1)$ data owners collude with each other and the data owner $\mathcal{O}_i$ dose not participate in the collusion,
they can only figure out $a_i+r_i$ and $\sum_{c=1}^{n}{a_c}+r_i$ cannot obtain $a_i$ or $\sum_{c=1}^{n}{a_c}$.
When $(n-2)$ data owners collude with the model demander,
and $\mathcal{O}_{i-1}$ and $\mathcal{O}_{i}$ do not participate in the collusion,
they can only figure out $a_{i-1}+a_i$ cannot obtain $a_i$ or $a_{i-1}$.}

\emph{See Appendix \ref{app:Security Analysis for Building Blocks} for a complete proof.
}
\qed
\end{proofintuition}

\subsection{Building Blocks for Conversion}
Since multiple encryption schemes are used in \texttt{Heda},
we developed two protocols for converting ciphertexts from one encryption scheme to another while maintaining the underlying plaintexts.

\noindent$\sf{Building\ Block\ 7:\ Converting\ {\small Cloud-RSA}\ to\ Paillier}$ $([e^{\vec{a \cdot b}}]_R$ to $[e^{\vec{a \cdot b}}]_P)$.
\\\noindent
The model demander holds a ciphertext $[e^{\vec{a \cdot b}}]_R$ encrypted under a data owner's {\small Cloud-RSA} public key.
For continuing to the subsequent ciphertext computations,
the model demander needs to convert $[e^{\vec{a\cdot b}}]_R$ to the ciphertext encrypted by the data owner's Paillier public key.
During the conversion, $e^{\vec{a \cdot b}}$ is confidential to the model demander and the data owner.

\begin{algorithm}[!h]
\renewcommand{\algorithmcfname}{Building Block}
\small
\caption{Converting ${[e^\vec{a\cdot b}]}_R$ to ${[e^\vec{a\cdot b}]}_P$}\label{algorithm:Converting Ciphertext 1} 
\LinesNumbered
\SetKwInOut{KIN}{Data Owner $(\mathcal{O})$ Input}
\KIN{{\footnotesize$(\sf{PK}, \sf{SK})_{Paillier}$ and $(\sf{PK},\sf{SK})_{CloudRSA}$}}
\SetKwInOut{KIN}{Model Demander $(\mathcal{A})$ Input}
\KIN{${[e^\vec{a\cdot b} ]}_{R}$}
\SetKwInOut{KOUT}{Model Demander $(\mathcal{A})$ Output}
\KOUT{${[e^\vec{a\cdot b} ]}_{P}$}
$\mathcal{A}$ randomly picks $r $\;
$\mathcal{A}$: ${[e^{\vec{a\cdot b}+r} ]}_{R}:={[e^\vec{a\cdot b} ]}_{R}*[e^r]_R$\;
$\mathcal{A}$ sends ${[e^{\vec{a\cdot b}+r} ]}_{R}$ to $\mathcal{O}$\;
$\mathcal{O}$: $e^{\vec{a\cdot b}+r}:=\sf{CloudRSA.Dec}$$({[e^{\vec{a\cdot b}+r} ]}_{R})$\;
$\mathcal{O}$: ${[e^{\vec{a\cdot b}+r} ]}_{P}:=\sf{Paillier.Enc}$$(e^{\vec{a\cdot b}+r})$\;
$\mathcal{O}$ sends ${[e^{\vec{a\cdot b}+r} ]}_{P}$ to $\mathcal{A}$\;
$\mathcal{A}$: ${[e^{\vec{a\cdot b}} ]}_{P}:=({[e^{\vec{a\cdot b}+r} ]}_{P})^{e^{-r}}$\;
\Return ${[e^{\vec{a\cdot b}} ]}_{P}$ to $\mathcal{A}$\;
\end{algorithm}

The correctness analysis is as follows:
The model demander randomly picks a random noise $r$,
generates {\small${[ e^{\vec{a \cdot b}+r} ]}_{R}$} by secure ciphertext-ciphertext multiplication.
Then the data owner decrypts {\small${[ e^{\vec{a \cdot b}+r} ]}_{R}$} with its {\small$\sf{SK}_{CloudRSA}$} and encrypts it with its {\small$\sf{PK}_{Paillier}$}, obtaining {\small${[ e^{\vec{a \cdot b}+r} ]}_{P}$}.
Using secure plaintext-ciphertext multiplication, the model demander multiplies {\small${[ e^{\vec{a\cdot b}+r} ]}_{P}$} by {\small$ e^{-r} $} to remove {\small$e^{r}$}.

\begin{mypro}\label{Propo:Converting Ciphertext 1}
Secure converting ${[e^\vec{a\cdot b}]}_R$ to ${[e^\vec{a\cdot b}]}_P$ is secure in the honest-but-curious model.
\end{mypro}

\begin{proofintuition}[for Proposition \ref{Propo:Converting Ciphertext 1}]
\emph{
Even though the data owner is able to decrypt and obtain $e^{\vec{a \cdot b}+r}$, the random noise $r$ hides $ \vec{a \cdot b}$ in an information-theoretic way (it is an one-time pad).
See Appendix \ref{app:Security Analysis for Building Blocks of Conversion} for a complete proof.
}
\qed
\end{proofintuition}

\noindent$\sf{Building\ Block\ 8:\ Converting\ one\ Paillier\ to\ Another\ Paillier}$ ($[m]_P^1$ to $[m]_P^2$).

\begin{algorithm}[!h]
\renewcommand{\algorithmcfname}{Building Block}
\small
\caption{Converting ${[m]}_P^1$ to ${[m]}_P^2$}\label{algorithm:Converting Ciphertext 2} 
\LinesNumbered
\SetKwInOut{KIN}{Data Owner $(\mathcal{O})$ Input}
\KIN{{\footnotesize$(\sf{PK}, \sf{SK})_{Paillier}^1$ and $(\sf{PK})_{Paillier}^2$}}
\SetKwInOut{KIN}{Model Demander $(\mathcal{A})$ Input}
\KIN{${[m]}_{P}^1$}
\SetKwInOut{KOUT}{Model Demander $(\mathcal{A})$ Output}
\KOUT{${[m]}_{P}^2$}
$\mathcal{A}$ uniformly picks $r \in\mathbb{Z}_N$\;
$\mathcal{A}$: $[m+r]_P^1:=[m]_P^1*[r]_P^1$\;
$\mathcal{A}$ sends $[m+r]_P^1$ to $\mathcal{O}$\;
$\mathcal{O}$: $m+r:= \sf{Paillier^1.Dec([m+r]_P^1)}$\;
$\mathcal{O}$: $[m+r]_P^2:= \sf{Paillier^2.Enc}$$(m+r)$\;
$\mathcal{O}$ sends $[m+r]_P^2$ to $\mathcal{A}$\;
$\mathcal{A}$: $[m]_P^2:=[m+r]_P^1*[r]_P^{-1}$\;
\Return $[m]_P^2$ to $\mathcal{A}$;
\end{algorithm}

\begin{figure*}[!ht]
\centering
\includegraphics[width=18cm]{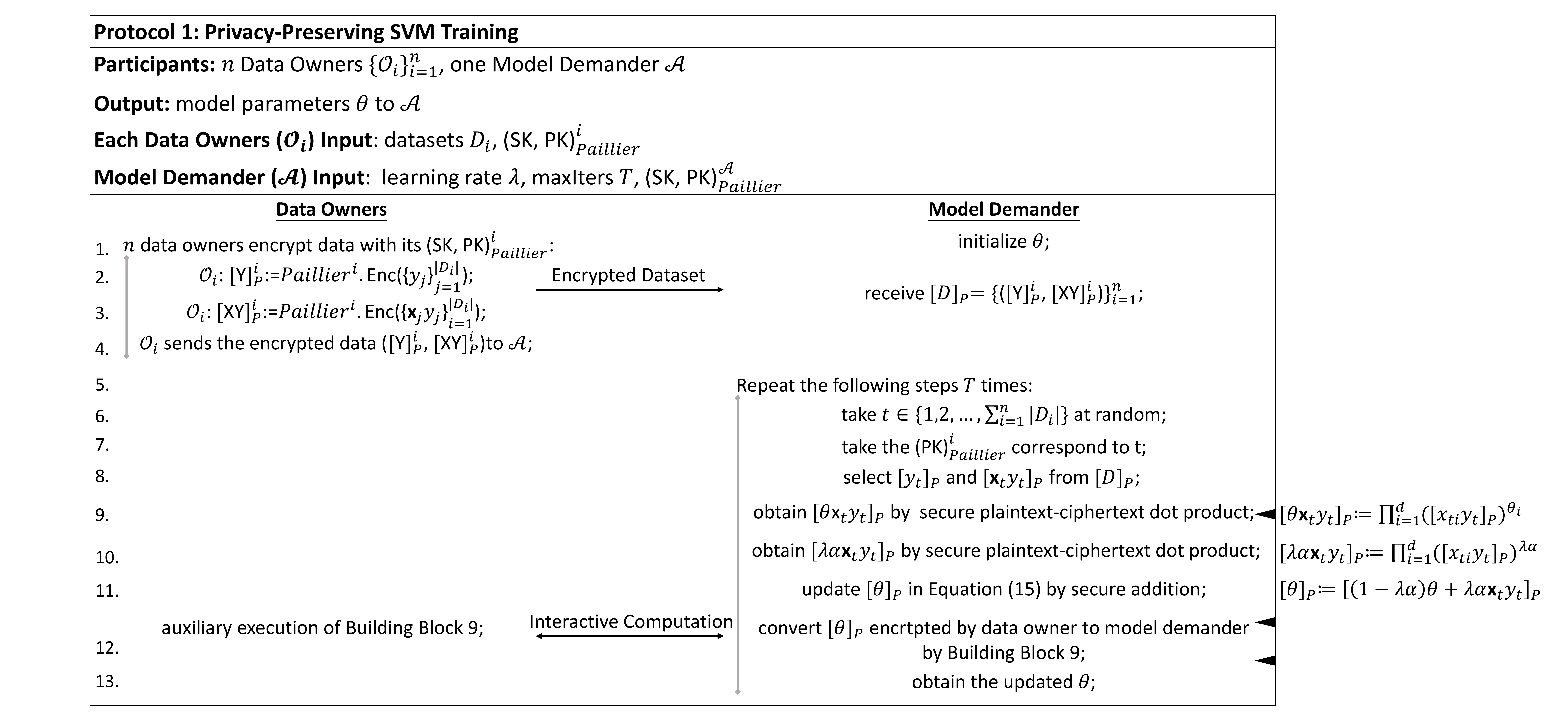}\\
\caption{Privacy-Preserving Support Vector Machine Training Protocol}
\label{fig:securesvm}
\end{figure*}

\noindent
The model demander holds a ciphertext $[m]_P^1$ encrypted by the Paillier encryption of data owner 1.
He wants to convert $[m]_P^1$ to the ciphertext $[m]_P^2$ encrypted by another data owner's Paillier encryption.
During the conversion, the message $m$ is revealed to the data owners or the model demander.
Building Block \ref{algorithm:Converting Ciphertext 2} achieves the switching of $[m]_P^1$ to $[m]_P^2$.

In Building Block \ref{algorithm:Converting Ciphertext 2},
the model demander adds a random noise $r$ to {\small$[m]_P^1$} by secure addition, obtaining {\small$[m+r]_P^1$}.
Then the data owner 1 decrypts the resulting value with his {\small$\sf{SK}_{Paillier}$} and encrypts it with the other data owner's Paillier key, obtaining {\small$[m+r]_P^2$}.
The model demander is able to remove $r$ from {\small$[m+r]_P^2$} by using secure subtraction.

\begin{mypro}\label{Propo:Converting Ciphertext 2}
Secure converting ${[m]}_P^1$ to ${[m]}_P^2$ is secure in the honest-but-curious model.
\end{mypro}

\begin{proofintuition}[for Proposition \ref{Propo:Converting Ciphertext 2}]
\emph{
The model demander adds random noise to the ciphertext $[m]_P^1$.
Even though the data owner is able to decrypt and obtain $m+r$, the random noise $r$ hides $ m$ in an information-theoretic way (it is an one-time pad).
See Appendix \ref{app:Security Analysis for Building Blocks of Conversion} for a complete proof.
}
\qed
\end{proofintuition}

\section{Privacy-Preserving Machine Learning Training with \texttt{Heda}}\label{sec:Privacy-Preserving Machine Learning Models Training with Heda}
In this section, we detail how to construct the privacy-preserving ML training protocols using the proposed building blocks.
Since these building blocks are designed in a modular way, so carrying the privacy-preserving ML training come down to invoking the right module.

For performing Stochastic Gradient Descent (SGD), the model demander randomly selects a small number of records from $\mathcal{D}$ in each iteration.
Since the dataset $\mathcal{D}$ is gathered sequentially, the model demander is able to locate the data owner that the selected records belong to, which facilitates subsequent ciphertext computations,
i.e., the model demander encrypts the correlative parameters using the picked data owner's public key.
ML preliminaries used in this paper can be found in Appendix \ref{app:Machine Learning Preliminary}.

Both Paillier and {\small Cloud-RSA} work with positive integers in a finite space,
while the three ML training algorithms manipulate floating point numbers.
Hence,
after detailed secure training algorithms,
we present the solution for dealing with floating point numbers.

\subsection{Secure Support Vector Machine Training}
This paper adopts Hinge Loss with SGD for solving SVM, which involves the less amount of vector multiplication than other optimization methods \cite{SGDSVM}.
We restate the iteration equation:
\begin{equation}
\small
 \theta := \theta-\lambda(\alpha\theta-\alpha y_i \hat{\vec{x}}_i).
\end{equation}

In iteration, the key points are two multiplication $\lambda\alpha y_i \hat{\vec{x}}_i$ and $\theta^{\sf{T}} y_i \hat{\vec{x}}_i$.
From the perspective of the model demander, $\lambda\alpha y_i \hat{\vec{x}}_i$ and $\theta^{\sf{T}} y_i \hat{\vec{x}}_i$ belong to the plaintext-ciphetext multiplication, where $y_i \hat{\vec{x}}_i$ is unseen but $\lambda\alpha$ and $\theta$ are known.
Thus, the two multiplications are coped with by secure plaintext-ciphertext multiplication.
Fig. \ref{fig:securesvm} specifies our secure SVM training protocol instantiated from \texttt{Heda},
which needs $3$ interactions (i.e. interactions between the model demander and data owners) throughout each iteration.

\begin{mypro}\label{Propo:SVM}
Privacy-preserving SVM training protocol is secure in the honest-but-curious model.
\end{mypro}

\begin{figure*}[t]
\centering
\includegraphics[width=18cm]{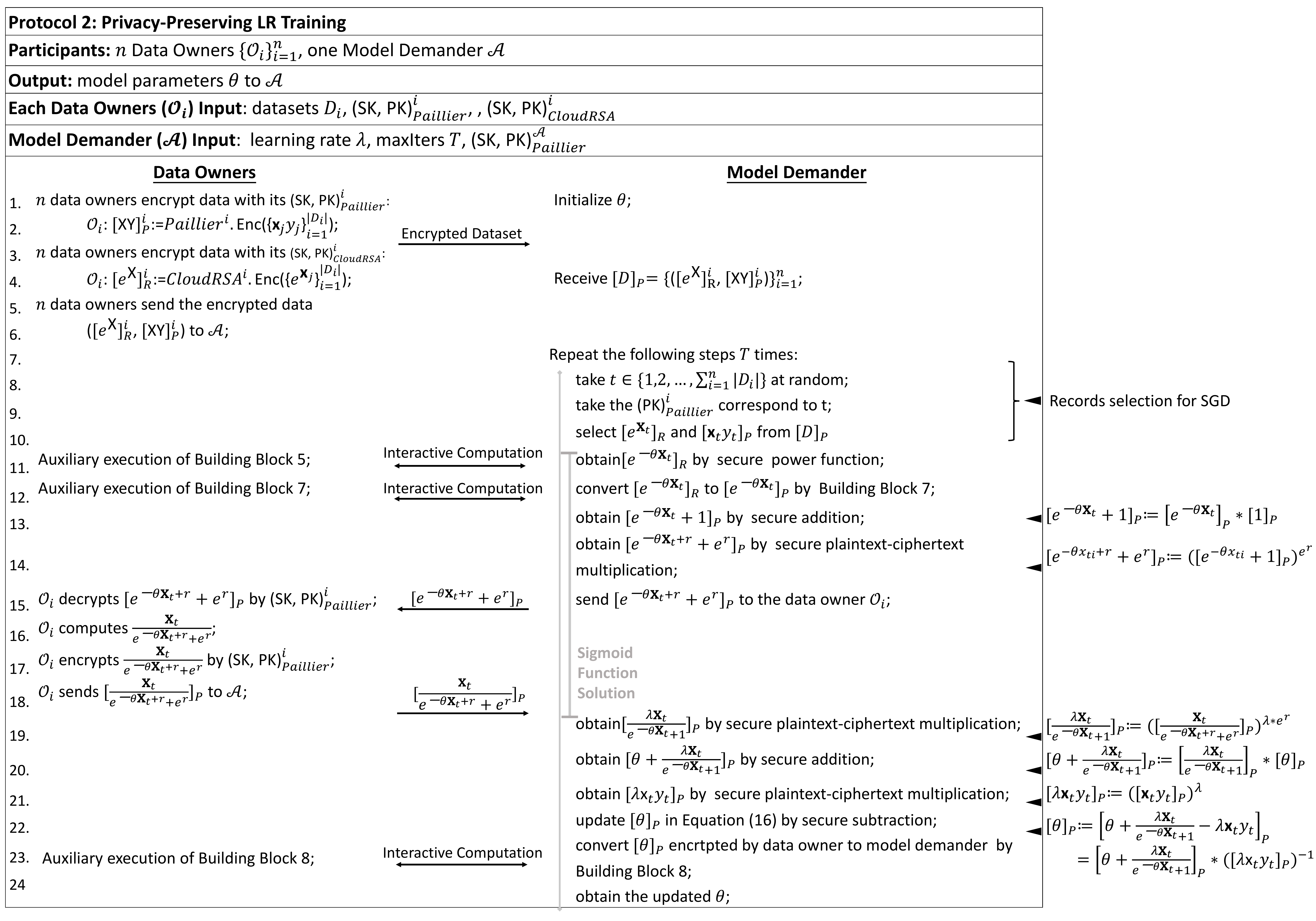}\\
\caption{Privacy-Preserving Logistic Regression Training Protocol}
\label{fig:secureLR}
\end{figure*}

\begin{proofintuition}[for Proposition \ref{Propo:SVM}]
\emph{
No collusion happen:
Data owners do not receive any message in the protocol except for the auxiliary execution of secure building blocks.
All the training data is ciphertext that is encrypted by corresponding data owners.
The model demander can only infer the training data from the model parameters, which is a ciphertext-only attack with perfectly secret \cite{1}.
As Building Block used in this protocol are secure in the honest-but-curious model, we obtain the security using modular sequential composition.}

\emph{As for the collusion situations,
since the training data and model parameters are encrypted by corresponding data holders with their own public key.
Without all the data owners participating in the collusion, data owners cannot figure out the complete model information.
Without the corresponding private key, the ciphertext indistinguishability of Paillier ensures that no bit of information is leaked from ciphertexts of other data owners who do not participating in the collusion.}

\emph{See Appendix \ref{app:Security Definition} for a complete proof and the definition of ciphertext-only attack and perfectly secret.}
\qed
\end{proofintuition}

\subsection{Secure Logistic Regression Training}
Here recalls the iteration equation of LR:
\begin{equation}\label{eq:iteration formula of LR}
\small
  \theta _{j}:=\theta _j-\lambda ({x_{ij}}y_i-\frac{{x_{ij}}}{1+e^{-\theta^{\sf{T}}\vec{x}_i}})
\end{equation}

\begin{figure*}[!ht]
\centering
\includegraphics[width=18cm]{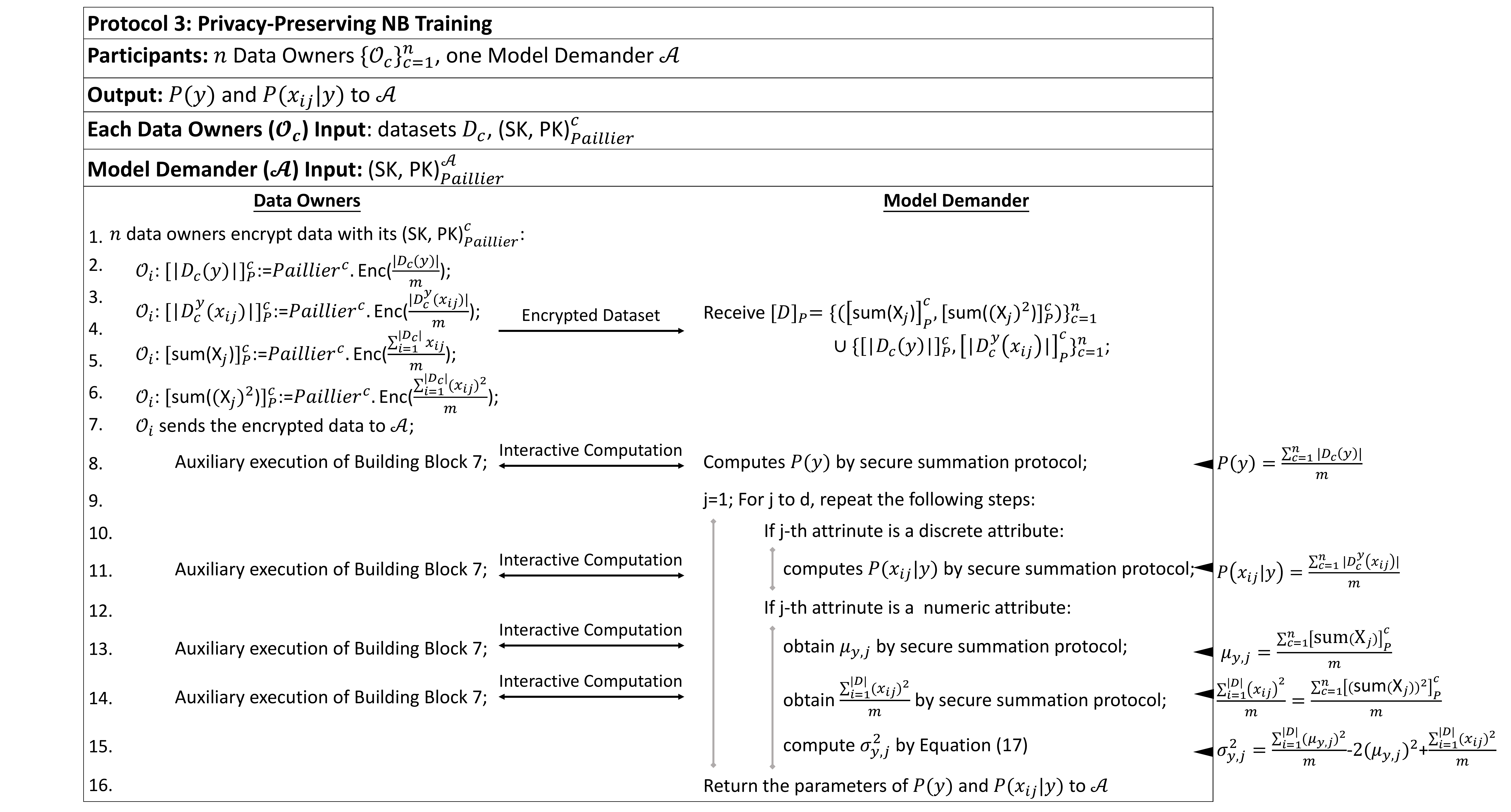}\\
\caption{Privacy-Preserving Logistic Regression Training Protocol}
\label{fig:secureNB}
\end{figure*}

The complex part in Equation (\ref{eq:iteration formula of LR}) is the Sigmoid function $\frac{x_i}{1+e^{-\theta^{\sf{T}}\vec{x}_i}}$. 
By adapting secure power function, the model demander obtains $[e^{-\theta^{\sf{T}}\vec{x}_i}]_R$ first.
Then, as specified in Fig. \ref{fig:secureLR}, the results in the operations of the model demander evolve as follows:
\[\footnotesize
[e^{-\theta^{\sf{T}}\vec{x}_i}]_R \rightarrow [e^{-\theta^{\sf{T}}\vec{x}_i}]_p \rightarrow [e^{-\theta^{\sf{T}}\vec{x}_i} +1]_p
\rightarrow [e^{-\theta^{\sf{T}}\vec{x}_i+r} +e^r]_p
\]
The model demander sends $[e^{-\theta^{\sf{T}}\vec{x}_i+r} +e^r]_p$ to data owners.
The data owner carries on computations as following:
\[\footnotesize
[e^{-\theta^{\sf{T}}\vec{x}_i+r} +e^r]_p \rightarrow e^{-\theta^{\sf{T}}\vec{x}_i+r} +e^r \rightarrow \frac{{\vec{x}}_i}{e^{-\theta^{\sf{T}}\vec{x}_i+r} +e^r}\]
The data owner returns the encrypted resulting value $\small [\frac{{\vec{x}}_i}{(e^{-\theta^{\sf{T}}{\vec{x}}_i}+1)*e^r}]_P$ to the model demander.
Finally, removing the noise $e^r$,
the model demander get $[\frac{{\vec{x}}_i}{1+e^{-\theta^{\sf{T}}{\vec{x}}_i}}]_P$.
Fig. \ref{fig:secureLR} specifies the privacy-preserving LR training protocol,
which needs 4 interactions (i.e. interactions between the model demander and data owners) throughout each iteration.

\begin{mypro}\label{Propo:LR}
Privacy-preserving LR training protocol is secure in the honest-but-curious model.
\end{mypro}

\begin{proofintuition}[for Proposition \ref{Propo:LR}]
\emph{
No collusion happen:
In the process of sigmoid function solution,
even though the data owners is able to decrypt {\small$e^{-\theta^{\sf{T}}{\vec{x}}_i+r}+e^r$}, the random noise $r$ hides $\theta$ in an information-theoretic way (it is an one-time pad).
All the training data is ciphertext that is encrypted by corresponding data owners.
The model demander can only infer the training data from the model parameters, which is a ciphertext-only attack with perfectly secret.}
\emph{The collusion situations is similar as Protocol 1, so we do not repeat them here.
See Appendix \ref{app:Security Analysis for ML Training Protocols} for a complete proof.}
\qed
\end{proofintuition}

\subsection{Secure Naive Bayes Training}
For learning a NB model from dataset $\mathcal{D}$,
the model demander needs to compute the the class prior probability {\small$P(y)$} and the conditional probability {\small$P(\vec{x}_i |y)$}.

\begin{equation}\label{eq:Secure subtraction}
\footnotesize
P(\vec{x}_i |y):=
\begin{cases}
\frac{|\mathcal{D}^{y}({x}_{ij})|}{m},&discrete\ attribute \\
\frac{1}{\sqrt{{2\pi}}\sigma_{y, j}}exp(-\frac{(x_{ij}-\mu_{y,j})^2}{2\sigma_{y, j}^2}),&numeric\ attribute
 \end{cases}
\end{equation}

If the j-th attribute is a discrete attribute,
the conditional probability is {\small$P({x}_{ij}|{y})= \frac{|\mathcal{D}^{y}({x}_{ij})|}{m}$},
where the total number of records $m$ is public \cite{112} without loss of generality.
{\small$|\mathcal{D}^{y}({x}_{ij})|$} is the number of records belonging to class $y$ with the attribute value $x_{ij}$ in $\mathcal{D}$.
Since there are $n$ data owners in the aggregation scenario,
{\small$|\mathcal{D}^{y}({x}_{ij})| = \sum_{c=1}^{n}{|\mathcal{D}_c^{y}({x}_{ij})|}$}.

The class prior probability i.e., {\small$P(y)= \frac{\sum_{c=1}^{n}{|\mathcal{D}_{c}(y)|}}{m}$} has a similar computation manner with the conditional probability of discrete attributes,
where $|\mathcal{D}_{c}(y)|$ is the number of records belonging to class $y$ in $\mathcal{D}_c$.

If the j-th attribute is a numeric attribute,
we assume that the j-th attribute obeys normal distribution,
where two parameters need to be estimated securely: mean $\mu_{y,j}$ and variance $\sigma_{y,j}^{2}$.
\begin{equation}\label{eq:NB mean variance}
\footnotesize
\begin{split}
\mu_{y, j}&=\frac{\sum_{i=1}^{|\mathcal{D}|}{x_{ij}}}{m}\\
\sigma_{y,j}^{2}&=\frac{\sum_{i=1}^{|\mathcal{D}|}{(\mu_{y,j}-x_{ij})^{2}}}{m}
= \frac{\sum_{i=1}^{|\mathcal{D}|}({\mu_{y, j}})^2+(x_{ij}^2-2\mu_{y,j} x_{ij})}{m}\\
&=\frac{\sum_{i=1}^{|\mathcal{D}|}(\mu_{y, j})^2}{m}
+\frac{\sum_{i=1}^{|\mathcal{D}|}(x_{ij})^2}{m}
-2\mu_{y,j}*\frac{\sum_{i=1}^{|\mathcal{D}|}(x_{ij})}{m}\\
\end{split}
\end{equation}

Employing secure summation protocol, we obtain $\sum_{c=1}^{n}{|\mathcal{D}_{c}(y)|}$, $\sum_{c=1}^{n}|\mathcal{D}_c^{y}({x}_{ij})|$, $\sum_{i=1}^{|\mathcal{D}|}x_{ij}$ and $\sum_{i=1}^{|\mathcal{D}|}(x_{ij})^2$ securely.
Fig. \ref{fig:secureNB} specifies our privacy NB training protocol.
Since the conditional probability of each feature can be figured out concurrently,
the number of interactions is depended on the number of involved secure summation algorithm, e.i., $n+1$ interactions.

\begin{mypro}\label{Propo:NB}
Privacy-preserving NB training protocol is secure in the honest-but-curious model.
\end{mypro}

\begin{proofintuition}[for Proposition \ref{Propo:NB}]
\emph{
No collusion happen:
Data owners do not receive any message in the protocol except for the auxiliary execution of secure building blocks.
All the training data is ciphertext that is encrypted by corresponding data holders.
As Building Block used in this protocol are secure in the honest-but-curious model, we obtain the security using modular sequential composition.}
\emph{The collusion situations is similar as Protocol 1, so we do not repeat them here.
See Appendix \ref{app:Security Analysis for ML Training Protocols} for a complete proof.}
\qed
\end{proofintuition}

\subsection{Dealing with Floating Point Numbers}\label{sec:Dealing with Floating Point Numbers}
The floating point number issues for PHE based secure ML training have been previously studied \cite{2,8}.
Using similar idea, we perform the format conversion by multiplying each floating point value $v$ by a big constant $10^e$ for fixed point representation.
Empirically, we retained two decimal places (i.e., $e=2$) for each floating point value involved in our implementation.
Suppose $v$ is a floating point number that is fixed to 2 decimal places in decimalism.
Then,
\begin{equation}\label{eq:floating point numbers}
 (v)_{10} = \frac{(\hat{v})_{10}}{10^2},\ \hat{v} \in \mathbb{N},\hat{v}>0
\end{equation}

Without loss of generality, when facing the same training task, we assume that features of records in training dataset have been locally preprocessed and restricted in range [0, 1].
Despite the data standardization on training dataset,
when the ciphertext is calculated many times,
the $\hat{v}$ may become very large integers, which might cause overflows errors.
We must ensure $log_{10}N > e^* $, where $N$ denotes the modulus for Paillier or {\small Cloud-RSA}'s cryptosystem,
and $e^*$ denotes $log_{10}\hat{v}$.
 $log_{10}N \approx 308$ when 1024-bit key length,
and $log_{10}N \approx 616$ when 2048-bit key length.

\textbf{\emph{For Protocol 1.}}
Referring to Fig. \ref{fig:securesvm},
at the end of each iteration,
$\theta$ is refreshed with the format of Equation (\ref{eq:floating point numbers}) $( e^*\leq2)$.
There are $2$ secure plaintext-ciphertext dot product for each record throughout an iteration.
2048-bit key length is enough to control these computations.

\textbf{\emph{For Protocol 2.}}
Referring to Fig. \ref{fig:secureLR},
at the end of each iteration,
$\theta$ is refreshed with the format of Equation (\ref{eq:floating point numbers}) $( e^*\leq2)$.
We discuss the situation of $e^*$ in secure power function where $[e^{\vec{x}_i\cdot \theta}]_R$ is obtained by {\small($\sum_{i=1}^d \theta_i + d-1$)} times multiplication of $[e^{\vec{x}_i}]_R$.
Since, $e^*\leq2$ for $[e^{\vec{x}_i}]_R$,
we must ensure:
\begin{equation}\label{equ:floatLR}
\small
log_{10}N > (\sum_{i=1}^d\theta_i + d-1)*2
\end{equation}

In the implementation of secure power function, we decompose $\theta_i$ into an integer part and a decimal part:
{\small$\theta_i= \theta_i^1+\frac{\theta_i^2}{10^2},\ \theta_i^1\geq0,100>\theta_i^2\geq1$}.
After obtaining {\small$[e^{\vec{x}_i}\theta_i^1]_R$} and {\small$[e^{\vec{x}_i}\theta_i^2]_R$},
we merge the respective results for recovering: {\small$e^{\theta_i \vec{x}_i} = e^{\theta_i^1 \vec{x}_i}* \sqrt[100]{e^{\theta_i^2 \vec{x}_i}}$}.
We provide a case study for explicating the recovering of {\small$e^{\theta_i \vec{x}_i}$} from {\small$[e^{\vec{x}_i}\theta_i^1]_R$} and {\small$[e^{\vec{x}_i}\theta_i^2]_R$} in Appendix \ref{app:A Case Study of Secure Power Function}.
The key length of {\small Cloud-RSA} in Protocol 2 can be set appropriately according to Equation (\ref{equ:floatLR}) and the number of features.

After secure power function, the $e^*$ is refreshed to 2 in Bilding Block 8.
Subsequent operations in an iteration, $3$ secure plaintext-ciphertext multiplication, $2$ secure addition, are controlled naturally.

\textbf{\emph{For Protocol 3.}}
As shown in Fig. \ref{fig:secureNB}, the main computation focus on the 4 operations of secure summation.
Referring to the 10 and 11 steps of Building Block 6, in secure summation $e^*$ is refreshed to 2 once a secure addition's computation ends.
Therefor, 2048-bit key length is enough to control these computations.

\section{Performance Evaluation}\label{sec:evaluation}
In this section, we present the evaluation of \texttt{Heda} from the following aspects:
(i) the performance overhead of \texttt{Heda},
(ii) the scalability of \texttt{Heda}.

\subsection{Preparations}\label{sec:evaluation-Preparations}
\textbf{\textit{Implementations.}}
In our design, each IoT data owner collects all pieces of data from the IoT devices in its own domain and then performs the following operations of \texttt{Heda}.
Data owners and the model demander have adequate computing resources generally.
Our experiments are run using a desktop computer with the configuration: single Intel i7 (i7-3770 64bit) processor for a total of 4 cores running at 3.40GHz and 8 GB RAM.
We implement all the proposed protocols in Java Development Kit 1.8.
Since multiple participants are involved in \texttt{Heda}, we mimic real network latency to be 30 ms for the round trip time of a packet.

\textbf{\textit{Datasets.}}
In our experiments, we use four datasets which are listed in Table \ref{table:Statistics of Datasets} from the UCI Machine Learning Repository.

\begin{table}[!h]
\centering
\renewcommand{\arraystretch}{1.1}
\caption{Datasets used in our experiments}\label{table:Statistics of Datasets}
\footnotesize{
\begin{tabular}{||p{1.2cm}<{\centering}|p{1.2cm}<{\centering}|p{1.2cm}<{\centering}|p{1.2cm}<{\centering}||}
\hline
Datasets	&Records&	Attributes &	Storage\\
\hline
\hline
BCWD&	699	&9&	172KB\\
\hline
Adult&	32561&	14&	11956KB\\
\hline
CAD	&690	&15&	271KB\\
\hline
Car	&1728&	6& 297KB\\
\hline
\end{tabular}}
\end{table}

\textbf{\textit{Key length setting.}}
The key length is not only related to the security guaranteed by cryptosystems but also to the plaintext space.
The underlying plaintext operated in the proposed protocols should be limited in a finite plaintext space that is defined by the key length.
As analyzed in Section \ref{sec:Dealing with Floating Point Numbers},
we use 2048-bit cryptographic keys in Paillier for the four datasets.
We use 2048-bit cryptographic keys in {\small Cloud-RSA} for BCWD and Car datasets and 4096-bit cryptographic keys for Adult and CAD datasets.

\subsection{Accuracy and Efficiency Evaluation of \texttt{Heda}}
\texttt{Heda} consists of a set of building blocks supporting the fundamental operations that underlie many ML training algorithms.
In this subsection, we present the performance of each building block first, following with the performance of the three privacy-preserving ML training protocols instantiated from \texttt{Heda}.

\begin{table}[!h]
\centering
\caption{Building Blocks Performance}\label{table:Building Blocks Evaluation}
\footnotesize{
\begin{tabular}{||c|p{0.65cm}|p{0.65cm}|c|c|c||}
\hline
\multirow{2}{*}{algorithm}&\multicolumn{2}{c|}{Computation}& {Total} & \multirow{2}{*}{Inter.} & \multirow{2}{*}{Comm.}\\
 &$\ \ \ \mathcal{O}$&\ \ $\mathcal{A}$ & Time& &\\
\hline
\hline
\multicolumn{1}{||c|}{{Addition}}&\ \ \ - &1ms &1ms &0 &- \\
\multicolumn{1}{||c|}{{Subtraction}}& 35ms& 49ms& 113ms& 1& 1.17kB\\
\multicolumn{1}{||c|}{{P-C Mult}}& \ \ \ -&150ms &150ms &0 &- \\
\multicolumn{1}{||c|}{{Dot Product}}&\ \ \ - & 630ms&630ms &0 &- \\
\multicolumn{1}{||c|}{{C-C Mult}}&\ \ \  -& 1ms& 1ms&0 &- \\
\multicolumn{1}{||c|}{{Power Function}}&23ms &3ms &56ms & 1& 0.33kB\\
\multicolumn{1}{||c|}{{Exchange 1}}&152ms &71ms &253ms &1 & 0.55kB\\
\multicolumn{1}{||c|}{{Exchange 2}}& 1ms&16ms &47ms &1 & 0.55kB\\
\hline
\end{tabular}}
\end{table}

\subsubsection{Performance of Building Blocks}
We run 50 times per building blocks and report the averages, with 5-dimensional vectors.
Table \ref{table:Building Blocks Evaluation} shows the results in terms of the time consumption at data owners $\mathcal{O}$ and the model demander $\mathcal{A}$, the number of interactions (round trips) and the communication overhead.
The total time consumption includes the time spend on encryption and decryption, the computation time consumption and the mimicked network latency.
We observe that all protocols of \texttt{Heda} are efficient, with a runtime on the order of milliseconds.

\begin{table*}[!t]
\centering
\caption{Performance of Privacy-preserving ML Training Protocols Instantiated from \texttt{Heda}}\label{table:Algorithms Evaluation}
\footnotesize{
\begin{tabular}{||c|c|c|c|c|c|c||}
\hline
\multirow{2}{*}{Algorithms}&\multirow{2}{*}{Accuracy} &
\multirow{2}{*}{Interactions}&\multirow{2}{*}{Comm.} &\multirow{2}{*}{Total Time}&\multicolumn{2}{c||}{Time per participant}\\
\cline{6-7}
 &	 &  & & &$\mathcal{O}$&	$\mathcal{A}$ \\
\hline
\hline
\multicolumn{1}{||c|}{\texttt{Heda}-LR}&	96.52\%& 4000& 514.01MB&	2239.70s& 1523.34s&	626.36s\\
\multicolumn{1}{||c|}{\texttt{Heda}-SVM}&	96.13\%&3000&532.55MB&	605.95s& 193.57s&	382.38s \\
\multicolumn{1}{||c|}{\texttt{Heda}-NB}&	95.99\%&	30& 7.92MB&	4.80s& 0.74s&	3.15s\\
\hline
\end{tabular}}
\end{table*}

\begin{figure*}
\begin{minipage}[t]{0.33\textwidth}
    \centering
    \includegraphics[height=4cm]{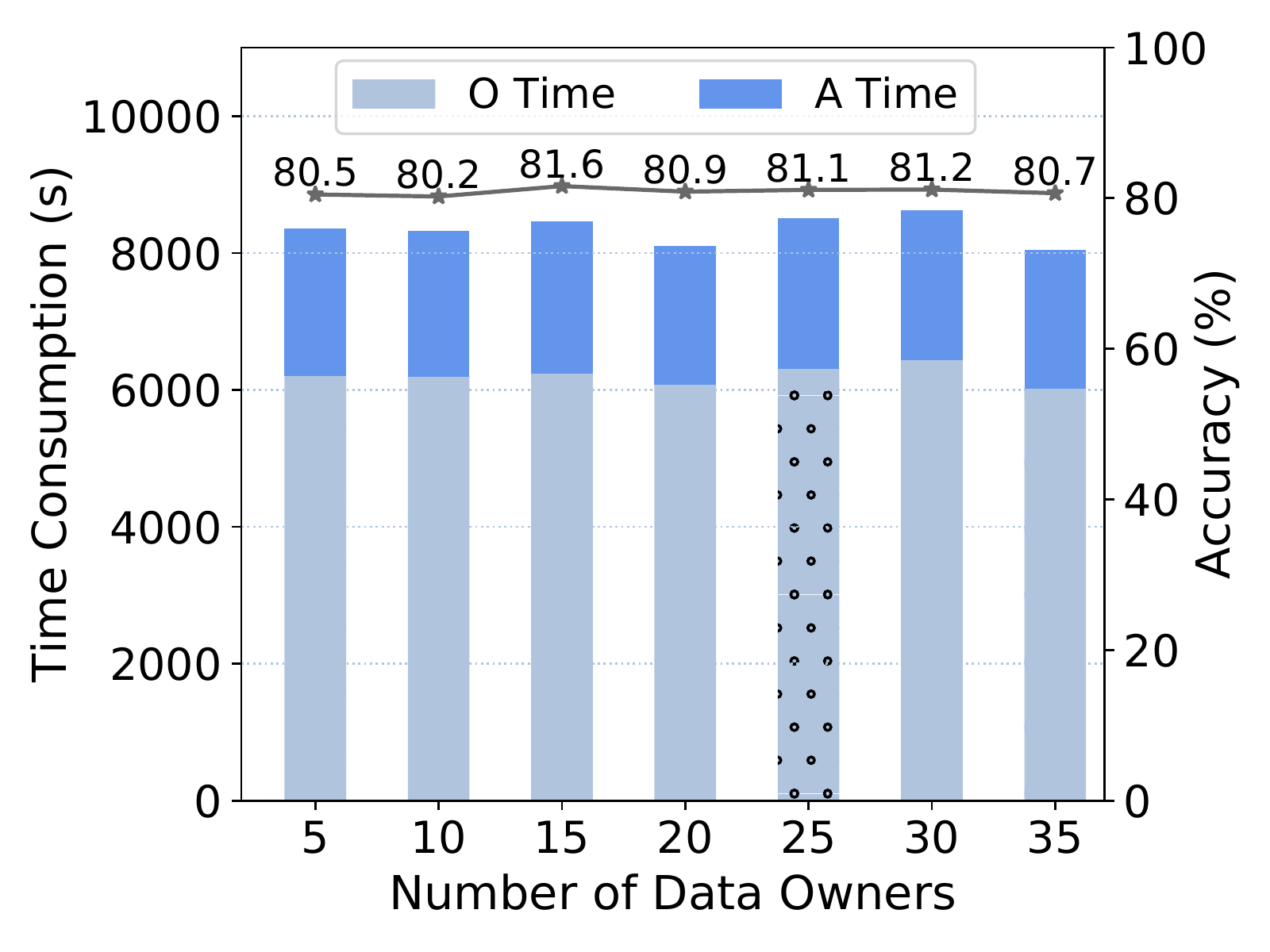}\\
    \scriptsize{(a) \texttt{Heda}-LR}
\end{minipage}%
\begin{minipage}[t]{0.33\textwidth}
    \centering
    \includegraphics[height=4cm]{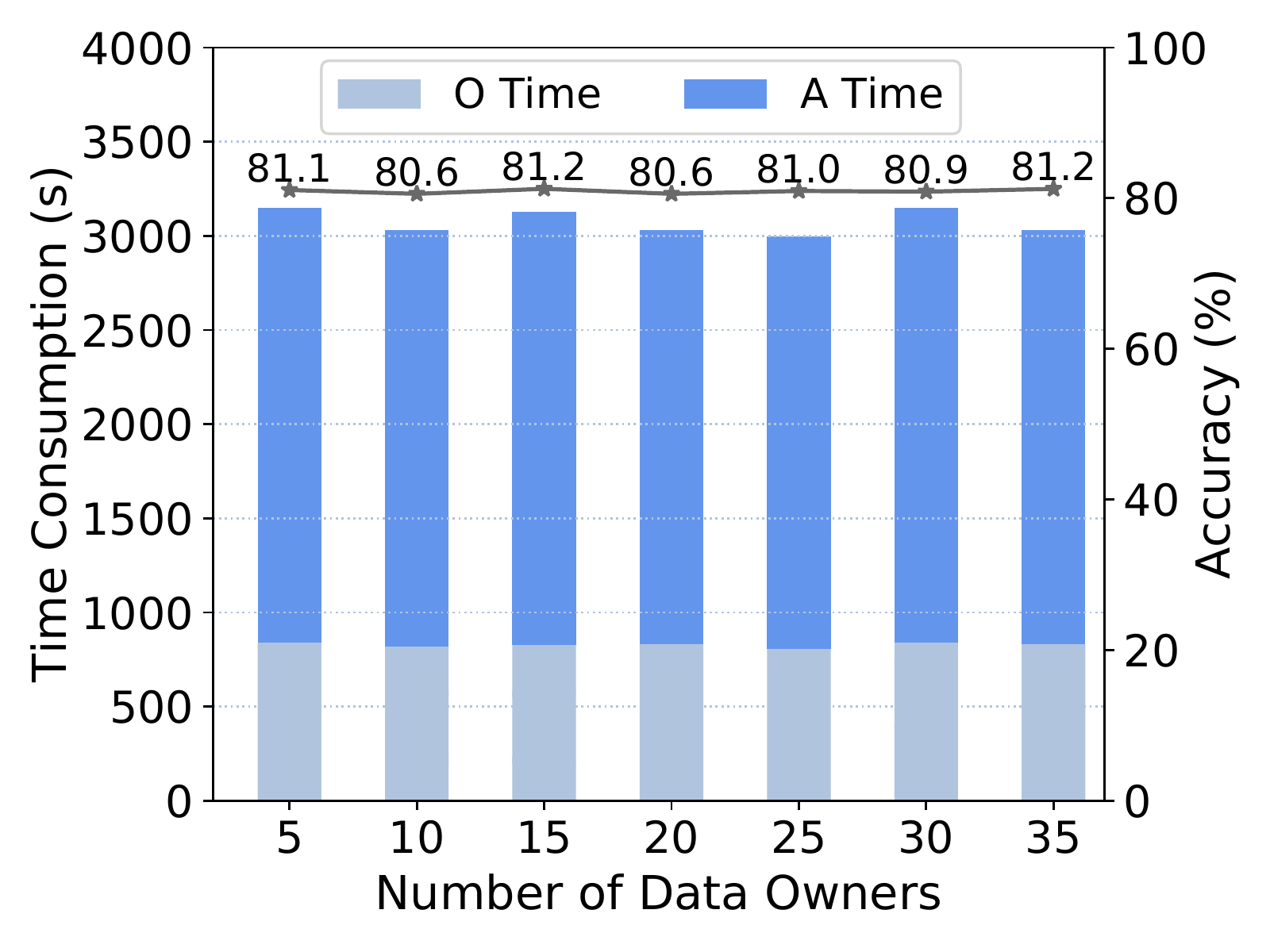}\\
    \scriptsize{(b) \texttt{Heda}-SVM}
\end{minipage}%
\begin{minipage}[t]{0.33\textwidth}
    \centering
    \includegraphics[height=4cm]{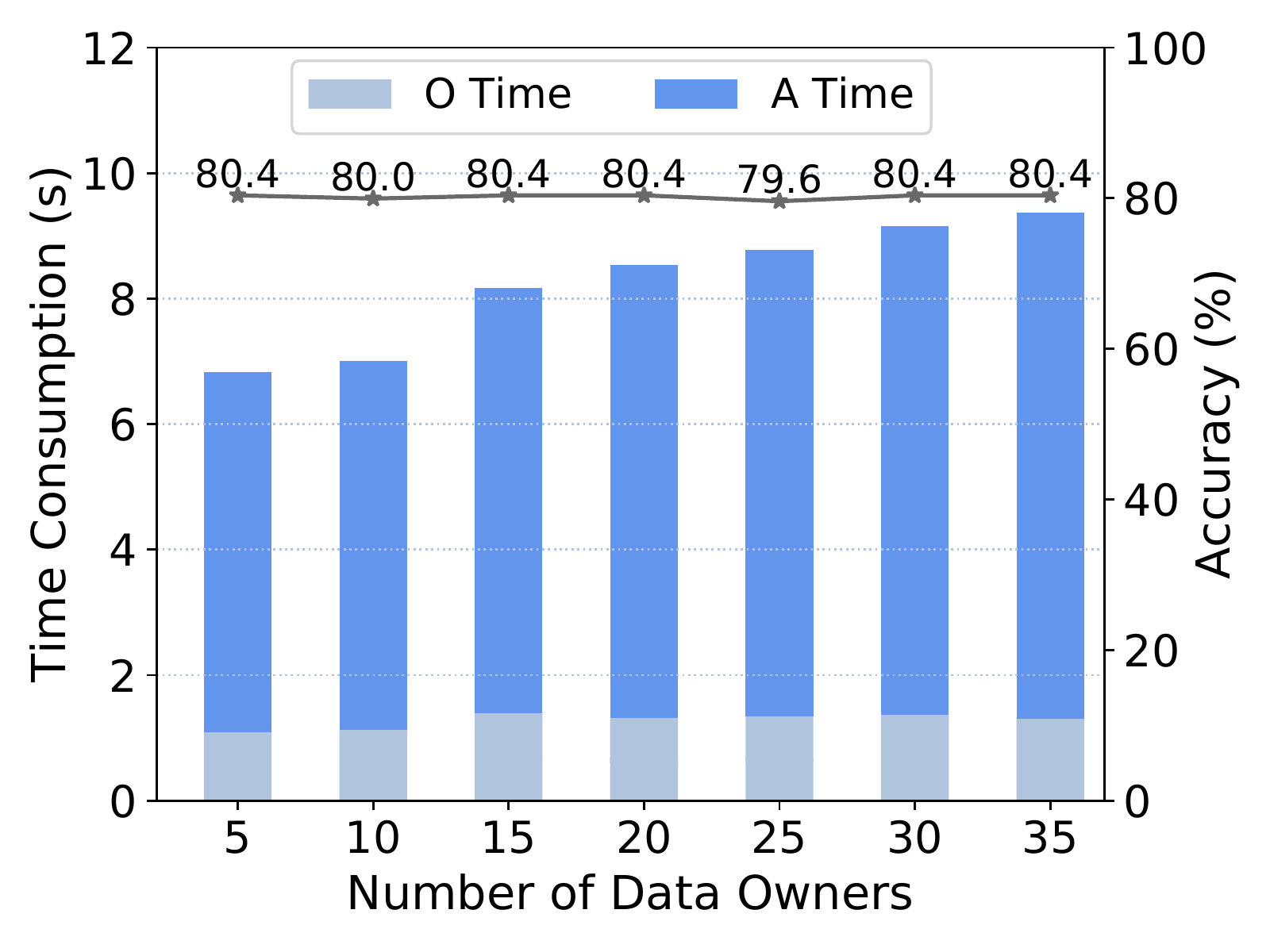}\\
    \scriptsize{(c) \texttt{Heda}-NB}
\end{minipage}%
\caption{Scalability evaluation of \texttt{Heda} when the number of participants changes, and the number of records in training datasets remains unchanged.
The abscissa shows the changes in the number of $\mathcal{O}$.
The bars represent the time consumption, and the line represents the accuracy.
}
\label{fig:Heda evaluation-Scalability_Change_Onum}
\end{figure*}

\begin{figure*}
\begin{minipage}[t]{0.33\textwidth}
    \centering
    \includegraphics[height=4cm]{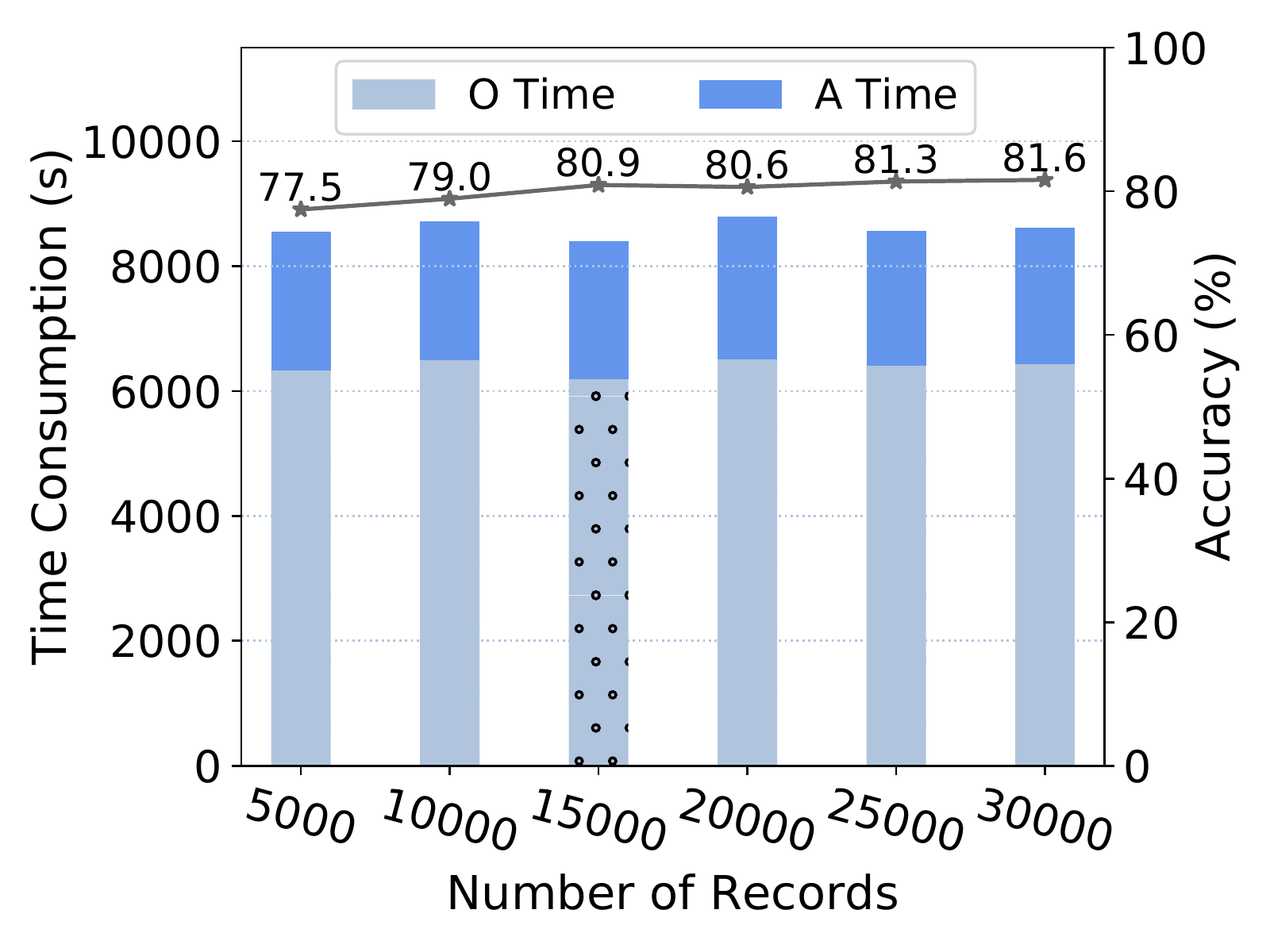}\\
    \scriptsize{(a) \texttt{Heda}-LR}
\end{minipage}%
\begin{minipage}[t]{0.33\textwidth}
    \centering
    \includegraphics[height=4cm]{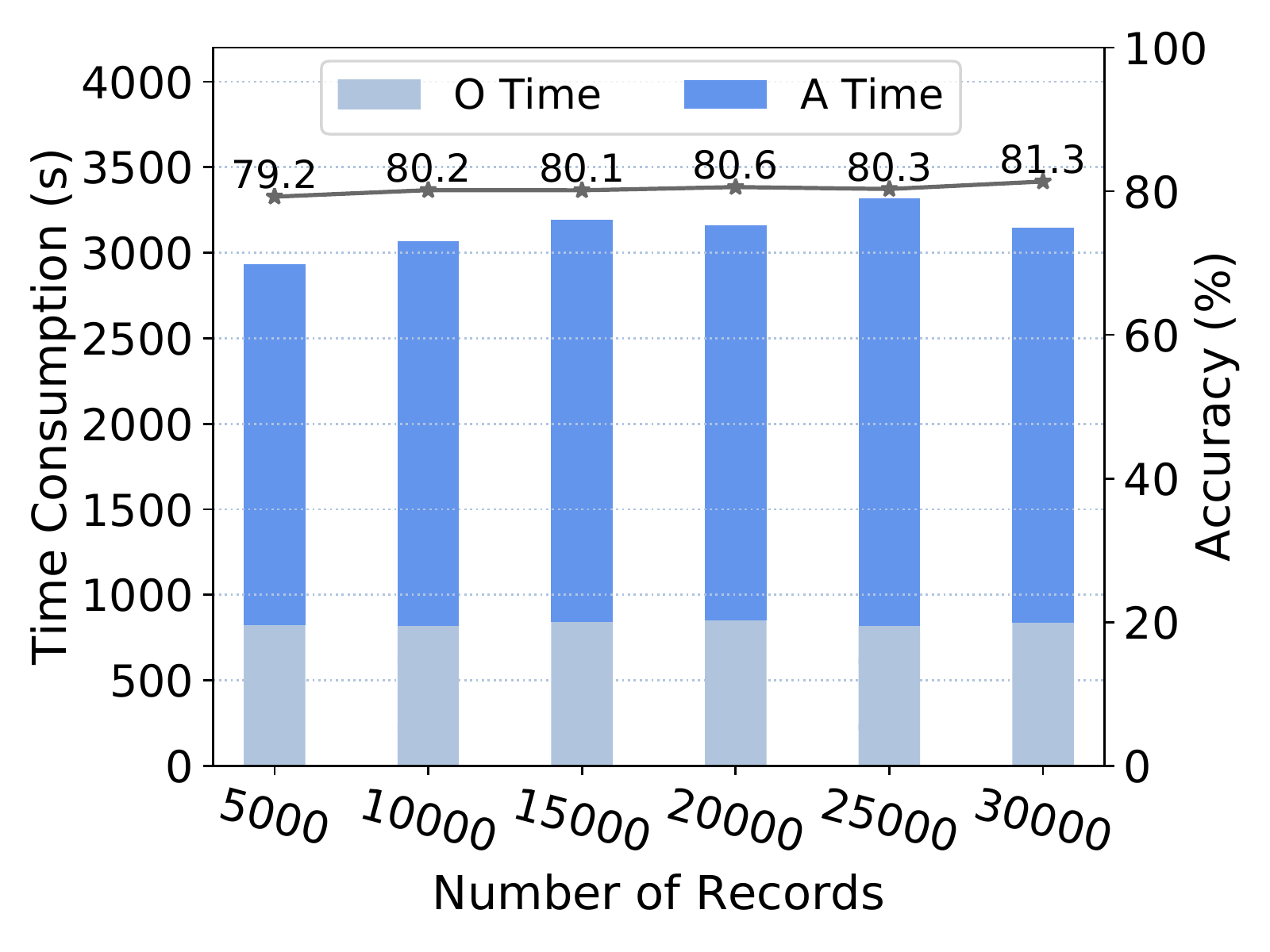}\\
    \scriptsize{(b) \texttt{Heda}-SVM}
\end{minipage}%
\begin{minipage}[t]{0.33\textwidth}
    \centering
    \includegraphics[height=4cm]{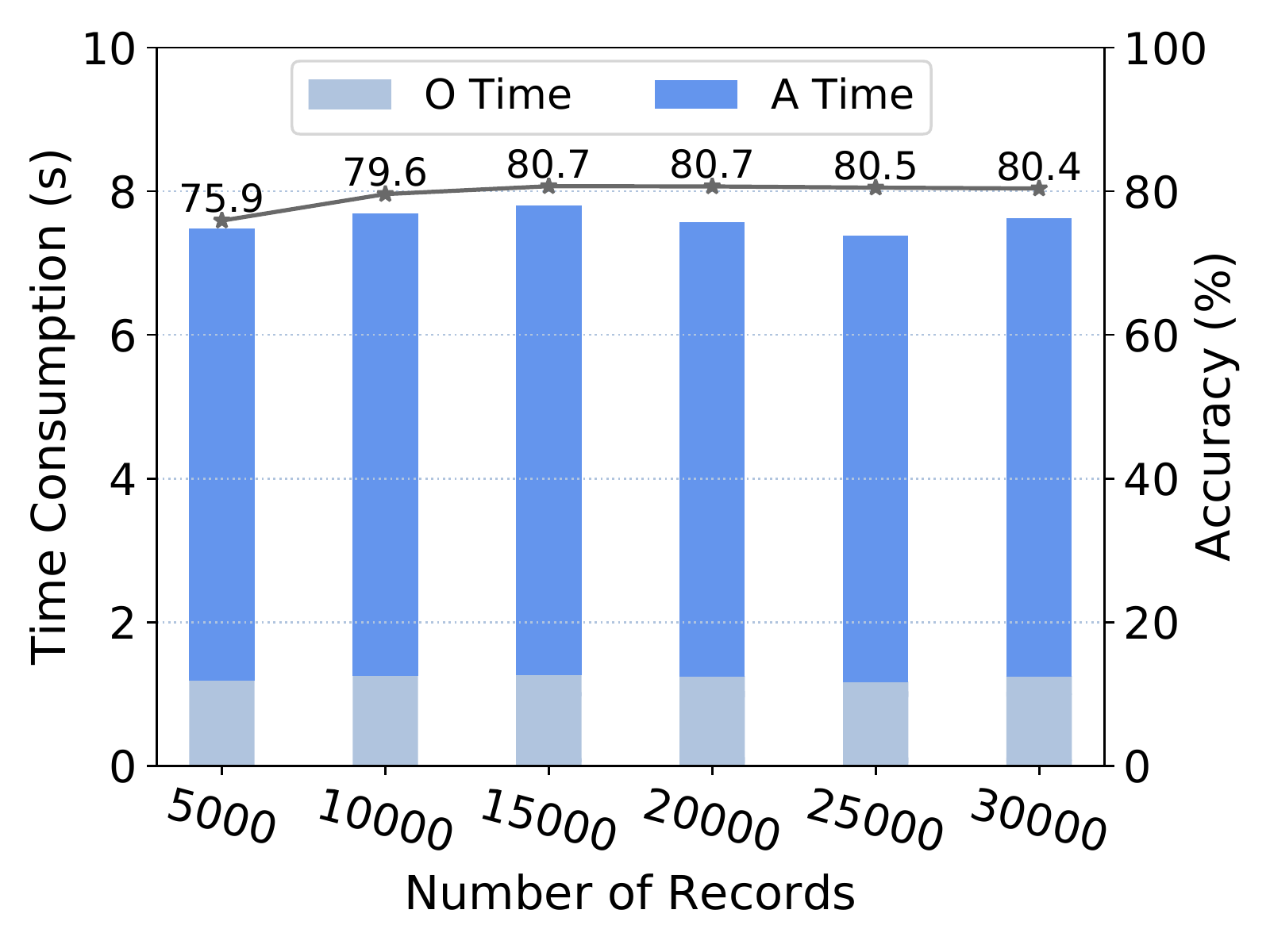}\\
    \scriptsize{(c) \texttt{Heda}-NB}
\end{minipage}%
\caption{
Scalability evaluation of \texttt{Heda} when the number of records changes, the number of participants remains unchanged.
The abscissa shows the changes in the number of records in training dataset, where $\mathcal{O}$'s number is fixed to $5$.
The bars and the line represent the time consumption and the accuracy respectively.}
\label{fig:Heda evaluation-Scalability_Change_dataset}
\end{figure*}

\subsubsection{Performance of Privacy-preserving ML Training Protocols} \label{sec:Performance of privacy-preserving ML training algorithms}
We instantiate \texttt{Heda} to three protocols to illustrate the power of \texttt{Heda}: \textit{\texttt{Heda}-SVM} (Protocol 1), \textit{\texttt{Heda}-LR} (Protocol 2) and \textit{\texttt{Heda}-NB} (Protocol 3).
We run these three protocols on datasets \textit{BCWD}.
Table \ref{table:Algorithms Evaluation} providers the overall performance.
The accuracy is evaluated by a widely used criterion ($ accuracy = \frac{\#correctly\ classified\ records}{\#total\ records}$).
The total time consumption includes the time spend on encryption and decryption, the computation time consumption and the mimic network latency.

To observe whether \texttt{Heda} causes an accuracy loss when conducing privacy-preserving ML training,
the standard implementations for LR, SVM and NB where models are trained non-privately using scikit-learn\footnote{http://scikit-learn.org} are employed as the control group.
By comparing the accuracy,
it is clear that although the three protocols deal with the encrypted dataset, they have almost no loss of accuracy.
Since the three protocols perform operations on ciphertext, the total time consumption is higher than the non-privacy ML training.
But training a model is still in an acceptable time consumption, as shown in Table \ref{table:Algorithms Evaluation}.
We believe \texttt{Heda} to be practical for sensitive applications.

The maximum number of iterations for \texttt{Heda}-LR and \texttt{Heda}-SVM are set to 1000 and 1000 respectively.
The number of interactions is determined by the maximum number of iterations in these two protocols.
The bandwidth can reach over 10MB/s in most practical application,
thus the communication overhead shown in Table \ref{table:Algorithms Evaluation} is acceptable.

\subsubsection{Comparison with Prior Works}\label{sec:Performance Evaluation-Compares}
We implement the three latest and effective solutions that considered using PHE to solve privacy-preserving ML training:
secure LR training (PLR) \cite{36},
secure SVM training (PSVM) \cite{8},
and secure NB training (PNB) \cite{113} where the privacy budget was set as $1$ according to the authors' setting.
The comparison results between the proposed protocols instantiated from \texttt{Heda} and the three solutions are reported in Table \ref{table:Performance Comparison of three Specialized privacy-preserving ML training Algorithms}.

\begin{table}[!h]
\centering
\setlength{\abovecaptionskip}{0pt}%
\setlength{\belowcaptionskip}{0pt}%
\caption{Performance Comparison with Other Schemes}\label{table:Performance Comparison of three Specialized privacy-preserving ML training Algorithms}
\footnotesize{
\subtable[Logistic Regression]{
\begin{tabular}{c|c|p{0.9cm}<{\centering}|p{0.9cm}<{\centering}|p{0.9cm}<{\centering}|p{0.9cm}<{\centering}||}
\cline{2-6}
 &\multicolumn{3}{c|}{Accuracy}&\multicolumn{2}{c||}{Time Consumption}\\
\hline
\multicolumn{1}{||c|}{Dataset}  &Original&Ours &PLR 	& Ours&	PLR  \\
\hline
\hline
\multicolumn{1}{||c|}{BCWD}&96.85\%&96.85\%&95.99\% &	2239.70s	&1849.81s	\\		
\multicolumn{1}{||c|}{Adult}&81.44\% &81.07\% &78.87\%	&8539.36s & 7581.73s\\	
\multicolumn{1}{||c|}{CAD}&85.65\% &86.23\%&83.76\%	&4382.94s &3644.10s\\
\multicolumn{1}{||c|}{Car}&72.39\% &71.99\% &70.37\%	&1793.66s &	1287.81\\
\hline
\end{tabular}}
\subtable[Support Vector Machine]{
\begin{tabular}{c|c|p{0.9cm}<{\centering}|p{0.9cm}<{\centering}|p{0.9cm}<{\centering}|p{0.9cm}<{\centering}||}
\cline{2-6}
 &\multicolumn{3}{c|}{Accuracy}&\multicolumn{2}{c||}{Time Consumption}\\
\hline
\multicolumn{1}{||c|}{Dataset}  &Original& Ours &PSVM 	& Ours&	PSVM  \\
\hline
\hline
\multicolumn{1}{||c|}{BCWD}&96.13\% &96.85\%&95.42\% &	605.95s	&650.03s	\\	
\multicolumn{1}{||c|}{Adult}&81.62\% &81.35\% &81.00\%	&3132.15s &	3269.03s\\			
\multicolumn{1}{||c|}{CAD}&86.81\% & 86.23\%&85.21\%	&1079.33s &	1153.51s\\
\multicolumn{1}{||c|}{Car}& 73.66\%& 73.03\% &72.33\%	&389.70s &	403.00s\\
\hline
\end{tabular}}
\subtable[Naive Bayes]{
\begin{tabular}{c|c|p{0.9cm}<{\centering}|p{0.9cm}<{\centering}|p{0.9cm}<{\centering}|p{0.9cm}<{\centering}||}
\cline{2-6}
 &\multicolumn{3}{c|}{Accuracy}&\multicolumn{2}{c||}{Time Consumption}\\
\hline
\multicolumn{1}{||c|}{Dataset}  &Original& Ours &PNB	& Ours&	PNB  \\
\hline
\hline
\multicolumn{1}{||c|}{BCWD}& 95.85\%&95.27\% &72.09\% &	5.01s	&6.82s	\\
\multicolumn{1}{||c|}{Adult}&80.38\% &80.70\% &70.02\%	&7.94s &	9.14s\\	
\multicolumn{1}{||c|}{CAD}& 80.57\%& 82.17\% &70.47\%	& 8.96s &	10.45s \\
\multicolumn{1}{||c|}{Car}& 78.93\%& 77.31\% &69.68\%	& 2.81s &	4.11s \\
\hline
\end{tabular}}}
\end{table}

The first thing worth to mention is that
PLR, PSVM, and PNB all need to introduce untrusted servers for completing the secure training computation,
while there are only two types of roles in \texttt{Heda}, i.e., the data owner and the model demander.
Additional roles may cause more communication delay and more privacy compromises because of the complex role scenarios.

PLR \cite{36} handles the sigmoid function by an approximate equation, which causes a loss of accuracy.
\texttt{Heda}-LR actually solve the sigmoid function without any approximate equation.
Thus the accuracy of \texttt{Heda}-LR is higher than PLR \cite{36}.
Since the sigmoid function asks for the power operation, \texttt{Heda}-LR spends more time than PLR does.
Actually, in many scenarios,
accuracy is a more important standard than time consumption,
such as in disease prediction system where experts are willing to obtain a more accurate prediction model at the expense of time consumption.

\texttt{Heda}-SVM and PSVM \cite{8} both rely on PHE,
thus there is little difference in accuracy.
While \texttt{Heda}-SVM do not need untrusted servers,
the overhead of communication and interactions is less than PSVM.
Moreover, the secure multiplication protocol in PSVM \cite{8} rely on the heavy interaction between the untrusted servers and data owners,
which cause more communication delay.
The model demander can complete the secure multiplication independently in our solution, after receiveing the encrypted data.


PNB \cite{113} protects the initial data by adding noise (i.e., DP mechanism),
which inevitably reduces the accuracy of the trained model.
\texttt{Heda}-NB does not harm for the accuracy, which is confirmed in Table \ref{table:Performance Comparison of three Specialized privacy-preserving ML training Algorithms}.

\subsection{Scalability Evaluation of \texttt{Heda}}\label{sec:Performance Evaluation-Scalability}
We evaluate the scalability of \texttt{Heda} in the two aspects:
(i) changing the number of participants, and keeping the number of records in training datasets unchanged.
(ii) changing the number of records in training datasets, and keeping the number of participants unchanged.

\subsubsection{Changing Participant Number}  \label{sec:Changing the number of participants}
We divide dataset \textit{Adult} into $n$ equal parts to simulate $n$ data owners participating in the same protocol, and varied the value of $n$.
A series of experiments are carried out respecting to different $n$ values.
The number of records in training datasets remains unchanged, i.e., the whole dataset \textit{Adult} was employed in each experiment.
Experimental results are plotted in Fig. \ref{fig:Heda evaluation-Scalability_Change_Onum}.

SGD is chose as the optimization method in \texttt{Heda}-LR and \texttt{Heda}-SVM.
Theoretically, in these two protocols,
after data owners send the encrypted datasets to the model demander,
the time consumption is only related to the iteration number and the number of the training samples selected at each iteration,
which is confirmed in Fig. \ref{fig:Heda evaluation-Scalability_Change_Onum}(a) and \ref{fig:Heda evaluation-Scalability_Change_Onum}(b).
On the other hand, when the data quality is unchanged,
the increasing number of data owners does not affect the accuracy of the trained model.
A slight fluctuation of accuracy shown in Fig. \ref{fig:Heda evaluation-Scalability_Change_Onum}(a) and Fig. \ref{fig:Heda evaluation-Scalability_Change_Onum}(b) is ascribed to the random selection of initialization parameters and SGD.

In \texttt{Heda}-NB,
after obtaining the noised summation,
the model demander interacts with data owners to eliminate the noise in \textit{Secure Summation}.
Thus, with the increasing number of data owners, the time consumption increases in \texttt{Heda}-NB.
As shown in Fig. \ref{fig:Heda evaluation-Scalability_Change_Onum}(c), the increase in time consumption is not violent (more moderate than linear growth).

\subsubsection{Changing Record Number} \label{sec:Changing the number of records in training datasets}
We vary the record number of training datasets from 5000 to 30000 at the interval of 5000, upon dataset \textit{Adult}.
Then, we conduct a series of experiments respecting to these training datasets with different record numbers,
where the number of data owners is fixed as 5 in these experiments.
Experimental results are visualized in Fig. \ref{fig:Heda evaluation-Scalability_Change_dataset}.

Experiment results confirm that supervised ML appreciates large and comprehensive training data \cite{61}.
As the number of records in training data increases, the accuracy of the trained model in \texttt{Heda}-LR and \texttt{Heda}-SVM increases to a certain extent.
In \texttt{Heda}-NB, the time consumption is related to the number of data owners and the number of attributes,
and in \texttt{Heda}-LR and \texttt{Heda}-SVM the time consumption is related to the number of iterations and the number of the training samples selected at each iteration,
thus the changes in the record number does not affect the time consumption.

\section{Conclusion}\label{sec:conclusion}
In this paper, we propose a novel privacy-preserving ML training framework named \texttt{Heda} for securely handling IoT data collected from diverse IoT devices.
By applying Paillier and {\small Cloud-RSA}, we develop a library of building blocks based on partial homomorphic encryption to support training multiple ML models in the aggregation scenario of IoT.
We demonstrate the efficiency and the security of \texttt{Heda} through rigorous security analysis and extensive experiments.
In the future work, we plan to explore a more outstanding solution enabling the balance between efficiency and accuracy for privacy-preserving ML training.


%
\appendix

\subsection{Problem Description}\label{app:related_work-Problem Description}

We devote to the privacy-preserving ML training in aggregation scenario,
while many studies exploring the privacy-preserving ML training fall in collaborative scenarios.
This section presents the formal description of the aggregation scenario and collaborative scenario, followed with the differences description between the two.

In aggregation scenario,
an untrusted model demander wishes to train a ML model over the training dataset gathered from multiple data owners (organizations).
Each data owner sends the encrypted dataset encrypted by its own private key to the model demander.
During the whole training process, the sensitive model information of the model demander is not revealed to the data owners.
And the dataset of each data owner is also confidential for the model demander.

In collaborative scenarios, a set of data owners want to train ML models on their joint data.
These data owners outsource the computation to several untrusted servers.
During the training process, each data owner cannot learn the sensitive data of other data owners, but the model information is known to both the data owners and the untrusted servers.

More specifically, they differ in
\begin{itemize}
\item Different Model Demanders:
the demander in aggregation scenario is the untrusted model demander,
and in collaborative scenario is each data owner.
\item Different Privacy Guarantee:
in aggregation scenario, the untrusted model demander cannot learn anything about the data owners' sensitive data, and each data owner learn neither the model nor the sensitive data of other data owners;
in collaborative scenario, the untrusted servers learning nothing about the data owners' sensitive data, while the model is known by every participant.
\item Different Compute Mode:
in aggregation scenario, the untrusted model demander interacts with the data owners for training models;
in collaborative scenario, the data owners generally outsource part or all computations to the untrusted servers.
\end{itemize}

\subsection{Security Definition}\label{app:Security Definition}
\begin{myDef}[Computational Indistinguishability \cite{1}]\label{def:Computational Indistinguishability}
Two probability ensembles $\mathcal{X}=\{X_n\}_{n\in \mathbb{N}}$ and $\mathcal{Y}=\{Y_n\}_{n\in \mathbb{N}}$ are computationally indistinguishable,
denoted $\mathcal{X} {\equiv}_c \mathcal{Y}$,
if for every probabilistic polynomial-time distinguisher $D$ there exists a negligible function $\mathsf{negl}$ such that:
\[ \underset{x\leftarrow X_n}{Pr}[D(1^n,x)=1]-\underset{y\leftarrow Y_n}{Pr}[D(1^n,y)=1] ||\leq \mathsf{negl}(n) \]
\end{myDef}
For more details of probabilistic polynomial-time distinguisher and negligible function, we refer the reader to \cite{1}.

\textbf{Modular Sequential Composition.}
Since all our protocols are designed and constructed in a modular way, we employ Modular Sequential Composition \cite{15} for justifying the security proofs of our protocols, the idea of which is that:

(i) A and B run a protocol $\pi$ and use calls to an ideal functionality $f$.
If we can show that $\pi$ respects privacy in the honest-but-curious model
and we have a protocol $\rho$ that privately computes $f$ in the same model,
then we can replace the ideal calls for $f$ by the execution of $\rho$ in $\pi$;
The new protocol, denoted ${\pi}^{\rho}$ is then secure in the honest-but-curious model.

(ii) A protocol $\pi$ is a hybrid model that uses calls to $f_1,\ldots,f_n$.
The hybrid model with ideal access to $f_1,\ldots,f_n$ is augmented with an incorruptible trusted party T.
The running of protocol $\pi$ contains calls to T for the execution of one of $f_1,\ldots,f_n$.
Since the proposed protocols in this paper are considered as sequential composition,
each party sends its input and wait until the trusted party sends the output back for each call.
Let $\rho_1,\ldots,\rho_n$ be real protocols in the semi-honest model securely computing $f_1,\ldots,f_n$.
In ${\pi}^{\rho_1,\ldots,\rho_n}$, $f_i$ is replaced by a real execution of $\rho_i$.
During the executing processes of ${\pi}^{\rho_1,\ldots,\rho_n}$, if a participates $P_i$ of ${\pi}^{\rho_1,\ldots,\rho_n}$ has to compute $f_j$ in the protocol $f_i$ with input $x_i$,
$P_i$ halts, calling T for starting an execution of $\rho_j$ with the other parties, and continues until T returns the results $\beta_j$.

\subsection{Machine Learning Preliminary}\label{app:Machine Learning Preliminary}
On the input of an unencrypted training dataset, ML training algorithms output the ML model parameters $\theta$.
After the objective optimization function is given, ML training algorithms updates the model parameters $\theta$ through a certain parameter optimization strategy.
In this paper, we employ Stochastic Gradient Descent (SGD) as the parameter optimization algorithm in SVM, LR and NB training algorithms.

\subsection{Stochastic Gradient Descent}
Gradient descent is an optimization algorithm for finding the local minimum of a function, which updates the parameters by using all the records in train dataset.
Given a loss function $\mathcal{J}(\theta)$,
the parameters updating equation of gradient descent is
$\theta := \theta - \lambda \frac{\partial \mathcal{J}(\theta)}{\partial\theta}$
,
where $\lambda$ is the learning rate and $\theta$ is the learnable parameters of models.

When the number of the records is large or the gradient calculation is time-consuming,
the updating process using all samples is slow.
Instead of using all samples, SGD randomly selects a small number of samples to update the gradient at each iteration,
which is typically fast in practice \cite{115}.
This paper uses SGD as the optimization algorithm for LR and SVM model training.

\subsection{Support Vector Machine}
SVM gives the maximum-margin hyperplane that might classify the test data.
The form of the hyperplane is expressed as {\small$f \{ \vec{x} \}=\theta^{\sf{T}} \hat{\vec{x}},\ \hat{\vec{x}}:=[\vec{x};1]$}.
For binary classification, if {\small$\theta^{\sf{T}} \hat{\vec{x}} \geq 1$}, the prediction label {\small$y_{i}= +1$};
otherwise the prediction label {\small$y_{i}=-1$}.
The optimization problem of the primary of SVM as Equation \eqref{eq:primary SVM}.
SVM can perform the non-linear classification by replacing the $\vec{x}$ in {\small$f \{ \vec{x} \}=\theta^{\sf{T}} \hat{\vec{x}}$} with a kernel trick $\kappa( \hat{\vec{x}})$.
In this paper, we consider the plain type of SVM, i.e., $\kappa( \hat{\vec{x}}):= \hat{\vec{x}}$.

\begin{equation}\label{eq:primary SVM}
\small
\begin{split}
     \mathop{\min }\limits_{\theta}&\  \frac{1}{2}{|| \theta ||^2}\\
     s.t.\ \ y(\theta^{\sf{T}} \hat{\vec{x}}) \ge 1&,\ i = 1,2, \ldots  \ldots |D|
\end{split}
\end{equation}

Employing Hinge Loss, we define a differentiable loss function which is shown in Equation \eqref{eq:objective function of SVM}.
\begin{equation}\label{eq:objective function of SVM}
\small
 \mathop{\min }\limits_{\theta} \sum_{i=1}^{|D|}max(0,1-y_i\theta^{\sf{T}} \hat{\vec{x}}_i)+ \lambda||\theta||^2
\end{equation}

We use SGD for solving the parameter $\theta$ in Equation \eqref{eq:objective function of SVM}.
For a records $\hat{\vec{x}}_i$ selected from the training dataset $D$,
the iteration formula of SVM training algorithm is shown in Equation \eqref{eq:iteration formula of SVM},
where $\alpha$ is the regularization coefficient.

\begin{equation}\label{eq:iteration formula of SVM}
\small
 \theta := \begin{cases}
 \theta-\lambda(\alpha\theta-\alpha y_i \hat{\vec{x}}_i),\ &y_i\theta^{\sf{T}} \hat{\vec{x}}_i< 1 \\
 \theta-\lambda(\alpha\theta),\ &y_i\theta^{\sf{T}} \hat{\vec{x}}_i\geq1 \\
 \end{cases}
\end{equation}

\subsection{Logistic Regression}
LR is a binary classifier that tries to learn parameters $\theta$,
where {\small$\theta=[\theta_1,\theta_2,\ldots,\theta_d]$} to satisfy {\small$f\{ \vec{x}_i \}=\theta^{\sf{T}}\vec{x}_i$} and {\small$f \{ \vec{x}_i \}\cong y_i$}.
$y_i$ is the probability that $\vec{x}_i$ belongs to the positive class.
LR uses the Sigmoid function to associate the true label {\small$y_i$} with the prediction label {\small$f \{ \vec{x}_i \}$}: {\small$h_\theta(\vec{x}_i) =\frac{e^{\theta^{\sf{T}}\vec{x}_i}}{1+e^{\theta^{\sf{T}} \vec{x}_i}}$}.
Choosing Cross Entropy as the cost function $J(\theta)$.
\begin{equation}
\small
 J(\theta)=-\frac{1}{m}[\sum_{i=1}^{m}(y_i log h_\theta(\vec{x}_i) + (1-y_i) log (1-h_\theta(\vec{x}_i)))]
\end{equation}
We update the parameters $\theta$ with SGD, solving the parameter values that can minimize $J(\theta)$.
The iteration formula of LR training algorithm is shown in Equation \eqref{eq:iteration formula of LR}.
\begin{equation}\label{eq:iteration formula of LR}
\small
  \theta _{j}:=\theta _j-\lambda {x_{ij}}(y_i-\frac{1}{1+e^{-\theta^{\sf{T}}\vec{x}_i}})
\end{equation}

\subsection{Naive Bayes}
The training process of NB is to compute the class prior probability {\small$P(y)$} and the conditional probability {\small$P(\vec{x}_i |y)$},
where {\small$y$} denotes a certain class category in $\mathcal{D}$.
The NB classifier is expressed as: {\small$f(\vec{x}_i) = argmax\;P( y )\mathop \prod \limits_{i = 1}^d P({\vec{x}_i}|{y})$}.

The class prior probability {\small$P(y)$} can be obtained through the maximum likelihood estimation,
i.e, {\small$P(y)=\frac{|\mathcal{D}(y)|}{m}$}, where $|\mathcal{D}(y)|$ is the number of the class $y$ in $\mathcal{D}$.

For discrete attributes,
the conditional probability for the discrete attributes is {\small$P(x_{ij}|y)=\frac{|{\mathcal{D}}^{y}(x_{ij})|}{m}$}
where $|{\mathcal{D}}^{y}(x_{ij})|$ is the number of the number of the j-th attribute having the value $x_{ij}$ with the class $y$ in $\mathcal{D}$.

As for the numeric attributes, probability density function is considered for computing the conditional probability {\small$P(\vec{x}_i |y)$}.
Assuming that {\small$P({x_{ij}}|{y})$} obeys normal distribution {\small$N(\mu_{y, j}, \sigma_{y, j}^{2})$}, where {\small$\mu_{y, j}$} and {\small$\sigma_{y, j}^{2}$} are mean and variance respectively for the j-th attributes of the class $y$ respectively.
Then, the conditional probability {\small$P(x_{ij}|y)$} can be obtained from Equation \eqref{eq:NB conditional probability}.
\begin{equation}\label{eq:NB conditional probability}
\small
  P(x_{ij}|y) = \frac{1}{\sqrt{{2\pi}}\sigma_{y, j}}exp(-\frac{(x_{ij}-\mu_{y,j})^2}{2\sigma_{y, j}^2})
\end{equation}

\subsection{Security Analysis for Building Blocks of Primitive Operations}\label{app:Security Analysis for Building Blocks}
When facing the honest-but-curious adversaries,
we follow secure two-party computation for guaranteeing the security of the proposed building blocks.
Here, we present our security proofs according to the ideas of secure two-party computation.
That is, all the message that can be computed out by a participant from the intermediate result must be included in the message that can be computed from its input and output.

\begin{mypro}\label{Propo:four building blocks}
Secure addition,
secure subtraction,
secure plaintext-ciphertext multiplication,
and secure piphertext-ciphertext multiplication is secure in the
honest-but-curious model.
\end{mypro}

\begin{myproof}[for Proposition \ref{Propo:four building blocks}]
The input of the data owner is ({\small$(\sf{SK},\sf{PK})_{Paillier}$},\ a).
As the data owner does not receive any message and call to any other protocols or algorithms during the execution processes, his view only consists in its input.
Hence, the simulator $S_{\mathcal{O}}$ simply generate random coins:
\[
  S_{\mathcal{O}}( a )= (a, \textsf{coins}) = view_{\mathcal{O}}( a,b)
\]

The input of the model demander is ({\small$b$}).
{\small$view_{\mathcal{A}}(a,b)=\{b,[b],[a];Output_{\mathcal{A}} \}$}.
{\small$Output_{\mathcal{A}} = [f(a,b)]$}.
The simulator $S_{\mathcal{A}}$ does the following:
\begin{itemize}
  \item Generates the random coins necessary for re-randomization and put them in $\hat{\textsf{coins}}$.
  \item Generates an encryption using the public key of the data owner: $[c]$.
  \item Outputs $\{[b],[c],\hat{\textsf{coins}}; [f(c,b)]\}$.
\end{itemize}
$\textsf{coins}$ and $\hat{\textsf{coins}}$ come from the same distribution, independently from other parameters.
Thus,
\[
  \{[b],[c],\hat{\textsf{coins}}; [f(c,b)]\} = \{[b],[c],\textsf{coins};[f(a,b)]\}
\]
and by ciphertext indistinguishability against chosen plaintext attacks of Paillier,
\[
  \{[b],[c],\textsf{coins}; [f(c,b)]\} {\equiv}_c \{[b],[a],\textsf{coins}; [f(a,b)]\}
\]

$[a]$ and $[f(a,b)]$ is encrypted by the public key of the data owner,
the confidentiality of which are equivalent to the cryptosystem.
By semantic security of Paillier,
{\small$S_{\mathcal{A}}( b,F( a,b ) )\ {\equiv}_c\ view_{\mathcal{A}}( a,b)$}.
\qed
\end{myproof}

\begin{mypro}\label{Propo:secure power function}
Secure power function is secure in the
honest-but-curious model.
\end{mypro}

\begin{myproof}[for Proposition \ref{Propo:secure power function}]
The input of data owner is ({\small$(\sf{SK},\sf{PK})_{CloudRSA},\ e^{\vec{a}}$}),
As the data owner does not receive any message and call to any other protocols during the execution processes, his view only consists in its input.
Hence, the simulator $S_{\mathcal{O}}$ simply generate random coins:
\[
  S_{\mathcal{O}}( e^{\vec{a}} )= (e^{\vec{a}}, \textsf{coins}) = view_{\mathcal{O}}( e^{\vec{a}},\vec{b})
\]

The input of model demander is ({\small$\vec{b}$}).
{\small$ view_{\mathcal{A}}(a,b)=\{\vec{b},[e^{\vec{a}}]_R,\{\prod_{i=1}^j[e^{a_ib_i}]_R\}_{j=1}^d;Output_{\mathcal{A}}\}$}.
{\small$Output_{\mathcal{A}} = [e^{\vec{a\cdot b}}]_R$}.
The simulator $S_{\mathcal{A}}$ does the following:
\begin{itemize}
  \item  Picks $2d$ random numbers that are limited to $\mathbb{Z}_N$: $(\hat{a_1},...,\hat{a_d},c_1,...,c_d)$.
  \item Encrypts $(\hat{a_1},...,\hat{a_d})$ by {\small$\sf{PK}_{CloudRSA}$}: $[e^{\hat{\vec{a}}}]_R:=\{[\hat{a_1}]_R,...,[\hat{a_d}]_R\}$.
  \item Encrypts $(c_1,...,c_d)$ by {\small$\sf{PK}_{CloudRSA}$}.
  \item Generates the random coins necessary for re-randomization and put them in $\hat{\textsf{coins}}$.
  \item Outputs $\{\vec{b},[e^{\hat{\vec{a}}}]_R,\hat{\textsf{coins}},\{[c_i]_R\}_{i=1}^d;[e^{\vec{\hat{a}\cdot b}}]_R\}$.
\end{itemize}
$\textsf{coins}$ and $\hat{\textsf{coins}}$ come from the same distribution, independently from other parameters.
Thus,
\[
\begin{split}
  \{\vec{b},[e^{\hat{\vec{a}}}]_R,\hat{\textsf{coins}},\{[c_i]_R\}_{i=1}^d;&[e^{\vec{\hat{a}\cdot b}}]_R\}\\
  = \{\vec{b},[e^{\hat{\vec{a}}}]_R,{\textsf{coins}},\{[c_i]_R\}_{i=1}^d;&[e^{\vec{\hat{a}\cdot b}}]_R\}
\end{split}
\]
and by ciphertext indistinguishability against chosen plaintext attacks of {\small Cloud-RSA},
\[
\begin{split}
    \{\vec{b},[e^{\hat{\vec{a}}}]_R,{\textsf{coins}},\{[c_i]_R\}_{i=1}^d;&[e^{\vec{\hat{a}\cdot b}}]_R\} \\
  {\equiv}_c  \{\vec{b},[e^{\vec{a}}]_R,\textsf{coins},\{\prod_{i=1}^j[e^{a_ib_i}]_R\}_{j=1}^d;&[e^{\vec{a\cdot b}}]_R\} \\
  = \{\vec{b},[e^{\vec{a}}]_R,\textsf{coins},\{\prod_{i=1}^j[e^{a_ib_i}]_R\}_{j=1}^d;&Output_{\mathcal{A}}\}
\end{split}
\]

$[e^{\vec{a}}]_R$, $\prod_{i=1}^d[e^{a_ib_i}]_R$ and $[e^{\vec{a\cdot b}}]_R$ is encrypted by {\small${\sf{PK}_{CloudRSA}}$},
the confidentiality of which are equivalent to the cryptosystem.
By semantic security of {\small CloudRSA},
$S_{\mathcal{A}} {\equiv}_c  view_{\mathcal{A}}$

\qed
\end{myproof}

\begin{mypro}\label{Propo:secure summation}
Secure summation is secure in the honest-but-curious model.
\end{mypro}

\begin{myproof}[for Proposition \ref{Propo:secure summation}]
We prove the security of the secure summation without collusion first.

\noindent\textbf{For the first $n-1$ data owners:}\\
The input of $\mathcal{O}_i$ is $(({\sf{SK},\sf{PK})_{Paillier}^i;} a_i,r_i,a_i+r_i,[r_i]_P^i)$.
Each $\mathcal{O}_i$'s view is
\[
view_{\mathcal{O}_i}=\{a_i,r_i,a_i+r_i,[r_i]_P^i;out_{block9} \}
\]
where $out_{block9}$ means the received output from Building Blocks 8 in secure summation.
\[
out_{block9}=(\sum_{c=1}^i a_c + \sum_{c=i}^n(a_c+r_c))
\]
We construct a simulator {\small$S_{\mathcal{O}}$} which runs as follows:
\begin{itemize}
  \item Picks a random $c \in \mathbb{Z}_N$;
  \item Outputs {\small$\{a_i,r_i,a_i+r_i,[r_i]_P^i;c\}$}.
\end{itemize}
The distribution of $c$ and $\sum_{c=1}^i a_c + \sum_{c=i}^n(a_c+r_c)$ are identical, so the real distribution $\{a_i,r_i,a_i+r_i,[r_i]_P^i;out_{block9} \}$ and the ideal distribution $\{a_i,r_i,a_i+r_i,[r_i]_P^i;c \}$ are statistically indistinguishable,
i.e., {\small$S_{\mathcal{O}_i}\ {\equiv}_c\ view_{\mathcal{O}_i}$}.

\noindent\textbf{For the last data owners $\mathcal{O}_n$:}\\
The input of $\mathcal{O}_n$ is $(({\sf{SK},\sf{PK})_{Paillier}^n};a_n,r_n,a_n+r_n,[r_n]_P^n)$.
$\mathcal{O}_n$'s view is
\[
view_{\mathcal{O}_n}=\{a_n,r_n,a_n+r_n,[r_n]_P^n;out_{block9} \}
\]
\[
out_{block9}=(\sum_{i=1}^n a_i + r_{\mathcal{A}})
\]
We construct a simulator {\small$S_{\mathcal{O}_n}$} which runs as follows:
\begin{itemize}
  \item Picks a random $c \in \mathbb{Z}_N$;
  \item Outputs {\small$\{a_n,r_n,a_n+r_n,[r_n]_P^n;c\}$}.
\end{itemize}
The distribution of $c$ and $\sum_{i=1}^n a_i + r_{\mathcal{A}}$ are identical, so the real distribution $\{a_n,r_n,a_n+r_n,[r_n]_P^n;out_{block9} \}$ and the ideal distribution $\{a_n,r_n,a_n+r_n,[r_n]_P^n;c\}$ are statistically indistinguishable,
i.e., {\small$S_{\mathcal{O}_n}\ {\equiv}_c\ view_{\mathcal{O}_n}$}.

\noindent\textbf{For the model demander $\mathcal{A}$:}\\
The input of model demander $\mathcal{A}$ is $(\sf{SK},\sf{PK})_{Paillier}^{\mathcal{A}}$.
The view of model demander is
\[
\begin{split}
  view_{\mathcal{A}}=\{r_{\mathcal{A}},\{(a_i+r_i)\}_{i=0}^n,\{[r_i]_P^i\}_{i=0}^n,&\\
  \{[\sum_{i=0}^na_i+\sum_{i=c}^nr_i]_P^c\}_{c=0}^n ;out_{block9}\}&
\end{split}
\]
where $out_{block9}$ means the received output from Building Blocks 8 in secure summation.
\[
out_{block9}:=
\begin{cases}
 [\sum_{c=1}^i a_c + \sum_{c=i}^n(a_c+r_c)]_P^i,\ &1\leq i < n \\
 \sum_{i=1}^n a_i,\ &i=n
\end{cases}
\]
We construct a simulator {\small$S_{\mathcal{A}}$} which runs as follows:
\begin{itemize}
  \item Picks $2n$ random numbers that are limited to $\mathbb{Z}_N$: $(\hat{a_1},...,\hat{a_n},\hat{r_1},...,\hat{r_n})$.
  \item Generates $\{(\hat{a}_i+\hat{r}_i)\}_{i=0}^n$.
  \item Generates the random coins necessary for re-randomization and put them in $\hat{\textsf{coins}}$.
  \item Encrypts $\{\hat{r_i}\}_{i=1}^{n}$ by corresponding {\small$\sf{PK}_{Paillier}^i$}.
  \item Encrypts $\{\sum_{i=0}^n\hat{a}_i+\sum_{c}^n\hat{r}_i\}_{c=1}^{n}$ by corresponding {\small$\sf{PK}_{Paillier}^i$}.
  \item Generates $\sum_{i=0}^n \hat{a_i}$.
  \item Outputs:
  \[\small
\begin{split}
  \{\hat{\textsf{coins}}, \{(\hat{a}_i+\hat{r}_i)\}_{i=0}^n,\{[\hat{r}_i]_P^i\}_{i=0}^n,\{[\sum_{i=0}^n\hat{a}_i+\sum_{i=c}^n\hat{r}_i]_P^c\}_{c=0}^n&\\
   ;\{[\sum_{c=1}^i \hat{a}_c + \sum_{c=i}^n(\hat{a}_c+\hat{r}_c)]_P^i\}_{i=0}^{n-1},\sum_{i=0}^n \hat{a_i}\}&
\end{split}
\]
\end{itemize}
$\textsf{coins}$ and $\hat{\textsf{coins}}$ come from the same distribution, independently from other parameters,
and the distribution of $(\hat{a_1},...,\hat{a_n},\hat{r_1},...,\hat{r_n})$ and $({a_1},...,{a_n},{r_1},...,{r_n})$ are identical, so the real distribution $({a_1},...,{a_n},{r_1},...,{r_n})$ and the ideal distribution $(\hat{a_1},...,\hat{a_n},\hat{r_1},...,\hat{r_n})$ are statistically indistinguishable.
\[\small\begin{split}
  \{\hat{\textsf{coins}}, \{(\hat{a}_i+\hat{r}_i)\}_{i=0}^n,\{[\hat{r}_i]_P^i\}_{i=0}^n,\{[\sum_{i=0}^n\hat{a}_i+\sum_{i=c}^n\hat{r}_i]_P^c\}_{c=0}^n&\\
   ;\{[\sum_{c=1}^i \hat{a}_c + \sum_{c=i}^n(\hat{a}_c+\hat{r}_c)]_P^i\}_{i=0}^{n-1},\sum_{i=0}^n \hat{a_i}\}&\\
   {\equiv}_c \{{\textsf{coins}}, \{({a}_i+{r}_i)\}_{i=0}^n,\{[\hat{r}_i]_P^i\}_{i=0}^n,\{[\sum_{i=0}^n\hat{a}_i+\sum_{i=c}^n\hat{r}_i]_P^c\}_{c=0}^n&\\
   ;\{[\sum_{c=1}^i \hat{a}_c + \sum_{c=i}^n(\hat{a}_c+\hat{r}_c)]_P^i\}_{i=0}^{n-1},\sum_{i=0}^n {a_i}\}&\\
\end{split}\]
by ciphertext indistinguishability against chosen plaintext attacks of Paillier
\[\small\begin{split}
  \{{\textsf{coins}}, \{({a}_i+{r}_i)\}_{i=0}^n,\{[\hat{r}_i]_P^i\}_{i=0}^n,\{[\sum_{i=0}^n\hat{a}_i+\sum_{i=c}^n\hat{r}_i]_P^c\}_{c=0}^n&\\
   ;\{[\sum_{c=1}^i \hat{a}_c + \sum_{c=i}^n(\hat{a}_c+\hat{r}_c)]_P^i\}_{i=0}^{n-1},\sum_{i=0}^n {a_i}\}&\\
   {\equiv}_c \{{\textsf{coins}}, \{({a}_i+{r}_i)\}_{i=0}^n,\{[{r}_i]_P^i\}_{i=0}^n,\{[\sum_{i=0}^n{a}_i+\sum_{i=c}^n{r}_i]_P^c\}_{c=0}^n&\\
   ;\{[\sum_{c=1}^i {a}_c + \sum_{c=i}^n({a}_c+{r}_c)]_P^i\}_{i=0}^{n-1},\sum_{i=0}^n {a_i}\}&\\
\end{split}\]
i.e., {\small$S_{\mathcal{A}}\ {\equiv}_c\ view_{\mathcal{A}}$}.

\textbf{Collusion situations.}

$\blacksquare$
We discuss the security when $\alpha\ (\alpha \leq n-1)$ data owners collude with each other to infer the summation $\sum_{i=0}^na_i$ and other data owners' data.
Considering the most extreme case, $\alpha = n-1$.
The $n-1$ data owners have information:
\[
\{ \{a_i,r_i,a_i+r_i,[r_i]_P^i\}_{i=1}^{n-1},\sum_{c=1}^{n-1} a_c + (a_n+r_n), a_n+r_n \}
\]
It is straightforward that the random noise $r_n$ in $\sum_{c=1}^{n-1} a_c + (a_n+r_n)$ hides $\sum_{i=0}^na_i$ in an information-theoretic way (it is an one-time pad).
Similarly, $a_n$ cannot be figured out too.

$\blacksquare$
We discuss the security when $\alpha\ (\alpha \leq n-2)$ data owners collude with model demander to infer other data owners' data.
Considering the most extreme case, $\alpha = n-2$,
where $\mathcal{O}_{n-1}$ and $\mathcal{O}_{n}$ do not participate in collusion.

The $n-2$ data owners have information:
\begin{equation}\label{equ:sum1}
\{ \{a_i,r_i,a_i+r_i,[r_i]_P^i\}_{i=1}^{n-2},\sum_{c=1}^{n-2} a_c + \sum_{c=n-2}^{n}(a_c+r_c) \}
\end{equation}

The model demander have information:
\begin{equation}\label{equ:sum2}
\begin{split}
  \{\{(a_i+r_i)\}_{i=0}^n,\{[r_i]_P^i\}_{i=0}^n,\{[\sum_{i=0}^na_i+\sum_{i=c}^nr_i]_P^c\}_{c=0}^n ,&\\
  \{[\sum_{c=1}^i a_c + \sum_{c=i}^n(a_c+r_c)]_P^i\}_{i=0}^n \}&
\end{split}
\end{equation}

Combining Equation (\ref{equ:sum1}) and Equation (\ref{equ:sum2}),
we obtain:
\[
\{(a_n+a_{n-1}),(a_{n-1}+r_{n-1}),(a_{n}+r_{n}),[r_{n-1}]_P^{n-1},[r_{n}]_P^{n}\}
\]
Ciphertext indistinguishability against chosen plaintext attacks of Paillier ensures that no bit of information is leaked from $[r_{n-1}]_P^{n-1},[r_{n}]_P^{n}$.
The random noise $r_n$ and $r_{n-1}$ hides $a_n$ and $a_{n-1}$ respectively in an information-theoretic way (it is an one-time pad).

The model demander and data owners may try to infer $a_n$ and $a_{n-1}$ from $a_n+a_{n-1}$.
In this case, we see $a_n+a_{n-1}$ as a ciphertext,
where the model demander and data owners infer information by observing $a_n+a_{n-1}$.

That is, the model demander and data owners just observes a ciphertext $a_n+a_{n-1}$ and attempts to determine information about the underlying plaintext (or plaintexts),
which is a ciphertext-only attack.
Observing $a_n+a_{n-1}$ has no effect on the knowledge regarding the actual message that was sent.
\qed
\end{myproof}

\subsection{Security Analysis for Building Blocks of Conversion} \label{app:Security Analysis for Building Blocks of Conversion}
\begin{mypro}\label{Propo:Converting Ciphertext 1}
Secure converting ${[e^\vec{a\cdot b}]}_R$ to ${[e^\vec{a\cdot b}]}_P$ is secure in the honest-but-curious model.
\end{mypro}

\begin{myproof}[for Proposition \ref{Propo:Converting Ciphertext 1}]
The input of data owner is \{{\small$(\sf{SK},\sf{PK})_{Paillier}$,$(\sf{SK},\sf{PK})_{CloudRSA}$}\}.
$\mathcal{O}$'s view is
\[
view_{\mathcal{O}}=(e^{\vec{a\cdot b}+r},[e^{\vec{a\cdot b}+r}]_R,[e^{\vec{a\cdot b}+r}]_P)
\]
We construct a simulator {\small$S_{\mathcal{O}}$} which runs as follows:
\begin{itemize}
  \item Picks two random vectors $(\vec{\hat{a}},\vec{\hat{b}})$.
  \item Picks a random $c$, computes $e^{c+\vec{\hat{a}}\cdot \vec{\hat{b}}}$.
  \item Generates the random coins necessary for re-randomization and put them in $\hat{\textsf{coins}}$.
  \item Encrypts $e^{c+\vec{\hat{a}}\cdot \vec{\hat{b}}}$ under $\sf{PK}_{Paillier}$ and $\sf{PK}_{CloudRSA}$ respectively.
  \item Outputs $( e^{c+\vec{\hat{a}}\cdot \vec{\hat{b}}},[e^{c+\vec{\hat{a}}\cdot \vec{\hat{b}}}]_P, [e^{c+\vec{\hat{a}}\cdot \vec{\hat{b}}}]_R,\hat{\textsf{coins}} )$.
\end{itemize}
$\textsf{coins}$ and $\hat{\textsf{coins}}$ come from the same distribution, independently from other parameters,
\[
\begin{split}
  ( e^{c+\vec{\hat{a}}\cdot \vec{\hat{b}}},[e^{c+\vec{\hat{a}}\cdot \vec{\hat{b}}}]_P, [e^{c+\vec{\hat{a}}\cdot \vec{\hat{b}}}]_R,&\hat{\textsf{coins}} )\\
  =( e^{c+\vec{\hat{a}}\cdot \vec{\hat{b}}},[e^{c+\vec{\hat{a}}\cdot \vec{\hat{b}}}]_P, [e^{c+\vec{\hat{a}}\cdot \vec{\hat{b}}}]_R,&{\textsf{coins}} )
\end{split}
\]
The distribution of {\small$(\vec{a\cdot b}+r)$} and {\small$(\vec{\hat{a}\cdot \hat{b}}+c)$} are identical, so the real distribution $(e^{\vec{a\cdot b}+r},[e^{\vec{a\cdot b}+r}]_R,[e^{\vec{a\cdot b}+r}]_P)$ and the ideal distribution $( e^{c+\vec{\hat{a}}\cdot \vec{\hat{b}}},[e^{c+\vec{\hat{a}}\cdot \vec{\hat{b}}}]_P, [e^{c+\vec{\hat{a}}\cdot \vec{\hat{b}}}]_R$ are statistically indistinguishable,
\[
\begin{split}
  ( e^{c+\vec{\hat{a}}\cdot \vec{\hat{b}}},[e^{c+\vec{\hat{a}}\cdot \vec{\hat{b}}}]_P, [e^{c+\vec{\hat{a}}\cdot \vec{\hat{b}}}]_R&)\\
  {\equiv}_c(e^{\vec{a\cdot b}+r},[e^{\vec{a\cdot b}+r}]_R,[e^{\vec{a\cdot b}+r}]_P&)
\end{split}
\]
Thus, {\small$S_{\mathcal{O}}\ {\equiv}_c\ view_{\mathcal{O}}$}.

The input of model demander is \{{\small$e^{\vec{a\cdot b}}$}\}.
\[view_{\mathcal{A}} = (r,[e^{\vec{a\cdot b}}]_R,[e^{\vec{a\cdot b}+r}]_R,[e^{\vec{a\cdot b}+r}]_P,[e^{\vec{a\cdot b}}]_P)\]
We construct a simulator {\small$S_{\mathcal{A}}$} which runs as follows:
\begin{itemize}
  \item Picks $2$ random numbers that are limited to $\mathbb{Z}_N$: $c_1,c_2$.
  \item Generates $e^{c_2},e^{c_1+c_2}$
  \item Generates the random coins necessary for re-randomization and put them in $\hat{\textsf{coins}}$.
  \item Encrypts $e^{c_1+c_2}$ by {\small$\sf{PK}_{Paillier}$}.
  \item Encrypts $e^{c_2}$ by {\small$\sf{PK}_{Paillier}$}.
  \item Encrypts $e^{c_1+c_2}$ by {\small$\sf{PK}_{CloudRSA}$}.
  \item Encrypts $e^{c_2}$ by {\small$\sf{PK}_{CloudRSA}$}.
  \item Outputs $(\hat{\textsf{coins}},c_1,[e^{c_2}]_R,[e^{c_1+c_2}]_R,[e^{c_1+c_2}]_P,[e^{c_2}]_P)$
\end{itemize}
$\textsf{coins}$ and $\hat{\textsf{coins}}$ come from the same distribution, independently from other parameters,
and by ciphertext indistinguishability against chosen plaintext attacks of Paillier and {\small Cloud-RSA},
\[
\begin{split}
  (c_1,[e^{c_2}]_R,[e^{c_1+c_2}]_R,[e^{c_1+c_2}]_P,[e^{c_2}]_P)&\\
  {\equiv}_c (r,[e^{\vec{a\cdot b}}]_R,[e^{\vec{a\cdot b}+r}]_R,[e^{\vec{a\cdot b}+r}]_P,[e^{\vec{a\cdot b}}]_P)&
\end{split}
\]
{\small $[e^{\vec{a\cdot b}}]_R,[e^{\vec{a\cdot b}+r}]_R$}, {\small $[e^{\vec{a\cdot b}}]_P$} and {\small $[e^{\vec{a\cdot b}+r}]_P$} are encrypted by data owners' public key,
the confidentiality of which is equivalent to the cryptosystem.
Thereby {\small$S_{\mathcal{A}}\ {\equiv}_c\ view_{\mathcal{A}}$}.
\qed
\end{myproof}

\begin{mypro}\label{Propo:Converting Ciphertext 2}
Secure converting ${[m]}_P^1$ to ${[m]}_P^2$ is secure in the honest-but-curious model.
\end{mypro}

\begin{myproof}[for Proposition \ref{Propo:Converting Ciphertext 2}]
The input of data owner is \{{\small$(\sf{SK},\sf{PK})_{Paillier}^1$, $(\sf{PK})_{Paillier}^2$}\}.
$\mathcal{O}$'s view is
\[
view_{\mathcal{O}_n}=(m+r,[m+r]_P^1,[m+r]_P^2)
\]
We construct a simulator {\small$S_{\mathcal{O}}$} which runs as follows:
\begin{itemize}
  \item Picks a random $c$, $c \in \mathbb{Z}_N$.
  \item Encrypts $c$ under $\sf{PK}_{Paillier}^1$ and $\sf{PK}_{Paillier}^2$ respectively.
  \item Outputs $( c,[c]_P^1,[c]_P^2 )$.
\end{itemize}
The distribution of $c$ and $m+r$ are identical, so the real distribution $(m+r,[m+r]_P^1,[m+r]_P^2)$ and the ideal distribution $( c,[c]_P^1,[c]_P^2 )$ are statistically indistinguishable,
\[
(m+r,[m+r]_P^1,[m+r]_P^2) {\equiv}_c ( c,[c]_P^1,[c]_P^2 )
\]
Thus, {\small$S_{\mathcal{O}}\ {\equiv}_c\ view_{\mathcal{O}}$}.

The input of model demander is \{{\small$[m]_P^1$}\}.
\[view_{\mathcal{A}} = \{r,[m]_P^1,[m]_P^2,[m+r]_P^1,[m+r]_P^2\}\]
$[m]_P^1,[m+r]_P^1,[m]_P^2,[m+r]_P^2$ are encrypted by data owners' public key,
the confidentiality of which is equivalent to the cryptosystem.
Thereby {\small$S_{\mathcal{A}}\ {\equiv}_c\ view_{\mathcal{A}}$}.
\qed
\end{myproof}

\subsection{Security Analysis for ML Training Protocols}  \label{app:Security Analysis for ML Training Protocols}

\begin{myDef}[Ciphertext-only Attack \cite{1}]
This is the most basic attack, and refers to a scenario where the adversary just observes a ciphertext (or multiple ciphertexts) and attempts to determine information about the underlying plaintext (or plaintexts).
\end{myDef}

\begin{myDef}[Perfectly Secret \cite{1}]
An encryption scheme $(\textsf{Gen}, \textsf{Enc}, \textsf{Dec})$ with message space $M$ is perfectly secret if for every probability distribution over $M$, every message $m \in M$, and every ciphertext $c \in C$ for which $Pr[C = c] > 0$:
\[
Pr[M = m | C = c] = Pr[M = m]
\]
\end{myDef}

Recalling that our security goal is achieving keeping privacy of each participant and computing model parameters securely when facing honest-but-curious adversaries.
Both the model of model demander $\mathcal{A}$ and the sensitive datasets from n data owners $\mathcal{O}_i$ are confidential throughout the execution of secure ML training protocols.
This section gives the security analyses for the three privacy-preserving ML training protocols instantiated from \texttt{Heda}.

We follow the ides of Modular Sequential Composition to present the security proofs.
If we can show that a protocol $\pi$ respects privacy in the honest-but-curious model using calls to a series of ideal
functionalities,
and these ideal functionalities can be replaced by real secure protocols in the semi-honest model,
$\pi$ is secure in the honest-but-curious model.
Then we discuss the security in situations of collusion.

\begin{mypro}\label{Propo:SVM}
Privacy-preserving SVM training protocol is secure in the honest-but-curious model.
\end{mypro}

\begin{myproof}[for Proposition \ref{Propo:SVM}]
We prove the security without collusion first.
Because the model demander and $n$ data owner perform interactive computations in turn, we analyze the security according to definition of Secure Two-party Computation.

\textbf{For each data owner $\mathcal{O}_i$.}

After sending the encrypted dataset to model demander,
the data owner does not receive any message and call to any other protocols during the execution processes, his view only consists in its input.
The input of the data owner is ({\small$(\sf{SK},\sf{PK})_{Paillier}^i$}, $\mathcal{D}_i$).
We use the trivial simulator that just outputs its inputs for the proof of security.
Hence, $S_{\mathcal{O}_i}= (\mathcal{D}_i, \textsf{coins}) = view_{\mathcal{O}_i}$.

\textbf{For model demander $\mathcal{A}$.}

The input of the model demander is the initialized model parameters $(\theta)$.
The model demander's view is
\[
\begin{split}
  view_{\mathcal{A}}=\{\theta,[\sf{Y}]_P^i,[\sf{XY}]_P^i,[\theta\vec{x}_ty_t]_P^i,[\lambda\alpha\vec{x}_ty_t]_P^i;&\\
  out_{block9}\}
\end{split}
\]
where $out_{block9}$ denote the received outputs from Building Blocks 7 respectively.

We construct a simulator {\small$S_{\mathcal{A}}$} which runs as follows:
\begin{itemize}
  \item Uniformly picks a value $v$ from $\{1,0\}$.
  \item Picks a random $\hat{\theta}$.
  \item Picks 4 random numbers $c_1,c_2,c_3,c_4$, $c_i \in \mathbb{Z}_N$.
  \item Encrypts $c_1,c_2,c_3,c_4$ by {\small$\sf{PK}_{Paillier}^i$}.
  \item Outputs {\small$\{\theta,[c_1]_P^i,[c_2]_P^i,[c_3]_P^i,[c_4]_P^i ;\hat{\theta},v\}$}.
\end{itemize}
by ciphertext indistinguishability against chosen plaintext attacks of Paillier:
\[
\begin{split}
  \{[c_1]_P^i,[c_2]_P^i,[c_3]_P^i,[c_4]_P^i\} {\equiv}_c&\\
  \{[\sf{Y}]_P^i,[\sf{XY}]_P^i,[\theta\vec{x}_ty_t]_P^i,[\lambda\alpha\vec{x}_ty_t]_P^i)\}&
\end{split}
\]
The distribution of $\{v,\hat{\theta}\}$ and $\{(1>?[\theta\vec{x}_ty_t]_P^i),  update(\theta)\}$ are come from the same distribution, independently from other parameters:
\[
  \{v,\hat{\theta}\} {\equiv}_c \{(1>?[\theta\vec{x}_ty_t]_P^i), update(\theta)\}
\]

The model demander may try to infer data owners' data from $\theta$ or $update(\theta)$.
In this case, we see ($\theta$, $update(\theta)$) as a ciphertext,
where model demander infers information by observing ($\theta$, $update(\theta)$).

That is the model demander just observes a ciphertext ($\theta$ and $update(\theta)$) and attempts to determine information about the underlying plaintext (or plaintexts) in training dataset $D$,
which is a ciphertext-only attack.
Besides,
observing $\theta$ and $update(\theta)$ have no effect on the the model demander's knowledge regarding the actual message that was sent.
For every probability data in training dataset $x_i \in \mathcal{D}$, and every ciphertext $\theta$ and $update(\theta)$:
\[
Pr[M = x_i | C = \theta\ or\ update(\theta)] = Pr[M = x_i]
\]
Thus, inferring data owners' data from $\theta$ or $update(\theta)$ is a ciphertext-only attack with perfectly secret.
Thereby {\small$S_{\mathcal{A}}\ {\equiv}_c\ view_{\mathcal{A}}$}.

In particular,
secure addition,
secure plaintext-ciphertext dot product
and Building Block 8 are secure in the honest-but-curious model.
We obtain the security of privacy-preserving SVM training protocol using Modular Sequential Composition.

\textbf{Collusion situations.}

$\blacksquare$
We discuss the security when $\alpha\ (\alpha \leq n-1)$ data owners collude with each other to infer the model parameters $\theta$ and other data owners' data.
Considering the most extreme case, $\alpha = n-1$.
The $n-1$ data owners have information:
\[
\{ ([\sf{Y}]_P^i,[\sf{XY}]_P^i)_{i=1}^{n-1} \}
\]

Since the $n-1$ data owners didn't receive the complete $\theta$,
and $\theta$ is encrypted by the model demander's public key.
It is straightforward that $\theta$ cannot be figured out.

Without the corresponding private key of the data owner who do not participating in the collusion, the ciphertext indistinguishability of Paillier ensures that no bit of information is leaked from ciphertexts of other data owners who do not participating in the collusion.

$\blacksquare$
We discuss the security when $\alpha\ (\alpha \leq n-2)$ data owners collude with the model demander to infer other data owners' data.
Considering the most extreme cases, $\alpha = n-2$.

The $n-2$ data owners have information:
\begin{equation}\label{equ:SVM1}
  \{ ([\sf{Y}]_P^i,[\sf{XY}]_P^i)_{i=1}^{n-2} \}
\end{equation}

The model demander have information:
\begin{equation}\label{equ:SVM2}
\{\theta,([\sf{Y}]_P^i,[\sf{XY}]_P^i,[\theta\vec{x}_ty_t]_P^i,[\lambda\alpha\vec{x}_ty_t]_P^i)_{i=0}^n\}
\end{equation}

Combining Equation (\ref{equ:SVM1}) and Equation (\ref{equ:SVM2}),
we deduce that:
\[
\begin{split}
  \{\theta,(\sf{Y},\sf{XY},\theta\vec{x}_ty_t,\lambda\alpha\vec{x}_ty_t)_{i=0}^{n-2}\}\ with&\\
  ([\sf{Y}]_P^i,[\sf{XY}]_P^i,[\theta\vec{x}_ty_t]_P^i,[\lambda\alpha\vec{x}_ty_t]_P^i)_{i=n-1}^n\}&
\end{split}
\]
Without the corresponding private key,
they cannot figure out $([\sf{Y}]_P^i,[\sf{XY}]_P^i)_{i=n-1}^n$.
\qed
\end{myproof}

\begin{mypro}\label{Propo:LR}
Privacy-preserving LR training protocol is secure in the honest-but-curious model.
\end{mypro}

\begin{myproof}[for Proposition \ref{Propo:LR}]
We prove the security without collusion first.
Because model demander and each data owner perform interactive computations in turn, we analyze the security according to definition of Secure Two-party Computation.

\textbf{For each data owner $\mathcal{O}_i$.}

The input of the data owner is ({\small$(\sf{SK},\sf{PK})_{Paillier}^i$}, {\small$(\sf{SK},\sf{PK})_{CloudRSA}^i$}, $\mathcal{D}_i$).
The data owner does not call to any other protocols or algorithms during the execution processes.
\begin{equation}\label{equ:LRO}
\small
view_{\mathcal{O}_i}=( \mathcal{D}_i, [ e^{ \theta \vec{x}_t+r}+e^r]_P,\frac{\vec{x}_t}{e^{ \theta \vec{x}_t+r}+e^r},[\frac{\vec{x}_t}{e^{ \theta \vec{x}_t+r}+e^r}]_P)
\end{equation}

We construct a simulator {\small$S_{\mathcal{O}_i}$} which runs as follows:
\begin{itemize}
  \item Uniformly picks a value $c$.
  \item Picks a random $\hat{\theta}$.
  \item Picks a record $\hat{\vec{x}_t}$ from $\mathcal{D}_i$.
  \item Generates $\frac{\hat{\vec{x}_t}}{e^{ \hat{\theta} \hat{\vec{x}_t}+c}+e^c}$ and $e^{ \hat{\theta} \hat{\vec{x}_t}+c}+e^c$.
  \item Encrypts $\frac{\hat{\vec{x}_t}}{e^{ \hat{\theta} \hat{\vec{x}_t}+c}+e^c}$ and $e^{ \hat{\theta} \hat{\vec{x}_t}+c}+e^c$ by {\small$\sf{PK}_{Paillier}^i$}.
  \item Outputs {\small$([\frac{\hat{\vec{x}_t}}{e^{ \hat{\theta} \hat{\vec{x}_t}+c}+e^c}]_P, [e^{ \hat{\theta} \hat{\vec{x}_t}+c}+e^c]_P,e^{ \hat{\theta} \hat{\vec{x}_t}+c}+e^c )$}.
\end{itemize}
The real distribution $(r,\theta, {\vec{x}_i})$ and the ideal distribution $(c,\hat{\theta},\hat{\vec{x}_i} )$ are come from the same distribution, independently from other parameters,
\[
\begin{split}
  ( [ e^{ \theta \vec{x}_t+r}+e^r]_P,\frac{\vec{x}_t}{e^{ \theta \vec{x}_t+r}+e^r},[\frac{\vec{x}_t}{e^{ \theta \vec{x}_t+r}+e^r}]_P) &\\
{\equiv}_c ([\frac{\hat{\vec{x}_t}}{e^{ \hat{\theta} \hat{\vec{x}_t}+c}+e^c}]_P, [e^{ \hat{\theta} \hat{\vec{x}_t}+c}+e^c]_P,e^{ \hat{\theta} \hat{\vec{x}_t}+c}+e^c )&
\end{split}
\]
Hence, $S_{\mathcal{O}_i} {\equiv}_c view_{\mathcal{O}_i}$.

\textbf{For model demander $\mathcal{A}$.}

The input of model demander the initialized model parameters $(\theta)$.
The model demander's view is
\[
\begin{split}
\footnotesize
  view_{\mathcal{A}}=\{\theta,{[{\vec{x}}_ty_t]}_P,[ e^{\vec{x}_t} ]_R,
  r,[ e^{-\theta\vec{x}_t+r}+e^r ]_P,[\lambda\vec{x}_ty_t]_P, &\\
[\frac{\vec{x}_t}{e^{-\theta\vec{x}_t+r}+e^r }]_P,[\frac{\lambda\vec{x}_t}{e^{-\theta\vec{x}_t+r}+e^r }+\theta]_P, [\frac{\lambda\vec{x}_t}{e^{-\theta\vec{x}_t+r}+e^r }+\theta]_P;&\\
  out_{block5};out_{block8};out_{block9}&\}
\end{split}
\]
\[
\begin{split}
 out_{block5} = [ e^{-\theta\vec{x}_t} ]_R;\ out_{block8} = [e^{-\theta\vec{x}_t} ]_P;& \\
 out_{block9} = update({\theta})&
\end{split}
\]

As you can see, except for the model parameters $\theta$,
all other data are encrypted by data owners public key.
As Paillier and {\small Cloud-RSA}'s cryptosystem is semantically secure,
model demander can only infer training dataset information from from $\theta$ or $update(\theta)$.

In this case, we see ($\theta$, $update(\theta)$) as a ciphertext,
where model demander infers information by observing ($\theta$, $update(\theta)$).
But the model demander just observes a ciphertext ($\theta$, $update(\theta)$) and attempts to determine information about the underlying plaintext (or plaintexts) of training dataset $\mathcal{D}$,
which is a ciphertext-only attack.
Besides,
observing $\theta$ and $update(\theta)$ have no effect on the the model demander's knowledge regarding the actual message that was sent.
That is for every probability data in training dataset $x_i \in \mathcal{D}$, and every ciphertext $\theta$ and $update(\theta)$:
\[
Pr[M = x_i | C = \theta\ or\ update(\theta)] = Pr[M = x_i]
\]
Thus, inferring data owners' data from $\theta$ or $update(\theta)$ is a ciphertext-only attack with perfectly secret.
Thereby {\small$S_{\mathcal{A}}\ {\equiv}_c\ view_{\mathcal{A}}$}.

In particular,
secure power function,
Bilding Block 8,
secure addition,
secure plaintext-ciphertext multiplication,
secure subtraction,
and Building Block 8 are secure in the honest-but-curious model.
We obtain the security of privacy-preserving LR training protocol using Modular Sequential Composition.

\textbf{Collusion situations.}

$\blacksquare$
We discuss the security when $\alpha\ (\alpha \leq n-1)$ data owners collude with each other to infer the model parameters $\theta$ and other data owners' data.
Considering the most extreme case, $\alpha = n-1$.
The $n-1$ data owners have information:
\[
\{( \mathcal{D}_i, [ e^{ \theta \vec{x}_t+r}+e^r]_P,\frac{\vec{x}_t}{e^{ \theta \vec{x}_t+r}+e^r},[\frac{\vec{x}_t}{e^{ \theta \vec{x}_t+r}+e^r}]_P)_{i=0}^{n-1}\}
\]
This case is almost the same as Equation (\ref{equ:LRO}) where each data owner without
collusion, the security of which that we have proved above.
Thus, $\theta$ cannot be figured out.

Without the corresponding private key of the data owner who do not participating in the collusion, the ciphertext indistinguishability of Paillier ensures that no bit of information is leaked from ciphertexts of other data owners who do not participating in the collusion.

$\blacksquare$
We discuss the security when $\alpha\ (\alpha \leq n-2)$ data owners collude with model demander to infer other data owners' data.
Considering the most extreme cases, $\alpha = n-2$.

The $n-2$ data owners have information:
\begin{equation}\label{equ:LR1}
\{ ( \mathcal{D}_i, [ e^{ \theta \vec{x}_t+r}+e^r]_P,\frac{\vec{x}_t}{e^{ \theta \vec{x}_t+r}+e^r},[\frac{\vec{x}_t}{e^{ \theta \vec{x}_t+r}+e^r}]_P)_{i=0}^{n-2} \}
\end{equation}

The model demander have information:
\begin{equation}\label{equ:LR2}
\begin{split}
\footnotesize
\{\theta,({[{\vec{x}}_ty_t]}_P^i,[ e^{\vec{x}_t} ]_R^i,
  r,[ e^{-\theta\vec{x}_t+r}+e^r ]_P^i,[\lambda\vec{x}_ty_t]_P, &\\
[\frac{\vec{x}_t}{e^{-\theta\vec{x}_t+r}+e^r }]_P^i,[\frac{\lambda\vec{x}_t}{e^{-\theta\vec{x}_t+r}+e^r }+\theta]_P^i, [\frac{\lambda\vec{x}_t}{e^{-\theta\vec{x}_t+r}+e^r }+\theta]_P^i,&\\
  [ e^{-\theta\vec{x}_t} ]_R^i,[e^{-\theta\vec{x}_t} ]_P^i)_{i=0}^n&\}
\end{split}
\end{equation}

Combining Equation (\ref{equ:LR1}) and Equation (\ref{equ:LR2}),
we obtain:
\[
\begin{split}
 \{\theta,({\vec{x}}_ty_t, e^{\vec{x}_t} ,
  r, e^{-\theta\vec{x}_t+r}+e^r ,\lambda\vec{x}_ty_t, &\\
\frac{\vec{x}_t}{e^{-\theta\vec{x}_t+r}+e^r },\frac{\lambda\vec{x}_t}{e^{-\theta\vec{x}_t+r}+e^r }+\theta, \frac{\lambda\vec{x}_t}{e^{-\theta\vec{x}_t+r}+e^r }+\theta,&\\
   e^{-\theta\vec{x}_t} ,e^{-\theta\vec{x}_t} )_{i=0}^{n-2}\}\ with&\\
  ({[{\vec{x}}_ty_t]}_P^i,[ e^{\vec{x}_t} ]_R^i,
  r,[ e^{-\theta\vec{x}_t+r}+e^r ]_P^i,[\lambda\vec{x}_ty_t]_P, &\\
[\frac{\vec{x}_t}{e^{-\theta\vec{x}_t+r}+e^r }]_P^i,[\frac{\lambda\vec{x}_t}{e^{-\theta\vec{x}_t+r}+e^r }+\theta]_P^i, [\frac{\lambda\vec{x}_t}{e^{-\theta\vec{x}_t+r}+e^r }+\theta]_P^i,&\\
  [ e^{-\theta\vec{x}_t} ]_R^i,[e^{-\theta\vec{x}_t} ]_P^i)_{i=n-1}^{n}&
\end{split}
\]
Without the corresponding private key, they cannot figure out $({[{\vec{x}}_ty_t]}_P^i,[ e^{\vec{x}_t} ]_R^i)_{i=n-1}^{n}$.
Therefor, $\alpha$ data owners collude with model demander cannot infer other data owners' data.
\qed
\end{myproof}

\begin{mypro}\label{Propo:NB}
Privacy-preserving NB training protocol is secure in the honest-but-curious model.
\end{mypro}

\begin{myproof}[for Proposition \ref{Propo:NB}]
We prove the security without collusion first.

\textbf{For each data owner $\mathcal{O}_i$.}

After sending the encrypted dataset to model demander,
the data owner does not receive any message and call to any other protocols during the execution processes, his view only consists in its input.
The input of the data owner is ({\small$(\sf{SK},\sf{PK})_{Paillier}^i$}, $\mathcal{D}_i$).
We use the trivial simulator that just outputs its inputs for the proof of security.
Hence, $S_{\mathcal{O}_i}= (\mathcal{D}_i, \textsf{coins}) = view_{\mathcal{O}_i}$.

\textbf{For model demander $\mathcal{A}$.}
The input of model demander is $((\sf{SK},\sf{PK})_{Paillier}^{\mathcal{A}})$.
The model demander's view is
\begin{equation}\label{equ:NB1}
  \begin{split}
  view_{\mathcal{A}}=\{&\\
  \mu_{y,j},\sigma_{y,j}^{2},&\sum_{c=1}^{n}{|\mathcal{D}_{c}(y)|}, \sum_{c=1}^{n}|\mathcal{D}_c^{y}({x}_{ij})|, \sum_{i=1}^{|\mathcal{D}|}x_{ij} , \sum_{i=1}^{|\mathcal{D}|}(x_{ij})^2\}
\end{split}
\end{equation}
\[
\begin{split}
\mu_{y, j}&=\frac{\sum_{i=1}^{|\mathcal{D}|}{x_{ij}}}{m}\\
\sigma_{y,j}^{2}&=\frac{\sum_{i=1}^{|\mathcal{D}|}(\mu_{y, j})^2}{m}
+\frac{\sum_{i=1}^{|\mathcal{D}|}(x_{ij})^2}{m}
-2\mu_{y,j}*\frac{\sum_{i=1}^{|\mathcal{D}|}(x_{ij})}{m}\\
\end{split}
\]
Model demander may try to infer data owners' data from his view.
In this case, we see the message in Equation (\ref{equ:NB1}) as a ciphertext,
where model demander infers information of $\mathcal{D}$ by observing these message.
The model demander just observes the message in Equation (\ref{equ:NB1}) and attempts to determine information about the underlying plaintext (or plaintexts),
which is a ciphertext-only attack.

Observing the message in Equation (\ref{equ:NB1}) have no effect on the the model demander's knowledge regarding the actual message that was sent.
That is for every probability data in training dataset $x \in \mathcal{D}$, and every ciphertext $c\ \in\ Equation$ (\ref{equ:NB1}):
\[
Pr[M = x | C = c] = Pr[M = x]
\]
Thus, inferring data owners' data from  message in Equation (\ref{equ:NB1}) is a ciphertext-only attack with perfectly secret.
Thereby {\small$S_{\mathcal{A}}\ {\equiv}_c\ view_{\mathcal{A}}$}.

In particular,
secure summation protocol is secure in the honest-but-curious model.
We obtain the security of privacy-preserving SVM training protocol using Modular Sequential Composition.

\textbf{Collusion situations.}

$\blacksquare$
We discuss the security when $\alpha\ (\alpha \leq n-1)$ data owners collude with each other to infer the model parameters $P(y)$ and $P(\vec{x}_i |y)$ and other data owners' data.
Considering the most extreme case, $\alpha = n-1$.
The $n-1$ data owners have information:
\[
\{ (\mathcal{D}_i)_{i=1}^{n-1} \}
\]

Since the $n-1$ data owners didn't even receive other information during Protocol 3,.
It is straightforward that $P(y)$ and $P(\vec{x}_i |y)$ cannot be figured out.
Without the corresponding private key of the data owner who do not participating in the collusion, the ciphertext indistinguishability of Paillier ensures that no bit of information is leaked.

$\blacksquare$
We discuss the security when $\alpha\ (\alpha \leq n-2)$ data owners collude with the model demander to infer other data owners' data.
Considering the most extreme cases, $\alpha = n-2$.

The $n-2$ data owners have information:
\begin{equation}\label{equ:NB2}
\{ (\mathcal{D}_i)_{i=1}^{n-2} \}
\end{equation}

The model demander have information:
\begin{equation}\label{equ:NB3}
\{ \mu_{y,j},\sigma_{y,j}^{2},\sum_{c=1}^{n}{|\mathcal{D}_{c}(y)|}, \sum_{c=1}^{n}|\mathcal{D}_c^{y}({x}_{ij})|, \sum_{i=1}^{|\mathcal{D}|}x_{ij} , \sum_{i=1}^{|\mathcal{D}|}(x_{ij})^2\}
\end{equation}

Combining Equation (\ref{equ:SVM1}) and Equation (\ref{equ:SVM2}),
we can figure out:
\begin{equation}\label{equ:NB4}
\begin{split}
  \{
  (|\mathcal{D}_{n-1}(y) |+|\mathcal{D}_{n}(y) |),(|\mathcal{D}_{n-1}^{y}({x}_{ij})|+|\mathcal{D}_{n}^{y}({x}_{ij})|),&\\
  \sum_{x_{ij} \in \mathcal{D}_{n-1}\cup\mathcal{D}_{n}}x_{ij} ,\sum_{x_{ij} \in \mathcal{D}_{n-1}\cup\mathcal{D}_{n}}(x_{ij})^2   &
  \}
\end{split}
\end{equation}
Model demander and data owners try to infer $x_{ij},y_i \in \mathcal{D}_{n-1}\cup\mathcal{D}_{n}$ from Equation \ref{equ:NB4}.

In this case, we see the message in Equation (\ref{equ:NB4}) as a ciphertext.
The model demander just observes a ciphertext and attempts to determine information about the underlying plaintext (or plaintexts),
which is a ciphertext-only attack.

Observing the message in Equation (\ref{equ:NB4}) have no effect on the the model demander's knowledge regarding the actual message that was sent.
That is for every probability data in training dataset $x \in \mathcal{D}_{n-1}\cup\mathcal{D}_{n}$, and every ciphertext $c\ \in\ Equation$ (\ref{equ:NB4}):
\[
Pr[M = x | C = c] = Pr[M = x]
\]
Thus, inferring data owners' data from message in Equation (\ref{equ:NB1}) is a ciphertext-only attack with perfectly secret.

Therefor, $\alpha$ data owners collude with model demander cannot infer other data owners' data.
\qed
\end{myproof}

\subsection{Complexity Analysis}   \label{app:Complexity Analysis}
Let $l$ be the length of attribute value.
Supposing $n$ is the number of data owners,
and $T$ is the maximum number of iterations.

It is straightforward that the number of ciphertext computation included in the following building blocks is constant:
secure addition,
secure subtraction,
secure plaintext-ciphertext multiplication,
secure piphertext-ciphertext multiplication
secure summation
and two converting building blocks.
Therefore, the time complexity of the above building blocks is $\mathcal{O}(1)$.

The number of the outermost loop is $d$ in secure power function and secure plaintext-ciphertext dot product,
which processes one attribute at each loop.
The time complexity of secure power function is $\mathcal{O}(ld)$.

Privacy-preserving SVM training protocol employs secure plaintext-ciphertext dot product in each iteration.
Thus, the time complexity is $\mathcal{O}(Tld)$.

Privacy-preserving LR training protocol adopts secure power function in each iteration.
Hence, the time complexity is $\mathcal{O}(Tld)$.

Privacy-preserving NB training protocol calls for other building blocks than secure power function and secure plaintext-ciphertext dot product.
And the number of the outermost loop of it is $d$.
Therefor, the time complexity of it is $\mathcal{O}(d\cdot ld)$.

\subsection{Homomorphic Definition}    \label{app:Homomorphic Definition}
Formalized definition of homomorphic is given in Definition \ref{def:homomorphic}.

\begin{myDef}[Homomorphic \cite{1}]\label{def:homomorphic}
 A public-key encryption scheme $( \text{Gen, Enc, Dec})$ is homomorphic if for all $n$ and all ($\sf{PK}$, $\sf{SK}$) output by $\text{Gen}( {{1}^{n}} )$, it is possible to define groups $\mathbb{M}$, $\mathbb{C}$ (depending on $\sf{PK}$ only) such that:
\begin{enumerate}
\item The message space is $\mathbb{M}$, and all ciphertexts output by $\text{Enc}_{\sf{PK}}$ are elements of $\mathbb{C}$.
\item For any ${{m}_{1}},{{m}_{2}}\in \mathbb{M}$, any ${{c}_{1}}$ output by $\text{Enc}_{\sf{PK}}( m_1)$, and any ${{c}_{2}}$ output by $\text{Enc}_{\sf{PK}}( {{m}_{2}} )$, it holds that $\text{Dec}_{\sf{SK}}(o( c_1,$\\$c_2 ) )=\sigma ( m_1,m_1 )$.
\end{enumerate}
\end{myDef}

\subsection{A Case Study of Secure Power Function}   \label{app:A Case Study of Secure Power Function}
This section provides a case study for explicating the recovering of {\small$e^{\theta_i \vec{x}_i}$} from {\small$[e^{\vec{x}_i}\theta_i^1]_R$} and {\small$[e^{\vec{x}_i}\theta_i^2]_R$}.
The input of the data owner is:
\[\vec{x}=\{0.1,0.2\},\ e^{\vec{x}} = \{1.10517091808, 1.22140275816\}\]
\[\hat{e^{\vec{x}}} = \{ 110,122\};\ [\hat{e^{\vec{x}}}]_R = \{ [110]_R,[122]_R\}\]
The input of model demander is:
\[\vec{\theta} = \{1.31,2.42\},\ \hat{\theta} = \{(1,31),(2,42)\}\]

$\triangle$
\emph{Secure power function.}
The data owner sends $[\hat{e^{\vec{x}}}]_R$ to the model demander.
The model demander computes:
{\small\[
\begin{split}
  &\{ ({[110]_R}^1,{[110]_R}^{31}) ,({[122]_R}^2,{[122]_R}^{42}) \} \Leftrightarrow\\
    &\{ ({[110]_R},{[1.9194342e+63]_R}) ,({[14884]_R},{[4.237531e+87]_R}) \} \\
    &\rightarrow \{ [1637240]_R ,[8.133662e+150]_R \}
\end{split}
\]}
At this time, the model demander has finished a secure power function {\small($ [1637240]_R *[\sqrt[100]{8.133662e+150}]_R = [\hat{e^{\vec{x}\vec{\theta}}}]_R$)}.
In our privacy-preserving LR training protocol, the model demander continues to Bilding Block 8.

$\triangle$
\emph{Conversion protocol (Bilding Block 8) for recovering.}
The model demander picks a $r=-2$ randomly, and computes:
\[e^{-2}=0.13533528323;\ \hat{e^{-2}}=1353;\ [e^{-2}]_R=[1353]_R\]
\[[\hat{e^{\vec{x}\vec{\theta}+1.01r}}]_R= \{ [2215185720]_R,[1.100484e+154]_R\}\]
The model demander sends $([\hat{e^{\vec{x}\vec{\theta}+1.01r}}]_R,(1+2)*2+4,(31+42)*2+4)$ to the data owner.
The data owner computes:
{\small\[
\begin{split}
  [\hat{e^{\vec{x}\vec{\theta}+1.01r}}]_R \rightarrow & \hat{e^{\vec{x}\vec{\theta}+1.01r}}=\frac{2215185720}{1e+10}*\sqrt[100]{\frac{1.100484e+154}{1e+150}}\\
  =& 0.221518572*0.97817337451=0.21668356909
\end{split}
\]}
\[0.21668356909 \rightarrow 22 \rightarrow [22]_P\]
The data owner sends $[22]_P$ to the model demander.
$e^{2.02} = 7.53832493366\rightarrow (7,54)$
The model demander obtains $[\hat{e^{\vec{x}\vec{\theta}}}]_P$ by:
{\small\[
\begin{split}
\{({[22]_P}^7,2*7),({[22]_P}^{54},2*54+2)\}\Leftrightarrow&\\
\{([154]_P,2),([1188]_P,4)&\}
\end{split}
\]}
Bilding Block 8 is end at here, where the plaintext of $\{([154]_P,2),([1188]_P,4)\}$ is $\frac{154}{100}+\frac{1188}{10000}=1.6588
\approx e^{0.1*1.31+0.2*2.42}=1.83$.

\bibliographystyle{IEEEtran}
\bibliography{IEEEabrv}

\begin{thebibliography}{10}
\providecommand{\url}[1]{#1}
\csname url@samestyle\endcsname
\providecommand{\newblock}{\relax}
\providecommand{\bibinfo}[2]{#2}
\providecommand{\BIBentrySTDinterwordspacing}{\spaceskip=0pt\relax}
\providecommand{\BIBentryALTinterwordstretchfactor}{4}
\providecommand{\BIBentryALTinterwordspacing}{\spaceskip=\fontdimen2\font plus
\BIBentryALTinterwordstretchfactor\fontdimen3\font minus
  \fontdimen4\font\relax}
\providecommand{\BIBforeignlanguage}[2]{{%
\expandafter\ifx\csname l@#1\endcsname\relax
\typeout{** WARNING: IEEEtran.bst: No hyphenation pattern has been}%
\typeout{** loaded for the language `#1'. Using the pattern for}%
\typeout{** the default language instead.}%
\else
\language=\csname l@#1\endcsname
\fi
#2}}
\providecommand{\BIBdecl}{\relax}
\BIBdecl

\bibitem{IoTsurvey1}
\BIBentryALTinterwordspacing
M.~binti {Mohamad Noor} and W.~H. Hassan, ``Current research on internet of
  things (iot) security: A survey,'' \emph{Computer Networks}, vol. 148, pp.
  283 -- 294, 2019. [Online]. Available:
  \url{http://www.sciencedirect.com/science/article/pii/S1389128618307035}
\BIBentrySTDinterwordspacing

\bibitem{IoTsurvey2}
\BIBentryALTinterwordspacing
M.~Ammar, G.~Russello, and B.~Crispo, ``Internet of things: A survey on the
  security of iot frameworks,'' \emph{Journal of Information Security and
  Applications}, vol.~38, pp. 8 -- 27, 2018. [Online]. Available:
  \url{http://www.sciencedirect.com/science/article/pii/S2214212617302934}
\BIBentrySTDinterwordspacing

\bibitem{61}
C.~Sun, A.~Shrivastava, S.~Singh, and A.~Gupta, ``Revisiting unreasonable
  effectiveness of data in deep learning era,'' in \emph{2017 IEEE
  International Conference on Computer Vision (ICCV)}, Oct 2017, pp. 843--852.

\bibitem{GDPR}
P.~Voigt and A.~von~dem Bussche, \emph{The EU General Data Protection
  Regulation (GDPR)}.\hskip 1em plus 0.5em minus 0.4em\relax Springer, 2017.

\bibitem{115}
P.~Mohassel and Y.~Zhang, ``Secureml: A system for scalable privacy-preserving
  machine learning,'' in \emph{2017 IEEE Symposium on Security and Privacy
  (SP)}, May 2017, pp. 19--38.

\bibitem{FederatedLearning}
J.~Konečný, B.~McMahan, F.~X. Yu, and D.~B. Peter~Richtárik, Ananda
  Theertha~Suresh, ``Federated learning: Strategies for improving communication
  efficiency,'' \emph{CoRR}, vol. abs/1610.05492, 2016.

\bibitem{40}
M.~Abadi, A.~Chu, I.~Goodfellow, H.~B. McMahan, I.~Mironov, K.~Talwar, and
  L.~Zhang, ``Deep learning with differential privacy,'' in \emph{Proceedings
  of the 2016 ACM SIGSAC Conference on Computer and Communications Security},
  ser. CCS '16.\hskip 1em plus 0.5em minus 0.4em\relax New York, NY, USA: ACM,
  2016, pp. 308--318.

\bibitem{112}
J.~Vaidya, M.~Kantarcioglu, and C.~Clifton, ``Privacy-preserving naive bayes
  classification,'' \emph{The VLDB Journal}, vol.~17, no.~4, pp. 879--898, Jul.
  2008.

\bibitem{8}
\BIBentryALTinterwordspacing
F.-J. Gonzlez-Serrano, n.~Navia-Vzquez, and A.~Amor-Martn, ``Training support
  vector machines with privacy-protected data,'' \emph{Pattern Recogn.},
  vol.~72, pp. 93--107, Dec. 2017. [Online]. Available:
  \url{https://doi.org/10.1016/j.patcog.2017.06.016}
\BIBentrySTDinterwordspacing

\bibitem{SALTDSC2020}
T.~Li, J.~Li, X.~Chen, Z.~Liu, W.~Lou, and T.~Hou, ``Npmml: A framework for
  non-interactive privacy-preserving multi-party machine learning,'' \emph{IEEE
  Transactions on Dependable and Secure Computing}, 2020.

\bibitem{PrivFL}
K.~Mandal and G.~Gong, ``Privfl: Practical privacy-preserving federated
  regressions on high-dimensional data over mobile networks,'' in
  \emph{Proceedings of the 2019 ACM SIGSAC Conference on Cloud Computing
  Security Workshop}, 2019, pp. 57--68.

\bibitem{28}
Y.~Rahulamathavan, R.~C.~W. Phan, S.~Veluru, K.~Cumanan, and M.~Rajarajan,
  ``Privacy-preserving multi-class support vector machine for outsourcing the
  data classification in cloud,'' \emph{IEEE Transactions on Dependable and
  Secure Computing}, vol.~11, no.~5, pp. 467--479, Sept 2014.

\bibitem{26}
T.~Graepel, K.~Lauter, and M.~Naehrig, ``Ml confidential: Machine learning on
  encrypted data,'' in \emph{Information Security and Cryptology -- ICISC
  2012}, T.~Kwon, M.-K. Lee, and D.~Kwon, Eds.\hskip 1em plus 0.5em minus
  0.4em\relax Berlin, Heidelberg: Springer Berlin Heidelberg, 2013, pp. 1--21.

\bibitem{30}
X.~Liu, R.~Lu, J.~Ma, L.~Chen, and B.~Qin, ``Privacy-preserving patient-centric
  clinical decision support system on naive bayesian classification,''
  \emph{IEEE Journal of Biomedical and Health Informatics}, vol.~20, no.~2, pp.
  655--668, March 2016.

\bibitem{31}
M.~Upmanyu, A.~M. Namboodiri, K.~Srinathan, and C.~V. Jawahar, ``Efficient
  privacy preserving k-means clustering,'' in \emph{Intelligence and Security
  Informatics}, H.~Chen, M.~Chau, S.-h. Li, S.~Urs, S.~Srinivasa, and G.~A.
  Wang, Eds.\hskip 1em plus 0.5em minus 0.4em\relax Berlin, Heidelberg:
  Springer Berlin Heidelberg, 2010, pp. 154--166.

\bibitem{aggregationIoT}
T.~Li, C.~Gao, L.~Jiang, W.~Pedrycz, and J.~Shen, ``Publicly verifiable
  privacy-preserving aggregation and its application in iot,'' \emph{Journal of
  Network and Computer Applications}, vol. 126, pp. 39--44, 2019.

\bibitem{aggregationsmartgrid}
S.~Li, K.~Xue, Q.~Yang, and P.~Hong, ``Ppma: Privacy-preserving multisubset
  data aggregation in smart grid,'' \emph{IEEE Transactions on Industrial
  Informatics}, vol.~14, no.~2, pp. 462--471, 2017.

\bibitem{aggregationsensingsystems}
H.~Jin, L.~Su, H.~Xiao, and K.~Nahrstedt, ``Inception: Incentivizing
  privacy-preserving data aggregation for mobile crowd sensing systems,'' in
  \emph{Proceedings of the 17th ACM International Symposium on Mobile Ad Hoc
  Networking and Computing}, 2016, pp. 341--350.

\bibitem{DPvsML}
N.~G. {Paterakis}, E.~{Mocanu}, M.~{Gibescu}, B.~{Stappers}, and W.~{van Alst},
  ``Deep learning versus traditional machine learning methods for aggregated
  energy demand prediction,'' in \emph{2017 IEEE PES Innovative Smart Grid
  Technologies Conference Europe (ISGT-Europe)}, 2017, pp. 1--6.

\bibitem{DPvsML2}
N.~Prakash, A.~Manconi, and S.~Loew, ``Mapping landslides on eo data:
  Performance of deep learning models vs. traditional machine learning
  models,'' \emph{Remote Sensing}, vol.~12, no.~3, p. 346, 2020.

\bibitem{DPvsML3}
C.~{Stanik}, M.~{Haering}, and W.~{Maalej}, ``Classifying multilingual user
  feedback using traditional machine learning and deep learning,'' in
  \emph{2019 IEEE 27th International Requirements Engineering Conference
  Workshops (REW)}, 2019, pp. 220--226.

\bibitem{DPSVM}
R.~Girshick, J.~Donahue, T.~Darrell, and J.~Malik, ``Rich feature hierarchies
  for accurate object detection and semantic segmentation,'' in
  \emph{Proceedings of the IEEE conference on computer vision and pattern
  recognition}, 2014, pp. 580--587.

\bibitem{DPSVM2}
G.~Ch{\'e}ron, I.~Laptev, and C.~Schmid, ``P-cnn: Pose-based cnn features for
  action recognition,'' in \emph{Proceedings of the IEEE international
  conference on computer vision}, 2015, pp. 3218--3226.

\bibitem{58}
W.~Wang, C.-M. Vong, Y.~Yang, and P.-K. Wong, ``Encrypted image classification
  based on multilayer extreme learning machine,'' \emph{Multidimensional Syst.
  Signal Process.}, vol.~28, no.~3, pp. 851--865, Jul. 2017.

\bibitem{20}
M.~D. Cock, R.~Dowsley, C.~Horst, R.~Katti, A.~Nascimento, S.~Truex, and W.-S.
  Poon, ``Efficient and private scoring of decision trees, support vector
  machines and logistic regression models based on pre-computation,''
  \emph{IEEE Transactions on Dependable and Secure Computing}, pp. 1--1, 2017.

\bibitem{2}
R.~Bost, R.~A. Popa, S.~Tu, and S.~Goldwasser, ``Machine learning
  classification over encrypted data,'' in \emph{Network and Distributed System
  Security Symposium}, 2014.

\bibitem{41}
\BIBentryALTinterwordspacing
R.~Shokri and V.~Shmatikov, ``Privacy-preserving deep learning,'' in
  \emph{Proceedings of the 22Nd ACM SIGSAC Conference on Computer and
  Communications Security}, ser. CCS '15.\hskip 1em plus 0.5em minus
  0.4em\relax New York, NY, USA: ACM, 2015, pp. 1310--1321. [Online].
  Available: \url{http://doi.acm.org/10.1145/2810103.2813687}
\BIBentrySTDinterwordspacing

\bibitem{relatedwork-MCS-DPtrainig}
G.~Jagannathan, K.~Pillaipakkamnatt, and R.~Wright, ``A practical
  differentially private random decision tree classifier,'' in \emph{2009 IEEE
  International Conference on Data Mining Workshops}, Dec 2009, pp. 114--121.

\bibitem{105}
Y.~Aono, T.~Hayashi, L.~T. Phong, and L.~Wang, ``Input and output
  privacy-preserving linear regression,'' \emph{IEICE TRANSACTIONS on
  Information and Systems}, vol. 100, no.~10, pp. 2339--2347, 2017.

\bibitem{36}
Y.~AONO, T.~HAYASHI, L.~T. PHONG, and L.~WANG, ``Privacy-preserving logistic
  regression with distributed data sources via homomorphic encryption,''
  \emph{IEICE Transactions on Information and Systems}, vol. E99.D, no.~8, pp.
  2079--2089, 2016.

\bibitem{9}
M.~d. Cock, R.~Dowsley, A.~C. Nascimento, and S.~C. Newman, ``Fast, privacy
  preserving linear regression over distributed datasets based on
  pre-distributed data,'' in \emph{Proceedings of the 8th ACM Workshop on
  Artificial Intelligence and Security}, ser. AISec '15.\hskip 1em plus 0.5em
  minus 0.4em\relax New York, NY, USA: ACM, 2015, pp. 3--14.

\bibitem{Introduction-aggregator-scenario-1}
E.~Shi, H.~Chan, E.~Rieffel, R.~Chow, and D.~Song, ``Privacy-preserving
  aggregation of time-series data,'' in \emph{Annual Network \& Distributed
  System Security Symposium (NDSS)}.\hskip 1em plus 0.5em minus 0.4em\relax
  Internet Society., 2011.

\bibitem{Introduction-aggregator-scenario-3}
G.~Acs and C.~Castelluccia, ``I have a dream! (differentially private smart
  metering),'' in \emph{Information Hiding}.\hskip 1em plus 0.5em minus
  0.4em\relax Berlin, Heidelberg: Springer Berlin Heidelberg, 2011, pp.
  118--132.

\bibitem{Relatedwork-aggregator-scenario-ccs17}
\BIBentryALTinterwordspacing
K.~Bonawitz, V.~Ivanov, B.~Kreuter, A.~Marcedone, H.~B. McMahan, S.~Patel,
  D.~Ramage, A.~Segal, and K.~Seth, ``Practical secure aggregation for
  privacy-preserving machine learning,'' in \emph{Proceedings of the 2017 ACM
  SIGSAC Conference on Computer and Communications Security}, ser. CCS
  '17.\hskip 1em plus 0.5em minus 0.4em\relax New York, NY, USA: ACM, 2017, pp.
  1175--1191. [Online]. Available:
  \url{http://doi.acm.org/10.1145/3133956.3133982}
\BIBentrySTDinterwordspacing

\bibitem{Relatedwork-aggregator-scenario-quadraticoptimization}
Y.~Shoukry, K.~Gatsis, A.~Alanwar, G.~Pappas, S.~Seshia, M.~Srivastava, and
  P.~Tabuada, ``Privacy-aware quadratic optimization using partially
  homomorphic encryption,'' in \emph{2016 IEEE 55th Conference on Decision and
  Control (CDC)}, Dec 2016, pp. 5053--5058.

\bibitem{Relatedwork-aggregator-scenario-SASkmeans}
D.~Mittal, D.~Kaur, and A.~Aggarwal, ``Secure data mining in cloud using
  homomorphic encryption,'' in \emph{2014 IEEE International Conference on
  Cloud Computing in Emerging Markets (CCEM)}, Oct 2014, pp. 1--7.

\bibitem{1}
J.~Katz and Y.~Lindell, \emph{Introduction to modern cryptography}, ser. CRC
  Cryptography and Network Security Series.\hskip 1em plus 0.5em minus
  0.4em\relax CRC press, 2014.

\bibitem{cloud-RSA}
K.~El~Makkaoui, A.~Ezzati, and A.~Beni-Hssane, ``Cloud-rsa: An enhanced
  homomorphic encryption scheme,'' in \emph{Europe and MENA Cooperation
  Advances in Information and Communication Technologies}.\hskip 1em plus 0.5em
  minus 0.4em\relax Cham: Springer International Publishing, 2017, pp.
  471--480.

\bibitem{15}
\BIBentryALTinterwordspacing
R.~Canetti, ``Security and composition of multiparty cryptographic protocols,''
  \emph{Journal of Cryptology}, vol.~13, no.~1, pp. 143--202, Jan 2000.
  [Online]. Available: \url{https://doi.org/10.1007/s001459910006}
\BIBentrySTDinterwordspacing

\bibitem{Introduction-18sec-GarbledCircuit}
\BIBentryALTinterwordspacing
C.~Juvekar, V.~Vaikuntanathan, and A.~Chandrakasan, ``Gazelle: A low latency
  framework for secure neural network inference,'' in \emph{27th USENIX
  Security Symposium (USENIX Security 18)}.\hskip 1em plus 0.5em minus
  0.4em\relax Baltimore, MD: {USENIX} Association, 2018, pp. 1651--1669.
  [Online]. Available:
  \url{https://www.usenix.org/conference/usenixsecurity18/presentation/juvekar}
\BIBentrySTDinterwordspacing

\bibitem{104}
Z.~Zhang, B.~I. Rubinstein, C.~Dimitrakakis \emph{et~al.}, ``On the
  differential privacy of bayesian inference.'' in \emph{AAAI}, 2016, pp.
  2365--2371.

\bibitem{Introduction-FHE-survey}
\BIBentryALTinterwordspacing
P.~Martins, L.~Sousa, and A.~Mariano, ``A survey on fully homomorphic
  encryption: An engineering perspective,'' \emph{ACM Comput. Surv.}, vol.~50,
  no.~6, pp. 83:1--83:33, Dec. 2017. [Online]. Available:
  \url{http://doi.acm.org/10.1145/3124441}
\BIBentrySTDinterwordspacing

\bibitem{46}
Y.~Aono, T.~Hayashi, L.~Trieu~Phong, and L.~Wang, ``Scalable and secure
  logistic regression via homomorphic encryption,'' in \emph{Proceedings of the
  Sixth ACM Conference on Data and Application Security and Privacy}, ser.
  CODASPY '16.\hskip 1em plus 0.5em minus 0.4em\relax New York, NY, USA: ACM,
  2016, pp. 142--144.

\bibitem{123}
Y.~Aono, T.~Hayashi, L.~Wang, S.~Moriai \emph{et~al.}, ``Privacy-preserving
  deep learning via additively homomorphic encryption,'' \emph{IEEE
  Transactions on Information Forensics and Security}, vol.~13, no.~5, pp.
  1333--1345, 2017.

\bibitem{10}
O.~Goldreich, \emph{Foundations of cryptography: volume 2, basic
  applications}.\hskip 1em plus 0.5em minus 0.4em\relax Cambridge university
  press, 2009.

\bibitem{dpbased2}
E.~Roth, D.~Noble, B.~H. Falk, and A.~Haeberlen, ``Honeycrisp: large-scale
  differentially private aggregation without a trusted core,'' in
  \emph{Proceedings of the 27th ACM Symposium on Operating Systems Principles},
  2019, pp. 196--210.

\bibitem{dpbased1}
M.~Abadi, A.~Chu, I.~Goodfellow, H.~B. McMahan, I.~Mironov, K.~Talwar, and
  L.~Zhang, ``Deep learning with differential privacy,'' in \emph{Proceedings
  of the 2016 ACM SIGSAC Conference on Computer and Communications Security},
  2016, pp. 308--318.

\bibitem{RSA-attack-1}
A.~Nitaj and T.~Rachidi, ``Factoring rsa moduli with weak prime factors,'' in
  \emph{Codes, Cryptology, and Information Security}, S.~El~Hajji, A.~Nitaj,
  C.~Carlet, and E.~M. Souidi, Eds.\hskip 1em plus 0.5em minus 0.4em\relax
  Cham: Springer International Publishing, 2015, pp. 361--374.

\bibitem{RSA-attack-2}
D.~Boneh, ``Twenty years of attacks on the rsa cryptosystem,'' \emph{NOTICES OF
  THE AMS}, vol.~46, pp. 203--213, 1999.

\bibitem{SGDSVM}
\BIBentryALTinterwordspacing
Z.~Wang, K.~Crammer, and S.~Vucetic, ``Breaking the curse of kernelization:
  Budgeted stochastic gradient descent for large-scale svm training,'' \emph{J.
  Mach. Learn. Res.}, vol.~13, no.~1, pp. 3103--3131, Oct. 2012. [Online].
  Available: \url{http://dl.acm.org/citation.cfm?id=2503308.2503341}
\BIBentrySTDinterwordspacing

\bibitem{113}
T.~Li, J.~Li, Z.~Liu, P.~Li, and C.~Jia, ``Differentially private naive bayes
  learning over multiple data sources,'' \emph{Information Sciences}, vol. 444,
  pp. 89 -- 104, 2018.

\end{thebibliography}

\begin{IEEEbiographynophoto}
{Liehuang Zhu} is a professor in the School of Computer Science, Beijing Institute of Technology.
He is selected into the Program for New Century Excellent Talents in University from Ministry of Education, P.R. China.
His research interests include Internet of Things, Cloud Computing Security, Internet and Mobile Security.
\end{IEEEbiographynophoto}

\begin{IEEEbiographynophoto}
{Xiangyun Tang} received the B.Eng degree in computer science from Minzu University of China, Beijing, China in 2016.
Currently she is a Ph.D student in the Department of Computer Science, Beijing Institute of Technology.
Her research interests include Differential Privacy and Secure Multi-party Computation.
\end{IEEEbiographynophoto}

\begin{IEEEbiographynophoto}
{Meng Shen} received the B.Eng degree from Shandong University, Jinan, China in 2009, and the Ph.D degree from Tsinghua University, Beijing, China in 2014, both in computer science. Currently he serves in Beijing Institute of Technology, Beijing, China, as an associate professor. His research interests include privacy protection for cloud and IoT, blockchain applications, and encrypted traffic classification.
He received the Best Paper Runner-Up Award at IEEE IPCCC 2014.
He is a member of the IEEE.
\end{IEEEbiographynophoto}

\begin{IEEEbiographynophoto}
{Jie Zhang} received the B.Eng degree in computer science from China University of Mining and Technology, Jiangsu, China in 2018.
Currently he is a master student in the Department of Computer Science, Beijing Institute of Technology.
His research interests include blockchain applications and machine learning privacy.
\end{IEEEbiographynophoto}

\begin{IEEEbiographynophoto}
{Xiaojiang Du} (S'99-M'03-SM'09-F'20) is a tenured professor in the Department of Computer and Information Sciences at Temple University, Philadelphia, USA. Dr. Du received his B.S. and M.S. degree in electrical engineering from Tsinghua University, Beijing, China in 1996 and 1998, respectively. He received his M.S. and Ph.D. degree in electrical engineering from the University of Maryland College Park in 2002 and 2003, respectively. His research interests are wireless communications, wireless networks, security, and systems. He has authored over 400 journal and conference papers in these areas, as well as a book published by Springer. Dr. Du has been awarded more than \$5 million US dollars research grants from the US National Science Foundation (NSF), Army Research Office, Air Force, NASA, the State of Pennsylvania, and Amazon. He won the best paper award at IEEE GLOBECOM 2014 and the best poster runner-up award at the ACM MobiHoc 2014. He serves on the editorial boards of three international journals. Dr. Du is a Fellow of IEEE and a Life Member of ACM.
\end{IEEEbiographynophoto}

\vfill




\end{document}